\documentclass[twocolumn ,amssymb,amsmath,nobibnotes,aps,nofootinbib,floatfix]{revtex4}
\usepackage[latin1]{inputenc}
\usepackage{graphics}
\usepackage{amsmath}
\usepackage{dcolumn}
\usepackage{amssymb}
\usepackage{times}
\usepackage{graphicx}
\usepackage{amsmath}
\usepackage{color}
\usepackage{latexsym}
\usepackage{slashed}
\usepackage{braket}
\usepackage{hyperref}
\usepackage{soul}

\newcommand{\me}{\mathrm{e}} 
\newcommand{\mi}{\mathrm{i}} 
\newcommand{\dd}{\ensuremath{\mathrm{d}}}
\newcommand{\bea}{\begin{eqnarray}}
\newcommand{\eea}{\end{eqnarray}}
\newcommand{\p}{\partial}

\newcommand{\eqn}[1]{Eq.~(\ref{#1})}

\newcommand{\dfd}[2]{\ensuremath{\frac{\mathrm{d}^{#2}#1}{(2\pi)^{#2}}}\,}

\newcommand{\nn}{\nonumber}
\newcommand{\be}{\begin{equation}}
\newcommand{\ee}{\end{equation}}
\newcommand{\bc}{\begin{center}}
\newcommand{\ec}{\end{center}}

\begin{document}
\title{Non-perturbative field theoretical aspects of graphene and related systems
\vspace{-6pt}}
\author{Juan Angel Casimiro Olivares}
\address{Instituto de F\'isica y Matem\'aticas, Universidad
Michoacana de San Nicol\'as de Hidalgo, Edificio C-3, Ciudad
Universitaria, Morelia, Michoac\'an 58040, M\'exico.}
\author{Ana Julia Mizher}
\address{
Instituto de F\'{\i}sica Te\'orica,  Estadual Paulista,
Rua Dr. Bento Teobaldo Ferraz, 271 - Bloco II, 01140-070 S\~ao Paulo, SP, Brazil}
\address{Centro de Ciencias Exactas, Universidad del B\'{\i}o-B\'{\i}o,\\ Avda. Andr\'es Bello 720, Casilla 447, 3800708, Chill\'an, Chile.}
\author{Alfredo Raya}
\address{Instituto de F\'isica y Matem\'aticas, Universidad
Michoacana de San Nicol\'as de Hidalgo, Edificio C-3, Ciudad
Universitaria, Morelia, Michoac\'an 58040, M\'exico}
\address{Centro de Ciencias Exactas, Universidad del B\'{\i}o-B\'{\i}o,\\ Avda. Andr\'es Bello 720, Casilla 447, 3800708, Chill\'an, Chile.}

\begin{abstract}
\vspace{1em} In this article, we review the dynamics of charge carriers in  graphene and related 2D systems from a quantum field theoretical point of view. By allowing the  electromagnetic fields to propagate throughout space and constraining fermions to move on a 2D manifold, the effective theory of such systems becomes a non-local version of quantum electrodynamics (QED) dubbed in literature pseudo or reduced QED. We review some aspects of the theory assuming the coupling arbitrary in strength. In particular, we focus on the  chiral symmetry breaking scenarios and the analytical structure of the fermion propagator in vacuum and under the influence of external agents like a heat bath, in the presence of a Chern-Simons term, anisotropy and in curved space. We briefly discuss the major advances and some new results on this field.\\

En este art\'{\i}culo revisamos la din\'amica de los portadores de carga en grafeno y otros sistemas 2D relacionados desde un punto de vista de la teor\'{\i}a de campos cu\'anticos. Permitiendo que los campos electromagn\'eticos se propaguen en el espacio, pero restringiendo a los fermiones a moverse en una variedad bidimensional, la teor\'{\i}a efectiva para estos sistemas se vuelve una versi\'on no local de la electrodin\'amica cu\'antica (QED por sus siglas en ingl\'es) llamada en la literatura pseudo QED o QED reducida. Hacemos un recuento de algunos aspectos de la teor\'{\i}a asumiento un acomplamiento de intensidad arbitraria. En particular, nos enfocamos en los escenarios de rompimiento din\'amico de la simetr\'{\i}a quiral y la estructura anal\'{\i}tica del propagador del fermi\'on en el vac\'{\i}o y bajo la influencia de agentes externos como un ba\~no t\'ermico, en la presencia de un t\'ermino de Chern-Simons, anisotrop\'{\i}a y en espacio curvo. Discutimos brevemente los mayores avances y algunos nuevos resultados en este campo.
\vspace{1em}

\end{abstract}

\maketitle

\section{Introduction}

Two dimensional relativistic Dirac fermions have been the subject of active research for many decades. These are by no means merely flat cousins of quarks and leptons on the high energy physics realm, but rather these particles offer an opportunity to explore vast and intriguing phenomenology in other branches of physics as well (see, for instance, Ref.~\cite{Miransky:2015ava}). The pioneer work of Wallace~\cite{Wallace} on the band structure of graphene paved the way to consider relativistic Dirac fermions in solid state physics realms. The realization of the quantum anomaly emerging from the electromagnetic dynamics of Dirac fermions and {\em photons} in two spatial dimensions was first simulated in the seminal work of Semenoff~\cite{Semenoff} under condensed-matter considerations, whereas one the first glimpses into topological matter was achieved from the dynamics of relativistic Dirac excitations in a Quantum Hall set up without Landau levels, where the role of parity anomaly is seen in the quantization of the conductivity as discussed by Haldane~\cite{Haldane:1988zza}. 
High temperature superconductivity in layered cuprates~\cite{supercond1,supercond2,supercond3,supercond4,supercond5}  has also been naturally explained in therms of the dynamics of planar Dirac fermions in these systems.  
All of these ideas have highlighted the intricacies and interesting features of the interactions among Dirac particles constrained to move on a plane. But the first isolation of graphene membranes by the Cambridge group~\cite{graphene1} and others~\cite{graphene3} with the overwhelming evidence of the {\em relativistic} nature of charge carriers in this material~\cite{relativistic} have indeed boosted the interest toward these systems.  

\begin{figure}
\includegraphics*[scale=0.3]{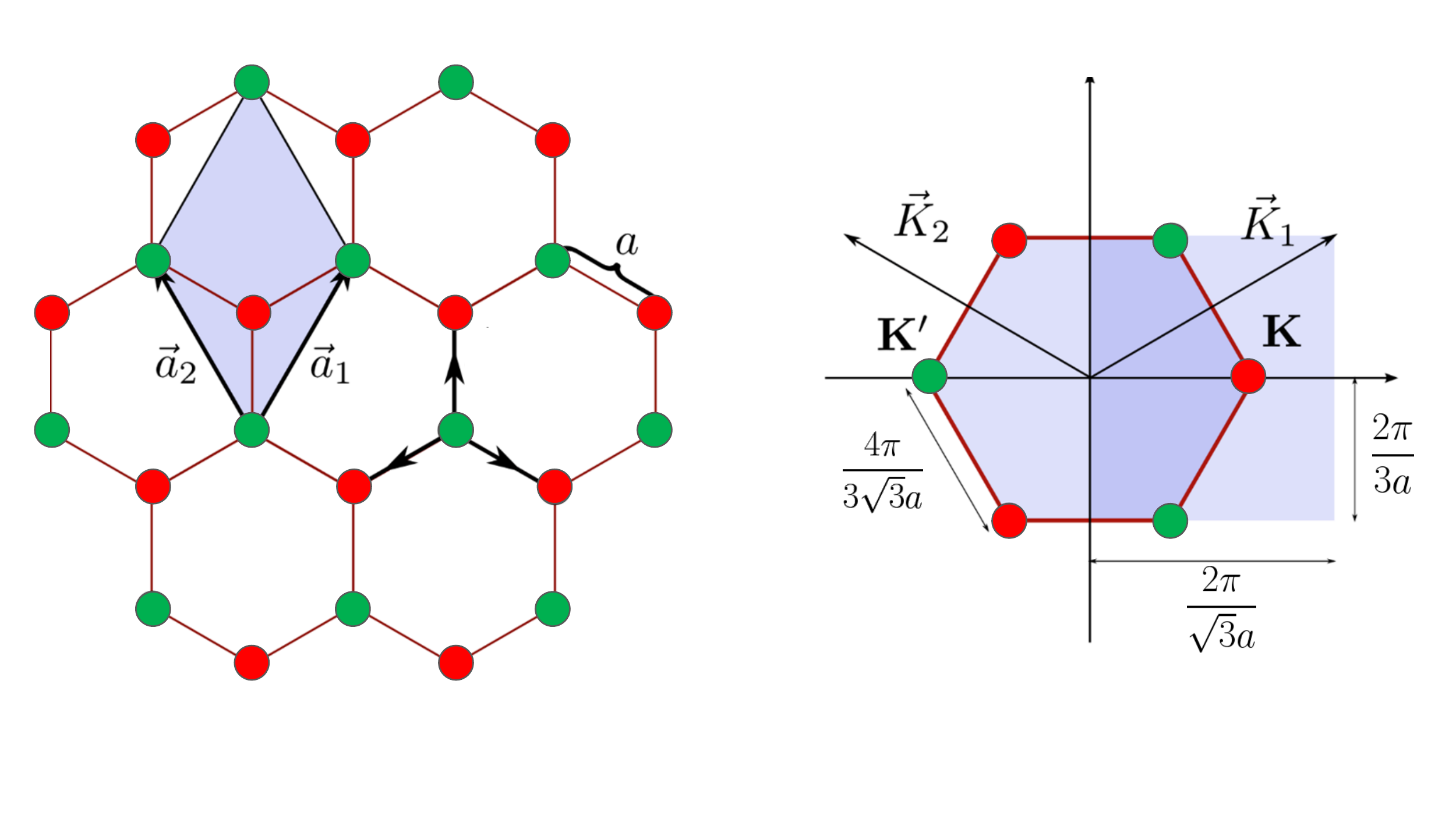}
\caption{Crystallographic structure of the honeycomb array. In the left panel, the two overlapping triangular lattices are represented by the red and green points. The primitive vectors $\vec{a}_1$ and $\vec{a}_2$ connecting all points in the crystal structure are shown, and the unit cell of the honeycomb array highlighted in blue. In the right panel, the first Brillouin zone is shown. The reciprocal lattice also shows a hexagonal structure. ${\bf K}$ and ${\bf K'}$ represent the inequivalent Dirac points. }
\label{fig:graphenelattice}
\end{figure}

The {\em relativistic} behavior of this material is rooted in the organization of the carbon atoms in honeycomb lattices (see, for instance, \cite{rise}). This structure is conveniently represented by two overlapped triangular sublattices (with a bipartite unit cell), as represented in Fig.\ref{fig:graphenelattice}

\vspace{10pt}

\noindent
The index labelling these sublattices constitute a degree of freedom for the charge carriers. In the tight-binding approach, electrons are firmly bounded to the atom except for one electron per atom belonging to the $p$ orbital. Considering that the charge carrier can only hop to the nearest neighbor (which means that only inter-lattice hopping is allowed) 
one gets an expression for the energy that vanishes in six points, two of them non-equivalent. Those are called the Dirac points ${\bf K}$ and ${\bf K'}$,  located precisely where the valence band touches the conduction band and where the Fermi level is found. One can easily verify that in the vicinity of each of the Dirac points, the dispersion relation becomes linear, a feature typically associated to relativistic fermions. Besides that, the operators associated to the sublattice degree of freedom acquire a Pauli matrices structure in such a way that the system can be represented by a Dirac equation, where the speed of the light is replaced by the Fermi velocity $v_F$, which turns out to be $v_F\simeq 1\times 10^{6}{\rm m/s}$~\cite{rise}. For this reason, an extensive work to characterize and explore graphene has been developed in the framework of quantum field theory (QFT). 
Moreover, the analytic structure of the Dirac equation unveils the fact that electrons in graphene have a chirality degree of freedom, which is another feature associated to relativistic particles and which has important implications for the electronic transport phenomena in this material.

Although in graphene the charge carriers are constrained to the plane, the external electromagnetic fields fermions may interact with are not. Because of that, usual Quantum Electrodynamics fully defined in (2+1)D (QED$_3$), does not provide a suitable description. Among other inaccuracies, QED$_3$ yields to a logarithmic static interaction (see, for instance, Ref.~\cite{craig}) rather than the expected Coulomb interaction \cite{Marino,Marino:1992xi,Gonzalez:1993uz} between electrons in graphene. Instead of that, in a QFT approach, electromagnetic fields must be represented by Abelian gauge boson fields in (3+1)D and the interacting theory therefore must deal with the subtlety of combining particles moving in different dimensionality. 

Mixed dimensional gauge theories have been proposed in the nineties aiming to explain the then recent discovery of the Quantum Hall effect \cite{QHE_tong}. After that, several aspects of these theories were developed, in both formal and applied contexts. The first appearance of theories that provide a suitable background to explore fermions in two-space dimension interacting with gauge fields in three-space dimension occurred almost simultaneously in \cite{Marino:1992xi} and \cite{Gonzalez:1993uz}. The procedure adopted in both approaches was to perform a dimensional reduction of the gauge field, integrating out its third spatial component, obtaining an effective interaction in $(2+1)$D. Among other features, this dimensional reduction softens the infrared behavior of the gauge boson propagator, which behaves as $\sim 1/q$ rather than the usual pole-dependence $1/q^2$. In \cite{Marino:1992xi} the authors obtain an effective Lagrangian in $(2+1)$D which for all intents and purposes corresponds to the mixed dimensional theory described above, naming the theory pseudo-quantum Electrodynamics (PQED). In particular, the authors focus on tracing a correspondence between PQED and the Chern-Simons theory. Further formal aspects of PQED were developed in \cite{doAmaral:1992td,Marino:2014oba}. On the other side, in \cite{Gonzalez:1993uz}, the discussion is focused  on the renormalization group aspects of graphene, and it is shown that the mixed dimensional theory possess a fixed point when the Fermi velocity evolves to the speed of light. The renormalization group analysis of this theory applied to graphene was posteriorly extended in \cite{RG2,RG3,RG4,Vozmediano:2010fz}. Recently, renormalization group techniques applied to PQED could successfully describe experimental data on the renormalization of the band gap in other two-dimensional materials like diselenide (WSe$_2$) and molybdenumm
disulfide (MoS$_2$) \cite{Fernandez:2020wne}.

A more general procedure that could also be applied to systems where the charged particles are constrained to a lower dimensional string was proposed a few years later in \cite{Gorbar:2001qt}. The theory resulting from this method was named reduced-quantum electrodynamics (RQED). Although the framework developed in \cite{Gorbar:2001qt} could be used to construct theories in arbitrary dimensions for the fermions $d_e$ and gauge bosons $d_\gamma$ (which then allow to label the extended theory as RQED$_{d_\gamma,d_e}$), the major goal in \cite{Gorbar:2001qt} was to analyse the dynamical chiral symmetry breaking in systems where two-spatially dimensional fermions interact with three-spatially dimensional gauge fields, focusing on condensed matter systems~\footnote{We refer to this scenario throughout the work simply as RQED.}. For this purpose, the authors solve the Schwinger-Dyson equations within an improved rainbow-ladder approximation and found that the dynamics of chiral symmetry breaking is rich and nontrivial. Posteriorly, the renormalization of RQED to one- and two-loops was deeply investigated \cite{Teber:2012de,Kotikov:2013eha,Teber:2014hna,Teber:2018goo}. The scale invariance obtained in previous approaches was confirmed in the context of RQED.

Following \cite{Gorbar:2001qt}, some activity involving the chiral symmetry breaking of dimensionally reduced theories took place. In \cite{Alves:2013bna} the Schwinger-Dyson formalism within the rainbow approximation was applied to conclude that there is a critical coupling $\alpha=\pi/8$ (in the conventions adopted in this work) above which there is room for dynamical mass generation in the theory. The existence of a critical coupling was confirmed in \cite{Kotikov:2016yrn}. Additionally, a critical number of fermion families was determined from the similar structure of the gap equation in QED$_3$ and RQED.

Chiral symmetry breaking was also explored in different external conditions: in \cite{Nascimento:2015ola,Baez:2020dbe} the dynamical symmetry breaking was studied at finite temperature. In \cite{Olivares:2020eko, CarringtonRQED,Magalhaes:2020nlc} the dynamical mass generation was studied in RQED and PQED coupled to a Chern-Simons term. In \cite{Carrington:2020qfz}, anisotropy associated to strained graphene was shown to slightly affect the critical coupling. Moreover, in \cite{RQEDcurved} perturbative aspects of RQED in curved space was analysed.

Additionally, analysis of the Landau-Khalatnikov-Fradkin transformations (LKFT) \cite{Landau:1955zz,Fradkin:1955jr} was performed in RQED. These are non-perturbative transformations that connect the Green's functions in different gauges and give valuable information about the renormalization coefficients in multi-loop calculations. Using the knowledge previously obtained in QED$_4$ and QED$_3$ \cite{Bashir:2000,Bashir:2002sp,Bashir:2004rg} the LKFT for the fermion propagator in RQED were derived in \cite{AftabLKFT}.

In this manuscript we review the foundations of Pseudo- and Reduced-QED and focus on several aspects of dynamical chiral symmetry breaking explored via Schwinger-Dyson equations. In section \ref{sec:historical} we deduce the Lagrangian of the theory reviewing formal development as locality, unitarity and scale invariance. In section \ref{sec:CSB} we review the formalism of Schwinger-Dyson equations, analyse general aspects of the gap equations, deduce the gap equations for RQED in the vacuum, analysing the wave function renormalization and  renormalization group results. Section \ref{sec:medium} is dedicated to study chiral symmetry breaking in a medium, including finite temperature effects, interaction with a Chern-Simons term and possible effects of curved space. In section \ref{sec:LKF} we discuss how  LKF transformations can provide information about renormalizability of the theory. Finally in section \ref{sec:outlook} we present an outlook of the subjects discussed in this work.

\section{Reduced or Pseudo QED: A historical recount}
\label{sec:historical}
\subsection{The Pseudo- or Reduced-QED Lagrangian}

The derivation of PQED in \cite{Marino:1992xi} and RQED in \cite{Gorbar:2001qt} follow a similar procedure, consisting of performing a dimensional reduction of standard QED, imposing the matter fields to be constrained to a lower dimensional space.  Although the general case, where an Abelian gauge field in arbitrary dimension interacts with fermions constrained to a generic manifold, has been explored in \cite{Gorbar:2001qt,Teber:2012de}, here we focus on the particular situation where four-dimensional QED is reduced to a three-dimensional theory that represents fermions in a plane. 

The starting point is the usual QED$_4$ Lagrangian, 
\be 
\mathcal{L}= -\frac{1}{4}F^{\mu\nu}F_{\mu\nu} - e j^\mu A_\mu + \mathcal{L}_M + \mathcal{L}_{GF},
\label{eq:4Dlagrangian}
\ee 
where the first term is the Maxwell term, $j^\mu$ is a matter current that couples to the gauge field $A$, $e$ is the electric charge, ${\cal L}_M$ is a generic matter kinetic lagrangian and ${\cal L}_{GF}$ is a gauge fixing term.

In order to describe matter confined to a plane, the current must be defined in such a way that there is no dynamics in the third spatial coordinate,
\begin{eqnarray}
j^\mu (\bf x) =\left\{ \begin{matrix} j^\mu (x^0,x^1,x^2)\delta(x^3) \ \ \ \mu&=&0,1,2 \\
0 \hspace{3cm}  \mu&=&\hspace{-0.6cm}3
\end{matrix} \right.
\label{eq:current}
\end{eqnarray}
An effective interaction can be obtained integrating Eq.(\ref{eq:4Dlagrangian}) over the gauge field $A_\mu$. This results in the generating functional (in Euclidean space)
\begin{eqnarray} \nonumber
Z_{eff}&=&{\rm exp}\left[\frac{e^2}{2}\int d^4 x \ d^4 x' j^\mu_{3+1}(x) G^{\mu\nu} (x-x') j^\nu_{3+1}(x')\right]\\ \nonumber
&=& {\rm exp}\left[\frac{e^2}{2}\int d^4 x \ d^4 x' j^\mu_{3+1}(x) \frac{1}{-\square} j^\nu_{3+1}(x')\right]\\
&\equiv& {\rm exp}\left[ -S_{eff}(j_{3+1}^\mu)\right],
\end{eqnarray}
where $\square$ denotes the d'Alambertian operator. 

Considering the definition in Eq.(\ref{eq:current}) and the usual photon propagator,
\be
G^{\mu\nu}= \left[-\square \delta^{\mu\nu} + \left(1-\frac{1}{\xi}\right)\partial^\mu \partial^\nu \right]\left[\frac{1}{(-\square)^2}\right],
\ee
the effective action can be written as
\be 
S_{eff}=-\frac{e^2}{2}\int d^3 x d^3 x' j^\mu (x) K_E(x-x') j^\mu (x'),
\label{eq:eff_action}
\ee 
where the (Euclidean) kernel is given by
\begin{eqnarray} \nonumber
K_E(x-x')&=&\int \frac{d^4 k}{(2\pi)^4}\frac{e^{i k\cdot (x-x')}}{k^2}\Bigg{|}_{x_3=x_3'=0}\\
&=&\frac{1}{8\pi^2 |x-x'|^2},
\label{eq:kernel}
\end{eqnarray}
and now $x$ and $x'$ are defined in three dimensions. 

It is possible to construct a theory fully defined in $(2+1)$D that mimics this effective action. This can be done noticing that the kernel in Eq.(\ref{eq:kernel}) can be written as a three-dimensional integral,
\begin{eqnarray} \nonumber
\frac{1}{8\pi^2|x-x'|^2}&=&\frac{1}{4}\int \frac{d^3k}{(2\pi)^2} \frac{e^{i(k\cdot (x-x'))}}{\sqrt{k^2}}\\
&\equiv& \frac{1}{4}\frac{1}{\sqrt{-\square_E}},
\end{eqnarray}
where here the d'Alambertian operator is defined in (2+1)D and the label $E$ denotes Euclidean space. This 
yields  the following effective action,
\be
S_{eff}=-\frac{e^2}{8}\int d^3x d^3x' j^\mu (x)\frac{1}{\sqrt{-\square_E}} j^\mu (x').
\label{eq:rqed_action}
\ee
It is straightforward to verify that the action in Eq.(\ref{eq:rqed_action}) can be obtained from the Lagrangian 
\be 
\mathcal{L}_{RQED} =-\frac{1}{4}F^{\mu\nu} \left[\frac{2}{\sqrt{\square}}\right]F_{\mu\nu}-e j^\mu A_\mu+\mathcal{L}_M+\mathcal{L}_{GF}.
\label{eq:Lagrangian}
\ee 
This is the Pseudo or Reduced-QED Lagrangian. 


A remarkable consequence of the dimensional reduction, is that the photon propagator of RQED becomes proportional to $1/q$ rather than the usual $1/q^2$ found in QED$_4$. This can be obtained writing the pure gauge sector of the Lagrangian Eq.(\ref{eq:Lagrangian}), including a proper gauge fixing term
\be
\mathcal{L}_G=-\frac{1}{2}F^{\mu\nu}\frac{1}{\sqrt{\Box}}F_{\mu\nu}+\frac{1}{2\xi}\left(\partial_a A^a\right)^2.
 \ee
Defining the most general form for the propagator,
\be 
\Delta^{\mu\nu}=a(k^2)g^{\mu\nu} + b(k^2)k^\mu k^\nu
\ee
it is possible to find the coefficients $a(k^2)$ and $b(k^2)$ (see e.g. \cite{Gorbar:2001qt}),
\be
\Delta^{\mu\nu}(q^2)=\frac{-i}{\sqrt{q^2}}\left[g^{\mu\nu}-\left(1-\xi\right)\frac{q^\mu q^\nu}{q^2}\right].
\label{eq:propagator}
\ee
Note that this propagator has a softer infrared behavior than the photon propagator in QED$_4$ and QED$_3$. We notice that several groups differ in conventions by a global factor of 1/2 in this propagator.

\begin{figure}
\includegraphics*[scale=0.6]{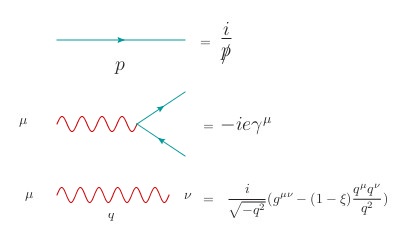}
\caption{Feynman rules for RQED or PQED.}
\label{fig:feynman_diag}
\end{figure}

The Feynman rules for RQED are represented in Fig.\ref{fig:feynman_diag}

\subsection{Coulomb interaction}

A crucial feature of this theory is that it correctly reproduces the Coulomb interaction expected for charged fermions, no matter if they are constrained to the plane. Naively considering full QED$_3$, where the gauge fields are also in lower dimension, yields to a logarithmic interaction that does not correspond to what is observed in graphene \cite{Marino,Marino:1992xi,Gonzalez:1993uz}.

This can be seen considering a pair of static point charges with associated current 
\begin{eqnarray}
j^\mu ({\bf r},t) =\left\{ \begin{matrix}\delta({\bf r}-x)+ \delta({\bf r}-y), \ \ \ \ \ \ \ \ \  \mu&=&\hspace{-0.3cm} 0 \\
0,  \hspace{3.5cm}\mu&=&1,2
\end{matrix} \right.
\label{eq:charges}
\end{eqnarray}
The interaction energy associated to the current defined above interacting with a gauge field is given by
\be 
E=\frac{e}{2}\int d^2r j^\mu({\bf r}) A_\mu({\bf r}).
\ee
Rewriting it in terms of the gauge field propagator,
\begin{eqnarray} 
E&=&\frac{e^2}{2}\int d^2 r d^2 r' dt' j^\mu({\bf r})\\
&&\Bigg[\int \frac{d^2 k}{(2\pi)^2}\int\frac{d\omega}{2\pi}e^{i{\bf k}\cdot({\bf r}-{\bf r'})-\omega(t-t')}G^{\mu\nu}(\omega,{\bf k})\Bigg]j^\nu({\bf r'}).\nonumber
\end{eqnarray}
Integrating over $t'$ and $\omega$ and taking $\mu=\nu=0$ to obtain the potential between static points,
\be 
E=\frac{e^2}{2}\int d^2 r d^2 r' \rho({\bf r})\left[\int \frac{d^2 k}{(2\pi)^2}e^{i{\bf k}\cdot({\bf r}-{\bf r'})}G^{00}(\omega=0,{\bf k})\right]\rho({\bf r'}).
\ee
Considering the current (\ref{eq:charges}) and inserting the propagator in Eq.(\ref{eq:propagator}), one obtains, apart from unphysical self-interaction terms, that
\be 
E=\frac{e^2}{4\pi|{\bf x}-{\bf y}|},
\ee
namely, the expected Coulomb interaction.

\subsection{Scale invariance}

In opposition to QED$_3$, in RQED the electromagnetic coupling $e^2$ is dimensionless, since it originates from the four-dimensional theory. This can also be seen doing a classical power counting of dimensions in Eq.(\ref{eq:Lagrangian}). The theory is therefore classically scale invariant.

In~\cite{Gonzalez:1993uz}, a 1-loop calculation in a graphene motivated model, similar to RQED, showed that the system flows towards a Lorentz covariant point in the infrared. Perturbative calculations performed in the $\overline{MS}$ scheme at 1 and 2-loops have shown that at this order the beta function vanishes and consequently the coupling in RQED does not run \cite{Teber:2012de,Teber:2018goo}, suggesting the absence of charge renormalization. A generalization of these previous results demonstrated that the RQED beta function vanishes at all orders \cite{Dudal:2018pta}.

The renormalized fields and parameters can be defined in terms of dimensionless renormalization constants,
\bea \nonumber
\psi&=&Z_\psi^{1/2} \psi_r, \ \ \ \ \ \ \ \tilde{A}=Z_A^{1/2}\tilde{A}_r, \\ \nonumber
\tilde{a}&=&Z_A\tilde{a}_r,\ \ \ \ \ \ \ \ \ \ e=Z_\alpha^{1/2}e_r,\\
&& \hspace{0.5cm}\Gamma^\mu=Z_\Gamma \Gamma^\mu_r,
\eea
where $\psi$ is the fermion field, $\tilde{A}$ is the reduced gauge field, {\it a} is a gauge fix parameter, $\Gamma^\mu$ is the fermion-photon vertex and the index $r$ refers to renormalized. Because of finiteness of the vertex and electron charge, there is a relation between the renormalization constants,
\be 
Z_\alpha=(Z_\Gamma Z_\psi)^{-2}Z_A^{-1}.
\label{eq:Z_relation}
\ee

\begin{figure}
\begin{center}
\includegraphics*[scale=0.6]{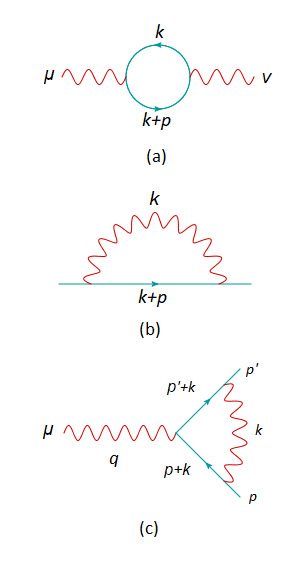}
\caption{RQED 1-loop corrections for (a) the photon propagator, (b) the fermion propagator and (c) the vertex.}
\label{fig:1loop_diagrams}
\end {center}
\end{figure}



%

In what follows we briefly review the 1-loop calculation of the $\beta$-function performed in \cite{Teber:2012de}. The 1-loop corrections to RQED propagators and vertex is represented in Fig.\ref{fig:1loop_diagrams}. For fermion fields constrained to a space in $d_e$ dimensions and gauge fields in $d_\gamma$ dimensions, the diagrams in Fig.\ref{fig:1loop_diagrams} are defined as,
\bea \nonumber
i\Pi^{\mu\nu}_1(q)&=&-\int \frac{d^{d_e}k}{(2\pi)^{d_e}} {\rm Tr} \left[(-ie\gamma^\mu)\frac{i(\slashed{k}+\slashed{q})}{(k+q)^2}(-ie\gamma^\nu)\frac{i\slashed{k}}{k^2}\right],\\ \nonumber
\Sigma_1(p)&=&\int \frac{d^{d_e}k}{(2\pi)^{d_e}}(-ie\gamma^\mu)\frac{i(\slashed{p}+\slashed{k})}{(p+k)^2}(-ie\gamma^\nu)\\ \nonumber
&&\times \frac{i}{(4\pi)^{\varepsilon_e}}\frac{\Gamma(1-\varepsilon_e)}{(-k^2)^{1-\varepsilon_e}}\left(g_{\mu\nu}-\tilde{\xi}\frac{k_\mu k_\nu}{k^2}\right),\\ \nonumber
-ie\Lambda^\mu_1 &=& \int \frac{d^{d_e}k}{(2\pi)^{d_e}}\frac{i}{(4\pi)^{\varepsilon_e}}\frac{\Gamma(1-\varepsilon_e)}{(-k^2)^{1-\varepsilon_e}}\left(g^{\nu\rho} -\tilde{\xi}\frac{k^\nu k^\rho}{k^2}\right)\\ 
&&\times(-ie\gamma^\nu)\frac{i\slashed{k}}{k^2}(-ie\gamma^\mu)\frac{i\slashed{k}}{k^2}(-ie\gamma^\rho),
\label{eq:1loopvertex}
\eea
where $\varepsilon_\gamma=\frac{4-d_\gamma}{2}$, $\varepsilon_e=\frac{d_\gamma - d_e}{2}$, $\tilde{\xi}=\xi(1-\varepsilon_e)$, $\varepsilon=1-a$ and $a$ is a gauge fixing parameter. Integrating out the expressions (\ref{eq:1loopvertex}), after some algebra one obtains \cite{Teber:2012de},
\bea \nonumber
\Pi_1(q^2)&=&-\frac{d e^2 (-q^2)^{-\varepsilon_\gamma-\varepsilon_e}}{(4\pi)^{d_e/2}}\frac{1-\varepsilon_\gamma-\varepsilon_e}{3-2\varepsilon_\gamma-2\varepsilon_e} G(1,1),\\ \nonumber
\Sigma_{V1}(p^2)&=&\frac{-e^2 \Gamma(1-\varepsilon_e)(-p^2)^{-\varepsilon_\gamma}}{(4\pi)^{d_\gamma/2}}\Bigg{[}\frac{2(1-\varepsilon_\gamma-\varepsilon_e)^2}{2-2\varepsilon_\gamma-\varepsilon_e},\\ \nonumber 
&-&\xi(1-\varepsilon_\gamma-\varepsilon_e)\Bigg{]}G(1,1-\varepsilon_e)\\ \nonumber
\Lambda^\mu_1&=&\frac{e^2 \Gamma(1-\varepsilon_e)m^{-2\varepsilon_\gamma}}{(4\pi)^{d_\gamma/2}}\\ 
&&\times \gamma^\mu \left[\frac{2(1-\varepsilon_\gamma-\varepsilon_e)^2 }{2-\varepsilon_\gamma - \varepsilon_e}-\bar{\xi}\right]\frac{\Gamma(\varepsilon_\gamma)}{\Gamma(2\varepsilon_e)},
\label{eq:1loopvertex2}
\eea
where $G(\nu_1,\nu_2)$ is the massless one-loop propagator defined by
\be 
G(\nu_1,\nu_2)=\frac{\Gamma(-d_e/2+\nu_1+\nu_2)\Gamma(d_e/2-\nu_1)\Gamma(d_e/2-\nu_2)}{\Gamma(\nu_1)\Gamma(\nu_2)\Gamma(d_e-\nu_1-\nu_2)}.
\ee
For the configuration relevant to graphene and further two-dimensional Dirac/Weyl materials, $d_\gamma = 4$, $d_e=3$, $\varepsilon_e=1/2$ and $\varepsilon\rightarrow 0$. The $\beta$-function is defined as 
\be 
\beta(\alpha(\mu))=\frac{d \log{\alpha(\mu)}}{d \log{\mu}},
\label{eq:beta_function}
\ee 
and the renormalized coupling constant $\alpha$ is defined as
\be 
\frac{\alpha(\mu)}{4\pi}= \mu^{-2\varepsilon_\gamma}\frac{e^2}{(4\pi)^{d_\gamma/2}} Z^{-1}_\alpha(\alpha(\mu)) e^{-\gamma\varepsilon_\gamma},
\label{eq:renorm_coupling}
\ee 
where $\mu^{-2\varepsilon_\gamma}$ is a factor to compensate the dimension of the coupling.

At this order, it can be obtained from Eqs. (\ref{eq:1loopvertex2}) that $Z_\Gamma Z_\psi=1$, which indicates that the Ward identity is satisfied. Therefore, according to (\ref{eq:Z_relation}), the charge renormalization constant $Z_\alpha$ depends exclusively on $Z_A$. The particular $\overline{MS}$ scheme chosen by the author is such that the finite terms are absorbed in the coupling and the renormalization constants reduces to unity for finite theories. 
Conversely, divergent theories are written as a Laurent series in $\varepsilon_\gamma$. Within this framework, it can be extracted from Eqs.(\ref{eq:1loopvertex2}) that for $d_\gamma=4$ and $d_e=3$, $\Pi_1$ is independent of $\epsilon_\gamma$, which implies that
\be 
Z_A=1 + O(\alpha^2).
\ee
Together with Eqs. (\ref{eq:beta_function}) and (\ref{eq:renorm_coupling}), it yields to
\be 
\beta(\alpha(\mu))=-2\varepsilon_\gamma + \gamma_A(\alpha(\mu)).
\ee 
Since this theory has no anomalous dimension associated to the gauge field, and $\varepsilon_\gamma\rightarrow 0$, the beta function vanishes.

A similar but more troublesome calculation can be performed at 2-loops \cite{Teber:2012de,Teber:2018goo}, showing that at this order the $\beta$-function remains vanishing. 

In order to generalize this result to all orders, it is argued in \cite{Teber:2018goo} that there is no renormalization of the gauge field since it would come from a non-local term in the free part of the action while counterterms must be local polynomials in the fields and their derivatives. Reminding that the charge renormalization depends only on $Z_A$, this would imply in scale invariance at all-order. It is pointed out in \cite{Dudal:2018pta} however that this argument does not hold for the present theory, since it is valid only for renormalizable theories, which does not happen to be the case. For a discussion on this topic we refer to \cite{Dudal:2018pta}.

An alternative procedure was proposed in order to generalize the result at 1-loop and show that it is valid at all orders \cite{Dudal:2018pta}. First of all an effective theory for the gauge fields is obtained integrating out the fermions in the action, namely,
\begin{eqnarray} \nn
\tilde{\Gamma}[A]&=&\int d^{3-\varepsilon}x \left[\frac{1}{2}Z_A^2F_{\mu\nu}\frac{1}{\sqrt{-\partial^2}}F_{\mu\nu}\right]\\
&+&\ln{ {\rm Det}(i\slashed{D})}+S_{gf}.
\end{eqnarray}
Invoking gauge symmetry, it is possible to show that $\tilde{\Gamma}(A)$ depends solely on the transverse projection of the gauge field,
\be 
A_\mu ^T=\left(\delta_{\mu\nu}-\frac{\partial_\mu\partial_\nu}{\partial^2}\right)A_\nu.
\ee 
Expressing $A_\mu ^T$ in terms of strength tensor,
\begin{eqnarray}\label{rqedc5}
A_\nu^T= \frac{\p_\mu}{{\Box}}F_{\mu\nu}=\int \frac{\dd^3r}{4\pi}\frac{(x-r)_\mu}{|x-r|^3}F_{\mu\nu}^{r}.
\end{eqnarray}
it is possible to define the all-order expansion of the effective action as,
\begin{eqnarray}\label{rqedc6}
\tilde\Gamma&=&\sum_{n\geq1} \int \dd^3x_1\ldots \dd^3 x_n A_1^T\ldots A_n^T\braket{j_{\mu_1}^{x_1}\ldots j_{\mu_n}^{x_n}}\\
&=&\sum_{n\geq1} \int \dd^3r_1 \ldots \dd^3r_n F_{\mu_1 \nu_1}^{r_1}\ldots F_{\mu_n \nu_n}^{r_n} \gamma_{\mu_1\nu_1,\ldots, \mu_n\nu_n}^{r_1,\ldots, r_n}\;,\nonumber
\end{eqnarray}
with
\begin{eqnarray}\label{rqedc7}
 \gamma_{\mu_1\nu_1,\ldots, \mu_n\nu_n}^{r_1,\ldots, r_n}&=& \int \frac{\dd^3x_1}{4\pi} \ldots \frac{\dd^3x_n}{4\pi}\frac{(x_1-r_1)_{\mu_1}}{|x_1-r_1|^3}\ldots\nonumber\\&&\times\frac{(x_n-r_n)_{\mu_n}}{|x_n-r_n|^3}\braket{ j_{\nu_1}^{x_1}\ldots j_{\nu_n}^{x_n}}\,.
\end{eqnarray}
By power counting and taking into account that the Furry theorem holds for RQED, the only contribution where divergences may arise is for $n=2$. This contribution corresponds to the usual transverse photon self-energy that leads to the following finite correction \cite{Gorbar:2001qt},
\begin{eqnarray}\label{rqedc8}
 \Pi_{\mu\nu}(q)=\frac{e^2}{8q}\left(\delta_{\mu\nu}-\frac{q_\mu q_\nu}{q^2}\right)\,.
\end{eqnarray}
The effective action for the gauge field then reads
\begin{eqnarray}\label{rqedd}
\tilde\Gamma[A]&=&\int \dd^{3-\epsilon} x \left[ \frac{1}{2}Z_A^2   F_{\mu\nu} \frac{1}{\sqrt{-\Box}} F_{\mu\nu}  \right.\\\nonumber&&+\left.\frac{e^2}{8}F_{\mu\nu} \frac{1}{\sqrt{-\Box}} F_{\mu\nu}+\mathcal{O}\left(\frac{e^4F^4}{\sqrt{-\Box}^5}\right)  \right]+S_{gf},
\end{eqnarray}
which shows that the contributions from higher orders in the field strength are sufficiently ultraviolet-suppressed and give only finite corrections. Given that, it is safe to say that the renormalization constant $Z_A=1$ in RQED to all orders.

\subsection{Renormalization group - RQED as a fixed point of a nonrelativistic model}

Although one of the main achievements of RQED is to correctly describe several aspects of graphene, one point that deserves special attention is that usually the calculations are performed in a relativistic framework, where the Fermi velocity is taken to be the speed of light, or in natural units $v_F\rightarrow1$. Since the actual value of the Fermi velocity $v_F/c \approx 1/300$, the standard procedure of recovering $v_F$ in the end of the calculation must be done with extreme caution.

Previous to \cite{Teber:2012de,Teber:2018goo}, a renormalization group analysis was performed considering a mixed dimensional non-relativistic model \cite{Gonzalez:1993uz}, where the Fermi velocity is  taken into account. This breaking of Lorentz invariance can be explicitly appreciated in the action,
\begin{eqnarray} \nonumber
S&=&\int d^3 r \bar{\psi} (-\gamma_0 \partial_0 + v {\bf \gamma}\cdot {\bf \nabla})\psi \\
&-& ie\int d^3 r (-\gamma_0 A_0 +v {\bf \gamma}\cdot {\bf A})\psi.
\label{eq:nonrela_action}
\end{eqnarray}
Here, a procedure similar to the one adopted in RQED was performed, where the gauge field was previously integrated out in the third space coordinate. As expected, this yields to the same dependence of the photon propagator on $1/q$ found before.

In order to fully renormalize this model, the Fermi velocity must be also renormalized, and one more renormalization constant must be introduced, $v_0=Z_v v_R$. This explicitly modifies the fermion self-energy as:
\be 
\Sigma(\omega,{\bf k})=Z_\psi (\omega,{\bf k})\left[\omega \gamma^0 - Z_v (\omega,{\bf k}) v {\bf \gamma}\cdot {\bf k}\right].
\ee 

The one-loop calculation within this model was the first to predict the absence of charge renormalization for this type of mixed dimensional theory. Besides that, the renormalization constant for the Fermi velocity was obtained \cite{Gonzalez:1993uz,Vozmediano:2010fz},
\be 
Z_v = 1 -\frac{1}{16\pi}\frac{e^2}{v}\ln{\Lambda}.
\ee 
Defining an effective coupling 
\be 
g=\frac{e^2}{4\pi v_F},
\ee 
where $g$ plays the role of a fine structure constant with $v$ replacing the speed of light. The Fermi velocities at different energies are related by
\be 
v(E) = v(E_0) \left[1-\frac{g}{4}\ln{\left(\frac{E}{E_0}\right)}\right].
\ee
It is possible to determine the existence of fixed points, imposing
\be 
\beta_v(v,e^2)=0.
\label{eq:fixedpoints}
\ee 
The Callan-Symanzik equation allows for obtaining an explicit form for the $\beta$-function~\cite{Gonzalez:1993uz}, and \eqref{eq:fixedpoints} is satisfied for $v_F=1$. 

From this, one can conclude that the Fermi velocity grows as the energy decreases. Furthermore, the $\beta$-function has a non-trivial zero when the Fermi velocity reaches the speed of light. This limit corresponds to a Lorentz invariant weak coupling model whose coupling corresponds to the fine structure constant of QED. RQED therefore can be seen as a fixed point of the renormalization group to which the system represented by (\ref{eq:nonrela_action}) flows. A detailed extension of renormalization group aspects of this mixed dimensional theory was performed in \cite{RG2,RG3,RG4,Vozmediano:2010fz}.

\subsection{Causality and unitarity}

As it usually happens to theories where some degrees of freedom are integrated out and the resulting theory is represented by an effective Lagrangian, RQED is nonlocal. Because of this, it becomes important to check if the theory respects causality and unitarity. 

In fact, it has been shown that this mixed dimension theory respects both conditions. In \cite{doAmaral:1992td}, the behavior of the classical Green's function, advanced and retarded, is investigated inside and outside the light-cone. Generalizing the theory, considering a generic power $\alpha$ for the d'Alambertian operator in the denominator of the pure gauge term in the Lagrangian (\ref{eq:Lagrangian}), it is shown that the Green's function vanishes outside the light-cone. This implies that this family of theories respects causality for any value of the parameter $\alpha$, including RQED, for which $\alpha=1/2$. Furthermore, for this theory the Green's function vanishes inside the light-cone as well, being finite only on its surface. This means that RQED is constrained by a more strict condition than causality, it obeys Huygens principle \cite{doAmaral:1992td,Bollini:1991fp}. While causality is found for any $\alpha$, the Huygens principle applies exclusively to RQED, not holding for its cousin QED$_3$.

Working with a more restrict generalization, where the power of the d'Alambertian operator in the RQED Lagrangian $\alpha=[0,1)$, the validity of unitarity was explored in \cite{Marino:2014oba}. To this end, the authors invoke the optical theorem as follows. The $S$-matrix, that relates the initial and the final state in a scattering process, must be unitary so that the theory is unitary. Representing the $S$-matrix as $S=1+iT$, this implies that 
\be 
i\left(T^\dagger - T \right)= T^\dagger T,
\label{eq:optical_theorem}
\ee
which is the optical theorem. Evaluating the operator in the above equation between initial and final states, it can be represented in terms of the propagator,
\be 
\langle i| T | f \rangle = (2\pi)^3 \delta^3 (k_i-k_f) D_ {if},
\label{eq:Toperator}
\ee 
where a phase space factor was introduced to ensure a correct dimensionality. For $i \rightarrow f$ the propagator in (\ref{eq:Toperator}) becomes the Feynman propagator and performing a Fourier transform, Eq.(\ref{eq:optical_theorem}) yields to 
\be 
D^*_F (\omega,{\bf k})-D_F (\omega,{\bf k})=-i\mathcal{T}^\gamma D^*_F (\omega,{\bf k})D_F (\omega,{\bf k}).
\ee 
Here $\mathcal{T}$ comes from integrating the phase space factor and is the characteristic time of the system, and $\gamma=-2(1-\alpha)$. The generalized Feynman propagator can be easily obtained from the generalized Lagrangian,
\be 
D_F(t,{\bf r})= \int \frac{d\omega}{2\pi}\int \frac{d^2 k}{(2\pi)^2}\frac{e^{-i\omega t}e^{i{\bf k}\cdot {\bf r}}}{(\omega^2 -{\bf k}^2 + i\varepsilon)^{1-\alpha}}.
\ee
A straightforward algebra shows that the optical theorem (\ref{eq:optical_theorem}) is obeyed exclusively for $\alpha=0$, which corresponds to QED$_3$, and for RQED, where $\alpha=1/2$, as long as a particular relation between $\mathcal{T}$ and $\varepsilon$ is respected. Therefore, unitarity holds for these two theories, while for other values of the parameter $\alpha$ it does not.

All this reasoning was obtained for the free theory. In the case of interacting theory a similar procedure can be performed considering a dressed propagator in the random phase approximation (RPA),
\begin{eqnarray} \nonumber
G_{\mu\nu}&=&G_{\mu\alpha}^{(0)}\Big[\delta_{\alpha,\nu}+\Pi^{\alpha\beta}G_{\beta\nu}^{(0)}\\
&+&\Pi^{\alpha\beta}G_{\beta\sigma}^{(0)}\Pi^{\sigma\gamma}G_{\gamma\nu}^{(0)}+...\Big].
\end{eqnarray}
Inserting in the above equation the 1-loop expression for $\Pi^{\mu\nu}$ calculated in \cite{Coste:1989wf}, yields to a result similar to the one for the free theory. Results beyond RPA were also obtained using the 2-loop expression calculated in \cite{Teber:2012de}. In this case, the functional form for the propagator remains the same, the only modification being the coefficients. Therefore, up to 2-loops the optical theorem still holds, fulfilling the unitarity condition.

\section{Chiral symmetry breaking}
\label{sec:CSB}
\subsection{Schwinger-Dyson equations}

Schwinger-Dyson equations (SDEs) are the field equations of a given quantum field theory (see, for instance~\cite{CraigReviewSDE}). These conform an infinite tower of relations among the Green functions involved: $n$-point functions are related to other $n$-point and higher-point functions, each verifying its own SDE. In their formal derivation, no assumption is made regarding the strength of the coupling constant(s) of the theory. Therefore, these equations are non-perturbative in nature and provide a useful tool to understand phenomena like bound-states, chiral symmetry breaking and confinement. The only systematic scheme to truncate the infinite tower of SDEs is perturbation theory, but in this scheme none of the above mentioned phenomena can be  addressed reliably. In gauge theories such as QCD and QED, symmetry-preserving truncations of SDEs have made a tremendous development in addressing several non-perturbative aspects of these theories, which would otherwise require other frameworks to be addressed, including lattice field theory and effective model considerations.

\begin{figure}
\includegraphics*[scale=0.4]{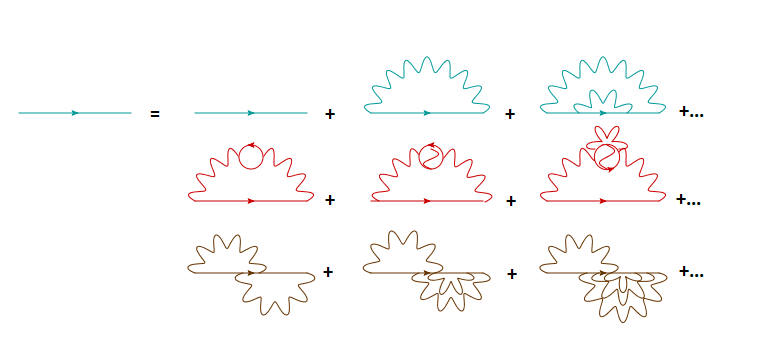}
\caption{Radiative corrections to the fermion propagator in QED. These include corrections to the fermion propagator itself, to the gauge boson propagator and  vertex corrections.}
\label{fig:Expansion}
\end{figure}

SDEs are formally derived from the observation of the vanishing of the functional derivatives of the connected Green functions generating functional with respect to the fields. An alternative diagrammatic derivation of these equations can be derived directly from the Feynman rules of the theory under study. In QED, for instance (see Ref.\cite{Mike}), the perturbative expansion of the fermion propagator, depicted in Fig.~\ref{fig:Expansion} shows three types of corrections, those to the fermion propagator itself (first row of corrections), those to the gauge boson propagator (second row of corrections) and the third kind corresponds to vertex corrections (third row of corrections). 

\begin{figure}[h!]
	\centering
	\includegraphics[width=0.2\textwidth]{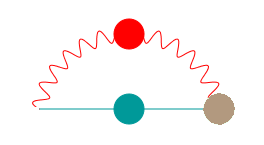}
	\caption{Fermion self-energy. Blobs over the different parts of the diagram indicate that all perturbative corrections to these pares are already taken into account.}\label{fig:Sigma}
\end{figure}

\noindent
The infinite resummation of diagrams is better carried out by defining the fermion self-energy $\Sigma(p)$, which is represented in the diagram in Fig.~\ref{fig:Sigma}
%
%



\begin{figure}[h!]
		\begin{center}
		\includegraphics[width=1\columnwidth]{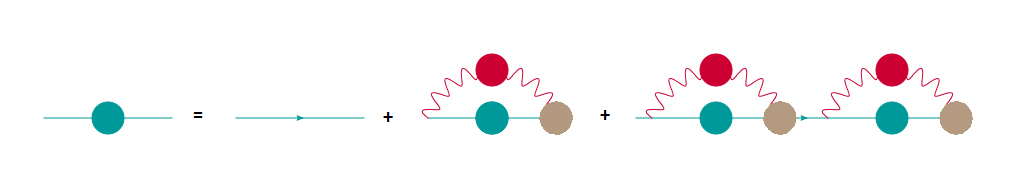}
		\caption{Perturbative expansion of the fermion propagator in terms of the fermion self-energy.}
		\label{fig:ExpSigma}
		\end{center}
	\end{figure}

\noindent
and upon which the perturbative expansion of the fermion propagator contains all the radiative corrections shown in Fig.~\ref{fig:Expansion}. In terms of $\Sigma(p)$, the perturbative expansion of the fermion propagator is shown in Fig.~\ref{fig:ExpSigma}.


	
\noindent
It corresponds to the expansion
	\begin{eqnarray}
		S(p)&=&S_0(p)+S_0(p)\Sigma(p)S_0(p)\nonumber\\
		&&+S_0(p)\Sigma(p)
		S_0(p)\Sigma(p)S_0(p)+\ldots\nonumber\\
		&=&S_0(p)+S_0(p)\Sigma(p)\left[S_0(p)+S_0(p)\Sigma(p)S_0(p)+\ldots\right]\nonumber\\
			&=&S_0(p)+S_0(p)\Sigma(p)S(p).
	\end{eqnarray} 

\begin{figure}[h!]
		\begin{center}
		\includegraphics[width=1\columnwidth]{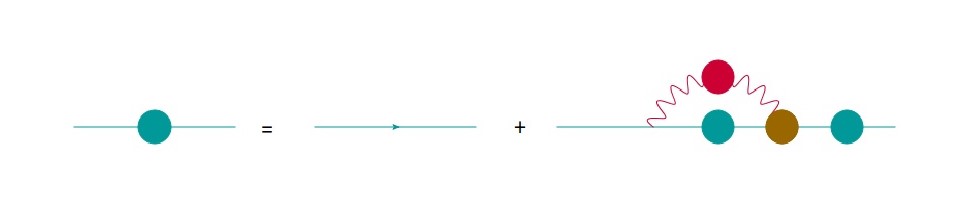}
		\caption {SDE for the fermion propagator.}
		\label{fig:SDEprop}
		\end{center}
	\end{figure}

\noindent
The last line corresponds to the SDE for the fermion propagator depicted in Fig.~\ref{fig:SDEprop}.

\begin{figure}[h!]
		\begin{center}
		\includegraphics[width=1\columnwidth]{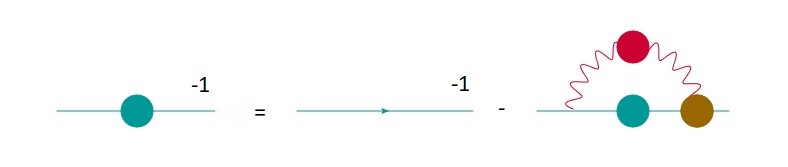}
		\caption{SDE for the inverse fermion propagator}
		\label{fig:SDEfp}
		\end{center}
	\end{figure}

It is convenient to re-write the above equation in terms of  the inverse fermion propagator, which then becomes
\begin{equation}
	S^{-1}(p)=(S_0(p))^{-1}-\Sigma(p),
\end{equation}
and diagrammatically can be depicted as in Fig.~\ref{fig:SDEfp}

\begin{figure}[h!]
	\begin{center}
		\includegraphics[width=0.8\columnwidth]{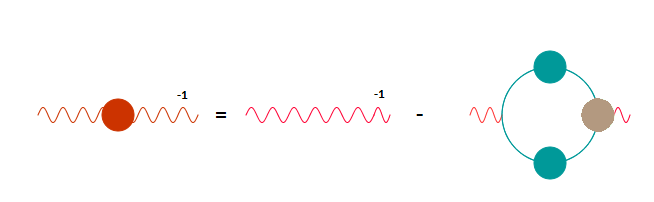}
		\caption{SDE for the gauge boson propagator.}
		\label{fig:SDEfoton}
		\end{center}
\end{figure}

This equation involves the full photon propagator and fermion-photon vertex. The former, by a similar reasoning, can be seen to obey its own SDE depicted in the diagram in Fig.~\ref{fig:SDEfoton}



\begin{figure}[h!]
	\centering
		\includegraphics[width=0.8\columnwidth]{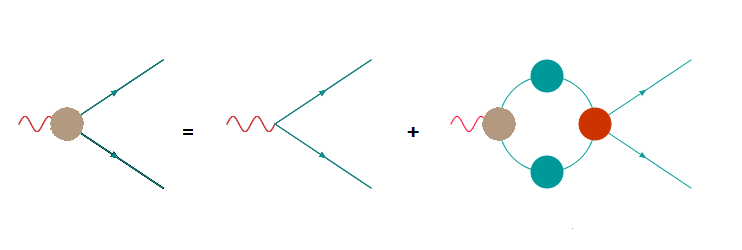}
		\caption{SDE for the fermion-boson vertex.}
		\label{fig:SDEvertex}
\end{figure}

\noindent
This two-point function is coupled to the fermion propagator and the fermion-photon vertex which verifies the SDE shown in Fig.~\ref{fig:SDEvertex}



\noindent
This diagram shows  that the three-point function is coupled to two-point and four-point functions, hence illustrating the structure of the infinite tower of SDE.

At first glance, it seems impossible to envisage a non-perturbative truncation of the tower of SDEs without compromising the reliability of the predictions hence derived. Nevertheless, because of the gauge symmetry, a number of relations exist in which a $(n+1)$-point Green function can be written on terms of $n$-point function. That is the case of Ward identities in QED\cite{ward,green,takahashi,twi1}. Thus, one can attempt to propose a symmetry-preserving truncation and explore general features of the non-perturbative solution to the infinite tower of equations.

Below we shall review a favorite truncation to the SDE for the fermion propagator or gap equation, the so-called rainbow-ladder truncation. We also review how to improve of this truncation by including vacuum polarization effects and vertex corrections.

\subsection{Structure of the gap equation}

\begin{figure}[h!]
	\begin{center}
		\includegraphics[width=0.8\columnwidth]{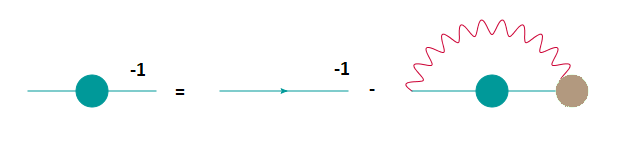}
		\caption{SDE for the fermion propagator in quenched approximation.}
		\label{fig:SDEquenched}
		\end{center}
\end{figure}

Let us consider the parity preserving ordinary version of QED. The SDE for the fermion propagator  in an arbitrary number space-time dimensions $d$ is given by
	\begin{eqnarray}
	S^{-1}(p)&=&S_{0}^{-1}(p)\nonumber\\
	&&\hspace{-10mm}+4\pi i \alpha\int\frac{d^{d}k}{(2\pi)^{d}}\Gamma^{\mu}(k,p)S(k)\gamma^{\nu}\Delta_{\mu\nu}(p-k),
	\label{sd}
	\end{eqnarray} 
where $\alpha=e^2/(4\pi)$ is the (dimensionless) fine-structure constant and $e$ is the electric charge. The most general form of this propagator is commonly expressed as
\be
S(p)=\frac{F(p^2)}{{\not \! p}-M(p^2)},
\ee
where $F(p^2)$ is the wavefunction renormalization function and $M(p^2)$ is the mass function. The tree level values of these functions are $F(p^2)=1$ and $M(p^2)=m_0$ where $m_0$ is the fermion bare mass. Furthermore, $\Delta_{\mu\nu}(p-k)$ represents the full gauge boson propagator and $\Gamma^\mu(k,p)$ the full fermion-boson vertex. By neglecting vacuum polarization effects, i.e., working within the {\em quenched} approximation, we replace the full photon propagator $\Delta_{\mu\nu}(q)$ by its bare counterpart 
\be
\Delta_{\mu\nu}^{(0)}(q)=\frac{1}{q^2}\left(g_{\mu\nu}+(\xi-1)\frac{q_\mu q_\nu}{q^2} \right),\label{eq:photonprop}
\ee
where $\xi$ is the covariant gauge parameter. In terms of diagrams, it amounts to consider Fig.~\ref{fig:SDEquenched}.



\begin{figure}[h!]
	\centering
		\includegraphics[width=0.8\columnwidth]{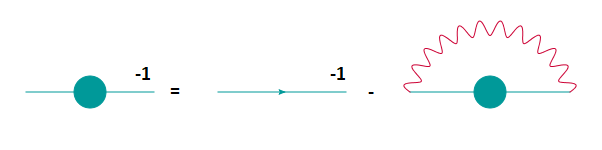}
		\caption{SDE for the fermion propagator in rainbow-ladder approximation.}
		\label{fig:SDErainbow}
\end{figure}

Under this approximation, the gap equation can be solved with a suitable choice of the fermion-boson vertex.  The rainbow-ladder approximation~(see \cite{Mike}) corresponds to the perturbation-theory inspired choice $\Gamma^\mu(k,p)=\gamma^\mu$, which allows to decouple the SDE for the fermion propagator from the infinite tower, as it corresponds to the diagram in Fig.~\ref{fig:SDErainbow} 



 Chiral symmetry breaking is usually approached starting with $m_0=0$. Thus, the natural consequence of this phenomenon corresponds to the dynamical appearance of fermion mass where there was none. Upon inserting the fermion and gauge boson propagators and fermion-photon vertex in the gap equation, after multiplying it by ${\not \! p}$ and 1, respectively, we obtain the coupled system of equations
\bea
\frac{1}{F(p^2)}&=& 1-\frac{4\pi i \alpha}{d\ p^2}\int \frac{d^dk}{(2\pi)^d} {\rm Tr}[{\not \! p}\gamma^\mu S(k) \gamma^\nu]\Delta_{\mu\nu}^{(0)}(p-k)\nn\\
\frac{M(p^2)}{F(p^2)}&=&\frac{4\pi i \alpha}{d}\int \frac{d^dk}{(2\pi)^d}{\rm Tr}[\gamma^\mu S(k) \gamma^\nu]\Delta_{\mu\nu}^{(0)}(p-k)\;.\label{eq:rainbow}
\eea
Let us explore the solutions to this coupled set of equations in different space-time dimensions.

\subsubsection{Gap equation in $d=4$}

It is interesting to consider the scenario of chiral symmetry breaking in ordinary QED. 
After taking the traces and Wick rotate to Euclidean space Eqs.~(\ref{eq:rainbow}), by switching to hyper-spherical coordinates, angular integrals can be performed analytically. Radial integrations are divergent. However, regulating these integrals with a hard cut-off in the momentum, we reach to the following pair equations to be solved self-consistently,
\bea
\frac{1}{F(p^2)}&=&1+\frac{\alpha\xi}{4\pi}\int_0^{\Lambda^2} dk^2 \frac{F(k^2)}{k^2+M^2(k^2)}\nn\\
&&\times\Bigg[ \frac{k^4}{p^2}\Theta(p^2-k^2)-\Theta(k^2-p^2)\Bigg]\;,\nn\\
\frac{M(p^2)}{F(p^2)}&=&  \frac{\alpha}{4\pi}(3+\xi) \int_0^{\Lambda^2} dk^2 \frac{F(k^2)M(k^2)}{k^2+M^2(k^2)}\nn\\
&&\times\Bigg[ \frac{k^2}{p^2}\Theta(p^2-k^2)-\Theta(k^2-p^2)\Bigg]
\eea
where $\Theta(x)$ is the Heaviside step function. In Landau gauge, $\xi=0$, we have that $F(p^2)=1$ and thus we have to solve the following non-linear integral equation for the mass function,
\bea
M(p^2)&=&\frac{3\alpha}{4\pi}\Big[\frac{1}{p^2}\int_{0}^{p^2}dk^2\frac{k^2M(k^2)}{k^2+M^2(k^2)}\nn\\
&&+\int_{p^2}^{\Lambda^2}dk^2\frac{M(k^2)}{k^2+M^2(k^2)}\Big].
\label{eq:diff}
\eea
This equation has no analytical solution. Nevertheless, we can gain some insight by transforming it into the  differential equation 
\be
\frac{d}{dp^2}\Big(p^4\frac{dM(p^2)}{dp^2}\Big)=-\frac{3\alpha}{4\pi}\frac{p^2M(p^2)}{p^2+M^2(p^2)}.\label{eq:diff2}
\ee
For $p^2\gg M^2(p^2)$, the above eq.\eqref{eq:diff2} linearizes to
\be
\frac{d}{dp^2}\Big(p^4\frac{dM(p^2)}{dp^2}\Big)= -\frac{3\alpha}{4\pi}M(p^2).\label{eq:lin4}
\ee
which is restricted to the boundary conditions
\be
\frac{d}{dp^2}\left[ p^2M(p^2)\right]\Bigg|_{p^2=\Lambda^2}=0,\qquad
\frac{d M(p^2)}{dp^2}\Bigg|_{p^2=\kappa^2}=0.
\ee
The infrared cut-off $\kappa^2$ is introduced to preserve the non-invariance of the original equation~\eqref{eq:diff} under $M(p^2)\to c M(p^2)$ with $c$ constant as $p^2\to 0$. Physically, it serves to quantify the amount of mass that is dynamically generated. We look for solutions to eq.~\eqref{eq:lin4} of the form
\begin{equation}
	M(p^2)=(p^2)^s
\end{equation}
which after substituting in~\eqref{eq:lin4} yields
\begin{equation}
	s(s+1)=-\frac{3\alpha}{4\pi}.
\end{equation}
Solving for $s$, we have that
\begin{equation}
	s_\pm=-\frac{1}{2}\pm\frac{\sqrt{1-\frac{3\alpha}{\pi}}}{2}.
\end{equation}
Thus, depending on the value of $\alpha$  compared to the value of $\alpha_c=\pi/3$, we could observe an oscillatory behavior or a power-law solution. Consistency with boundary conditions demands $\alpha>\alpha_c$. Moreover, the infrared and ultraviolet cut-offs are constrained as
\be
\frac{\Lambda}{\kappa}=\exp{\left(\frac{\pi}{\sqrt{\frac{\alpha}{\alpha_c}-1}}-2\right)}.
\ee
This behavior is called the {\em Miransky scaling law} which states that the chiral symmetry breaking in QED corresponds to a conformal phase transition~\cite{MiranskyCPT}. Thus, we conclude that in order for chiral symmetry breaking to be broken in ordinary QED, the coupling $\alpha$ must exceed a critical value.

\subsubsection{Gap equation in QED3}

\begin{figure}[h!]
	\centering
		\includegraphics[width=0.8\columnwidth]{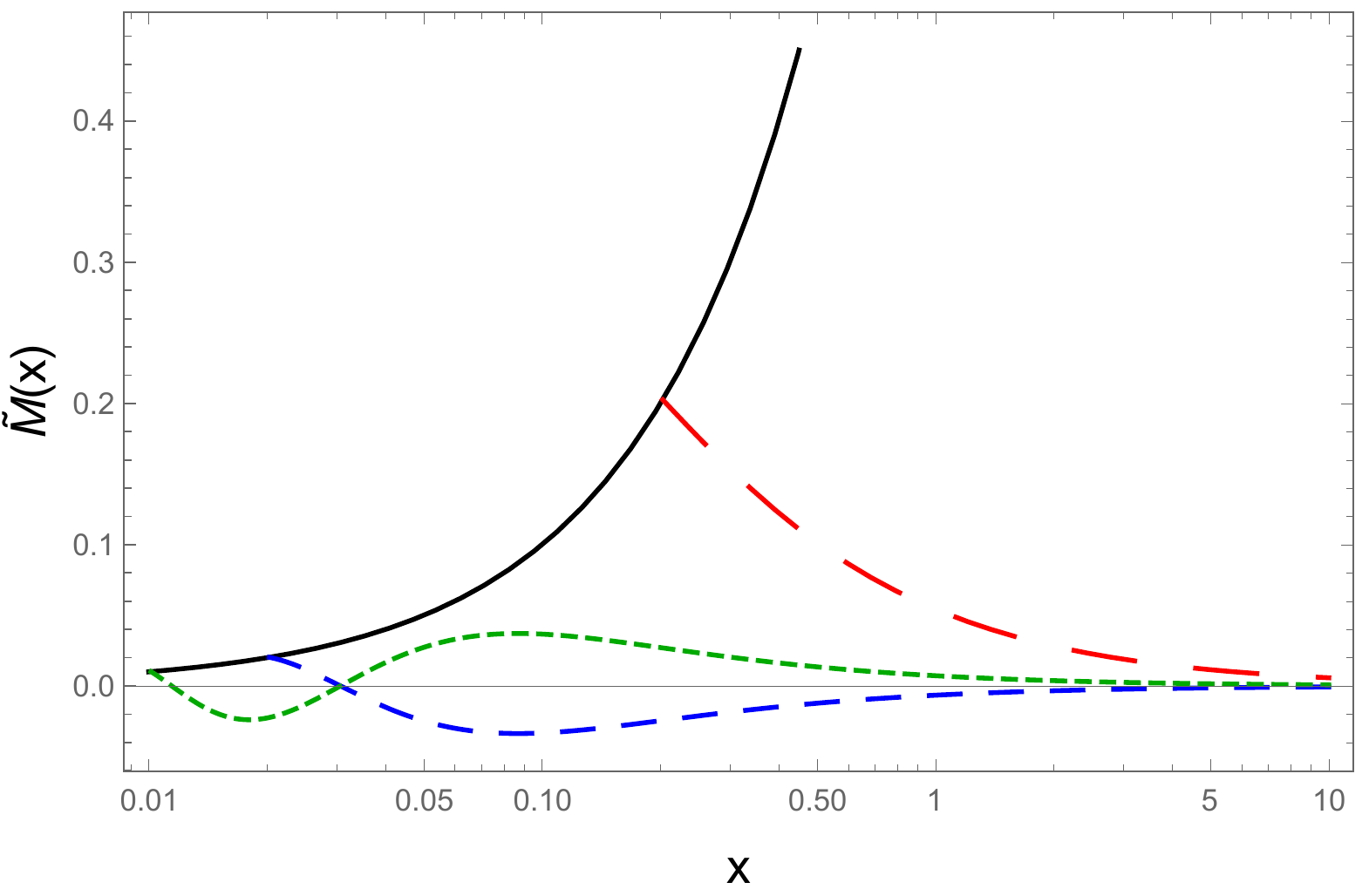}
		\caption{Multiple solutions to the gap equation in QED$_3$. The solid black curve corresponds to $\tilde{M}(x)=x$. The red long-dashed curve corresponds to a positive definite solution with $\kappa=\hat{\alpha}/2$, the dashed blue curve to $\kappa=\hat{\alpha}/20$ and the dotted green curve to $\kappa=\hat{\alpha}/40$.}
		\label{fig:multiple}
\end{figure}

A similar reasoning can be followed to explore the scenario of chiral symmetry breaking in QED restricted to a plane, namely, QED$_3$. The main difference in this case is that the electric charge $e^2$ has mass dimension one and hence serves as a natural scale for all mass scales in the theory. Furthermore, QED$_3$ is super-renormalizable, and thus there is no need to regulate the integrals. In the rainbow approximation and in Landau gauge the gap equation, reduces to (see Ref.~\cite{BRH})
\be
M(p)=\frac{e^2}{2\pi^2 p}\int_0^\infty dk\frac{k M(k)}{k^2+M(k)^2}\ln\left|\frac{k+p}{k-p}\right|.\label{eq:gapqed3}
\ee
At this point, although the integrals are finite, one can introduce an ultraviolet regulator $\Lambda$ in the integral with $\Lambda\gg e^2$ and consider at the end the limit $\Lambda\to\infty$ maintaining $e^2/\Lambda$ fixed. Proceeding in this way, after expanding the logarithm for $k\gg p$ and $p\gg k$, the integral reduces to
\be
M(p)=\frac{e^2}{\pi^2}\int_{0}^{p}dk\frac{M(k)}{k^2+M^2(k)}+\frac{e^2}{\pi^2p^2}\int_{p}^{\Lambda}dk\frac{k^2 M(k)}{k^2+M^2(k)}.
\ee
This equation can be transformed into the differential equation
\be
\frac{d}{dp}\Big[p^3\frac{dM(p)}{dp}\Big]=-\frac{2e^2}{\pi^2}\frac{p^2 M(p)}{p^2+M^2(p)}\;,\label{eq:diffqed3}
\ee
subjected to the boundary conditions
\be
\frac{d}{dp}\Big[p^3\frac{dM(p)}{dp}\Big]\Big|_{p=0}\rightarrow 0, \qquad M(p)\Big|_{p=\Lambda}\rightarrow 0.
\ee
Again, in the regime where $p^2\gg M^2(p)$, the differential equation~\eqref{eq:diffqed3} linearizes in the following form
\be
\frac{d}{dp}\Big[p^3\frac{dM(p)}{dp}\Big]=-\frac{2e^2}{\pi^2}M(p),
\ee
which admits the general solution 
\be
M(p)=\frac{4e^2}{\pi^2 p}\left[C_1 J_2\left(2\sqrt{\frac{2e^2}{\pi^2p}}\right)+C_2 Y_2\left(2\sqrt{\frac{2e^2}{\pi^2p}}\right)\right]
\ee
where $J_\nu(x)$ and $Y_\nu(x)$ are Bessel functions of the first and second kind of order $\nu$, respectively. UV boundary condition demands that $C_2=0$.
The constant $C_1$ cannot be fixed from the IR conditions. Nevertheless, noticing again that the nonlinear equation~\eqref{eq:gapqed3} is non-invariant under $M(p)\to cM(p)$ with $c$ a constant, we require to change that boundary condition by demanding that
\be
M(\kappa)=\kappa,
\ee
in the IR, where $\kappa$ is a regulator quantifying the amount of mass being generated. Such boundary condition allows to write the solution to the gap equation as~\cite{multiple}
\be
\tilde{M}(x)=n \frac{\hat{\alpha}}{x}J_2\left(\sqrt{\frac{2\hat{\alpha}}{x}} \right),
\ee
with
\be
\tilde{M}(x)=\frac{M(p/\Lambda)}{\Lambda}, \quad \hat{\alpha}=\frac{e^2}{4\pi \Lambda}, \quad n=\frac{\kappa^2}{\hat{\alpha}}\frac{1}{J_2\left(\sqrt{\frac{2\hat{\alpha}}{\kappa}} \right)}.
\ee
It is interesting that this solution is positive definite for $\kappa\sim \hat{\alpha}$, but when $\kappa\simeq \hat\alpha/20$, the solution develops a zero. Furthermore, when $\kappa\simeq\hat{\alpha}/40$ a new zero develops,  as illustrated in Fig.~\ref{fig:multiple}.



This pattern continues {\em ad infinitum} and emerges from the analytical properties of the gap equation, which corresponds to a Hammerstein equation of the first kind. This structure is particular of the truncation and is modified when some of the assumptions are removed from consideration. See Ref.~\cite{multiple} fir further discussion.

\subsubsection{Gap equation in QED3 including vacuum polarization effects}

\begin{figure}[h!]
	\centering
		\includegraphics[width=0.9\columnwidth]{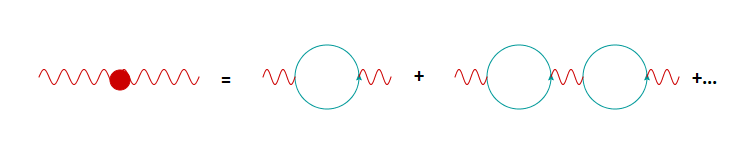}
		\caption{Vacuum polarization in the leading $1/N$ approximation.}
		\label{fig:loop1enN}
\end{figure}

\begin{figure}[h!]
	\centering
		\includegraphics[width=0.9\columnwidth]{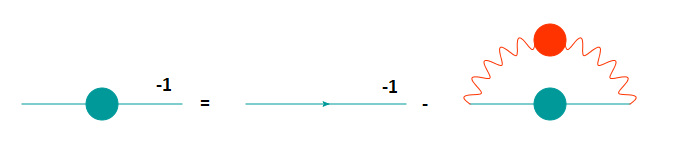}
		\caption{SDE for the fermion propagator in the leading $1/N$ approximation.}
		\label{fig:unquenched}
\end{figure}

Including vacuum polarization effects in this model turns out interesting. Let us consider the scenario where we include a large number $N$ of massless fermion families circulating in loops. This amounts to ressuming an infinite number of planar diagrams as shown in Fig.~\ref{fig:loop1enN} such that one considers the unquenched  approximation depicted in Fig.~\ref{fig:unquenched}.



\noindent



In this case, the non-pertubative gauge boson propagator can be calculated exactly. In an arbitrary covariant gauge, it reads~\cite{appel1,appel2,appel3}
\be
\Delta_{\mu\nu}(q)=\frac{1}{q^2 + \frac{\tilde{\alpha}q}{8}}\left( g_{\mu\nu}-\frac{q_\mu q_\nu}{q^2}\right) + \xi \frac{q_\mu q_\nu}{q^4},\label{eq:fotprop1n}
\ee
where $\tilde{\alpha}=e^2N$. We observe that the infrared behavior of this propagator softens as $1/q$ when $q\to 0$ as compared to the usual $1/q^2$ pole structure in Eq.~\eqref{eq:photonprop}. It turns out that this softening makes the theory to loose its confining properties. It also has consequences regarding the chiral transition, as we can see from the gap equation. In the leading order of the $1/N$ approximation, $F(p)=1+{\cal O}(1/N)$ and the full fermion photon vertex $\Gamma^\mu=\gamma^\mu+{\cal O}(1/N)$. Working in Landau gauge, the mass function obeys 
\be
M(p)=\frac{\tilde{\alpha}}{2\pi^2Np}\int_{0}^{\infty}dk\frac{k M(k)}{k^2+M^2(k)}\ln\Big|\frac{\tilde{\alpha}/8+8|k+p|}{\tilde{\alpha}/8+8|k-p|}\Big|
\ee
Once more, by expanding the logarithm for $k\gg p$ and $p\gg k$, we have
\bea
M(p)&=&\frac{\tilde{\alpha}}{\pi^2Np}\int_{0}^{p}dk \frac{k M(k)}{k^2+M^2(k)}\Big(\frac{k}{p+\tilde{\alpha}/8}\Big)\nn\\
&&\hspace{-5mm}+\frac{\tilde{\alpha}}{\pi^2Np}\int_{p}^{\infty}dk \frac{k M(k)}{k^2+M^2(k)}\Big(\frac{p}{k+\tilde{\alpha}/8}\Big).
\eea
This expression is equivalent to the differential equation 
\be
\frac{d}{dp}\Big[\frac{p^2(p+\tilde{\alpha}/8)^2}{2p+\tilde{\alpha}/8}\frac{dM(p)}{dp}\Big]=-\frac{\tilde{\alpha}}{\pi^2N}\frac{p^2M(p)}{p^2+M^2(p)}\;,
\label{eq:diff1ennfull}
\ee
subjected to the boundary conditions
\be
0\leq M(0)<\infty,\qquad \Big[p\frac{dM(p)}{dp}+M(p)\Big]_{p=\tilde{\alpha}}=0.
\ee
In the large-coupling regime $\tilde{\alpha}\gg p$, the differential equation~\eqref{eq:diff1ennfull} simplifies to
\be
\frac{d}{dp}\Big[p^2\frac{dM(p)}{dp}\Big]=-\frac{8}{\pi^2N}\frac{p^2M(p)}{p^2+M^2(p)}.
\ee
Moreover, in the momentum domain $p^2\gg M^2(p)$, the above equation linearizes and takes the simplified form
\be
\frac{d}{dp}\Big[p^2\frac{dM(p)}{dp}\Big]=-\frac{8}{\pi^2N}M(p). \label{eq:dif1enn}
\ee
Equation~\eqref{eq:dif1enn} admits a power law solution of the form
\be
M(p)=p^s,
\ee
which upon substitution into~\eqref{eq:dif1enn} yields
\be
s_\pm=-\frac12\pm\frac12\sqrt{1-\frac{32}{\pi^2N}}.
\ee
Thus, the solution to the linearized equation is
\be
M(p)= C_1 p^{-\frac{1}{2}+\frac12\sqrt{1-\frac{32}{\pi^2N}}}+C_2 p^{-\frac{1}{2}-\frac12\sqrt{1-\frac{32}{\pi^2N}}}.
\ee
Again, we notice the existence of a critical number of fermion families, $N_c=32/\pi^2$ that distinguishes an oscillatory from a decaying behavior of the mass function. Consistency with boundary conditions suggests that $N>32/\pi^2$ and the amount of dynamically generated mass, quantified as $M(p=0)$ also follows a Miransky scaling law of the form
\be
M(0)= \tilde{\alpha}\exp \left[\frac{-2\pi}{\sqrt{\frac{N_c}{N}}-1}+\delta \right].
\ee
Thus, this scenario has a similar scaling behavior as ordinary QED. Furthermore, vacuum polarization effects translate to the existence of a critical (large) number of fermion families above which chiral symmetry breaking is no longer possible and the chiral symmetry restoration corresponds to a conformal phase transition~\cite{MiranskyCPT}.

It is interesting that the gauge dependence of the mass function and wavefunction renormalization have been considered in covariant gauges~\cite{oldsaul} and up to the next-to-leading order of the approximation~\cite{critQED3T1,critQED3T2,critQED3T3,critQED3T4}. These ressumations carried out in an arbitrary non-local gauge do not spoil the existence of the critical value $N_c$ for the restoration of chiral symmetry and only refine the value of this critical number $N_c$. The conclusion up to date is that such number is an integer $N_c\le 3$.

\subsubsection{Gap equation in RQED}

So far we have reviewed different scenarios for chiral symmetry breaking in ordinary QED. The gap equation in this theory is such that the wavefunction renormalization is trivial in Landau gauge. The scenario of criticality emerges in $d=4$ as the need for the coupling to exceed a critical value for the theory to be able to break dynamically the chiral symmetry. In three dimensions, the quenched theory does not exhibit this feature. Chiral symmetry breaking can be broken for arbitrary values of the coupling. Nevertheless, including vacuum polarization effects, it is observed that if the number of fermion families circulating in loops exceeds a critical value, chiral symmetry cannot be broken. This statement is valid up to $1/N^2$ and in arbitrary non-local gauge.  An interesting observation is that when fermions in loops remain massless, there is a softening of the infrared pole in the propagator $1/q^2$ as the gauge boson momentum $q\to 0$ to a $1/q$ softer behavior. This behavior is similar to the tree-level gauge boson propagator in RQED
\be
\Delta_{\mu\nu}^{(0)}(q)=\frac{1}{2q}\Bigg[g_{\mu\nu}-(1-\zeta)\Bigg]\frac{q_{\mu}q_{\nu}}{q^{2}}.
\ee
Notice that in this case, the gauge parameter $\zeta$ is not the same as $\xi$ in the ordinary theory, but here it has to be renormalized because fermions and bosons live in different dimensions. In fact, $\zeta=\xi/2$, and hence $\zeta=0$ corresponds to Landau gauge.

Following the reasoning of the previous subsections, we truncate the gap equation in the rainbow approximation as in Ref.~\cite{Alves:2013bna}. In an arbitrary covariant gauge, it is equivalent to the following coupled system of equations (see~\cite{Juan})

\begin{eqnarray}
   \nonumber \frac{1}{F(p)}&=& 1+\frac{\alpha}{2\pi p^2}\int_{0}^{\Lambda}\frac{dk F(k) k^2}{k^2+M(k)^2}\times\\
   \nonumber &\times&\Bigg\{
  \frac{\theta(k-p)}{k}\left[-(2+\zeta)p^2 + (1-\zeta)\frac{p^4}{k^2} \right] \nn\\
  &&+ \frac{\theta(p-k)}{p} \left[-(2+\zeta)k^2+(1-\zeta)\frac{k^4}{p^2} \right] \Bigg\},\nonumber\\
   \frac{M(p)}{F(p)}&=&\frac{\alpha(2+\zeta)}{2\pi} \int_{0}^{\Lambda}dk\ \frac{k^2F(k)M(k)}{k^2+M^2(k)} \times \nn\\
   &\times&\left[\frac{\theta(k-p)}{k} +\frac{\theta(p-k)}{p} \right],\label{eq:gapRQED}
\eea
with the dimensionless coupling $\alpha=e^2/4\pi$ as usual.
It is evident from the above expressions that for $\zeta=0$,  $F(p)\ne 1$. Nevertheless, as a first approximation, let us take $F(p)=1$ and explore the solution of the gap equation. Then, the gap equation becomes the non-linear integral equation for the mass function~\cite{Alves:2013bna,Juan}
\bea
M(p)&=&\frac{2\alpha}{\pi p} \int_{0}^{p}dk\ \frac{k^2M(k)}{k^2+M^2(k)}\nn
\\
&&+\frac{2\alpha}{\pi }\int_{p}^{\Lambda}dk \frac{ k^2  M(k)}{[k^2+M^2(k)]k}.\label{eq:gap_MRQED}
\eea
This expression can be straightforwardly converted into the following differential equation
\be
p^2M''(p)+2pM'(p)+\frac{2\alpha}{\pi}\frac{p^2M(p)}{p^2+M^2(p)}=0,\label{eq:diffrqed}
\ee
restricted to the boundary conditions	
\bea
\lim_{p\to\Lambda}\left(p\frac{dM(p)}{dp}+M(p)\right)&&=0,\nonumber\\
\lim_{p\to 0}p^2\frac{dM(p)}{dp}&&=0.
\eea
Upon linearizing the differential equation~\eqref{eq:diffrqed} when $p\gg M(p)$, we can again write the resulting equation in the Euler-Cauchy form
\be
\frac{d}{dp}\left(p^2\frac{dM(p)}{dp}\right)+\frac{\alpha}{\pi}M(p)=0,
\ee
which,  admits a general solution of the form
\be
M(p)=C_1p^{n_+}+C_2p^{n_-},
\ee
where
\be
n_\pm=-\frac12\pm\frac12\sqrt{1-\frac{8\alpha}{\pi}}.
\ee
As in previous cases, the non-invariance of the gap equation under scalings $M(p)\to c M(p)$ where $c$ is a constant demands the introduction of an infrared cut-off $\kappa$ and the modification of the infrared boundary condition to
\be
M(\kappa)=\kappa.
\ee
Thus, defining $\alpha_c=\pi/{8}$, boundary conditions demand that $\alpha>\alpha_c$. Furthermore, the dynamical mass obeys the Miransky scaling
%
\be
\frac{\Lambda}{\kappa}=\exp\left(\frac{A}{\sqrt{\frac{\alpha}{\alpha_c}-1}}+\delta\right),\label{eq:MirRQED}
\ee
with $A=-2\pi$ and $\delta=-4$, 
hence indicating that the chiral symmetry is broken in a conformal phase transition provided the coupling exceeds a critical value.

In order to explore the effect of the wavefunction renormalization in the result \eqref{eq:MirRQED}, particularly in relation with the gauge dependence, below we explore some variants of the rainbow-ladder truncation following~\cite{Juan}. 


\subsubsection{Effect of the wavefunction renormalization}

To explore the sensitivity of the truncation to the effect of the wavefunction renormalization function in connection with the gauge parameter dependence of the critical coupling and the dynamical mass, we conduct the following exercise:

\begin{itemize}

\item First, we impose $F(p)=1$ in all covariant gauges and solve the equation 
\bea
M(p)&=&\frac{\alpha}{2\pi p} \int_{0}^{p}dk\ \frac{k^2M(k)}{k^2+M^2(k)}\frac{2k[2+\zeta]}{k}\nn
\\
&&+\frac{\alpha}{2\pi p}\int_{p}^{\Lambda}dk \frac{ k^2M(k)}{[k^2+M^2(k)]k}[2p](2+\zeta),\label{eq:gapRQEDF1}
\eea
for various values of the gauge parameter $\zeta$. 

\item As a second variant,  we solve the coupled system of equations in~\eqref{eq:gapRQED} for various values of the gauge fixing parameter. 

\item Next we impose the Ward-Takahashi identity (WTI), 
%
%
\be
(k-p)_\mu\Gamma^\mu(k,p)=S_F^{-1}(k)-S_F^{-1}(p),\label{eq:WTI}
\ee
which allow to split the vertex into its longitudinal and transverse pieces,
$$\Gamma^\mu(k,p)=\Gamma^\mu_L(k,p) + \Gamma^\mu_T(k,p),$$
where  $(k-p)_\mu\Gamma^{\mu}_T(k,p)=0$. We exploit this identity into the gap equation in the following way. Assuming that all the dependence on the gauge parameter arises only from the photon propagator, we split this two-point function into
 its transverse component and the gauge parameter dependent longitudinal part, 
\be
\Delta_{\mu\nu}^{(0)}(q)=\Delta_{\mu\nu}^{T}(q) + \zeta \frac{q_\mu q_\nu}{q^3},\label{eq:photon_split}
\ee
the gap equation can be written as
\bea
S^{-1}(p)&=&S_0^{-1}(p)+4 i\alpha\pi\int\frac{d^3k}{(2\pi)^3}\gamma^\mu S(k)\Gamma^\nu\Delta_{\mu\nu}^T(q)\nn\\
&&+4 i\alpha\pi\zeta\int\frac{d^3k}{(2\pi)^3}\gamma^\mu S(k)\Gamma^\nu\frac{q_\nu q_\mu}{q^3},
\eea
with $q=k-p$. In the final term we replace the identity~\eqref{eq:WTI} and
upon taking traces after multiplying by 1 and ${\not \! p},$ respectively, we find the gap equation to be equivalent to the coupled system of equations

\begin{eqnarray}
   \nonumber \frac{1}{F(p)}&=& 1+\frac{\alpha}{2\pi p^2}\int_{0}^{\Lambda}\frac{dk F(k)k^2}{k^2+M(k)^2}\times\\
   \nonumber &\times&\Bigg\{
  \frac{\theta(k-p)}{k}\left[-2p^2 -\zeta\frac{p^4}{k^2} \right] \nn\\
   &&+ \frac{\theta(p-k)}{p} \left[-2k^2-\frac{k^4}{p^2} \right] \Bigg\},\nonumber\\
   &+&\frac{\alpha\zeta}{4\pi p^2}\int_{0}^{\lambda}\frac{dk F(k)k^2}{F(p)(k^2+M^2(p)}\times\nonumber\\
 &\times&  \Bigg\{\frac{\theta(k-p)}{k}\left[\frac{2p^2k^2}{k^2}-\frac{p^4}{k^2}\right.\nn\\
 &&\left.-\frac{M(k)M(p)p^2}{k^2}+\frac{2p^2M(k)M(p)}{k^2}\right]\nn\\
   &+&\frac{\theta(p-k)}{p}\left[\frac{2p^2k^2}{p^2}-\frac{p^2k^2}{p^2}\right.\nn\\
   &&\left.-\frac{M(k)M(p)k^2}{p^2}+\frac{2p^2M(k)M(p}{p^2}\right]\Bigg\}\nn\\
   \frac{M(p)}{F(p)}&=&\frac{\alpha}{\pi} \int_{0}^{\Lambda}dk \frac{k^2F(k)M(k)}{k^2+M^2(k)} \times \nn\\
   &\times&\left[\frac{\theta(k-p)}{k} +\frac{\theta(p-k)}{p} \right]\nn\\
   &+&\frac{\alpha\zeta}{2\pi}\int_{0}^{\lambda}\frac{dk F(k)k^2}{F(p)(k^2+M^2(p)}\times\nn\\
   &\times&\Bigg\{\frac{\theta(k-p)}{k}\left[M(p)\Bigg(1-\frac{p^2}{k^2}\Bigg)-M(k)\frac{p^2}{2k^2}\right]\nn\\
   &+&\frac{\theta(p-k)}{p}\left[-M(k)\Bigg(1-\frac{k^2}{p^2}\Bigg)-M(p)\frac{k^2}{2p^2}\right]\Bigg\}\nn\\
\eea
Then we solve the coupled system of equations for various values of $\zeta$.

\item As a final variant of the gap equation, we  use the Ball and Chiu vertex~\cite{BallChiu}, which is explicitly constructed to verify the Ward identity and is written as follows
\bea
\Gamma^\mu_{BC}&=&\frac{\gamma^\mu}{2}\left[\frac{1}{F(k)}+\frac{1}{F(p)}\right]\nn\\
&&+\frac{1}{2}\frac{({\not \!k}+{\not \!p})(k+p)^\mu}{(k^2-p^2)}\left[\frac{1}{F(k)}-\frac{1}{F(p)}\right]\nn\\
&&+\frac{(k+p)^\mu}{(k^2-p^2)}\left[\frac{\mathcal{M}(k)}{F(k)}-\frac{\mathcal{M}(p)}{F(p)}\right].
\eea
This is a standard choice for the longitudinal piece of the vertex and it is explicitly constructed to satisfy these relations. Nevertheless, the additional terms to the central part exhibit spurious kinematic singularities as $k^2\to p^2$. These spurious singularities do not appear in perturbation theory and are unwanted in a non-perturbative construction of the vertex. These actually are cancelled by more educated {\em ansatze} of the vertex which include the transverse piece unconstrained by these identities. Thus, for our purposes, by keeping only the central part ,
which is
\begin{equation}
\Gamma^\mu_{CBC}=\frac{\gamma^\mu}{2}\left[\frac{1}{F(k)}+\frac{1}{F(p)}\right],
\end{equation}
we guarantee that the ward identities are satisfied up to spurious singular terms which are not expected to appear in a complete form of the vertex.

Then, the gap equation corresponds to the following coupled system of equations

\begin{eqnarray}
   \nonumber \frac{1}{F(p)}&=& 1+\frac{\alpha}{2\pi p^2}\int_{0}^{\Lambda}\frac{dk F(k)k^2}{k^2+M(k)^2}\left[\frac{1}{F(k)}+\frac{1}{F(p)}\right]\times\\
   \nonumber &\times&\Bigg\{
  \frac{\theta(k-p)}{k}\left[-(2+\zeta)p^2 + (1-\zeta)\frac{p^4}{k^2} \right] \nn\\
  &&+ \frac{\theta(p-k)}{p} \left[-(2+\zeta)k^2+(1-\zeta)\frac{k^4}{p^2} \right] \Bigg\},\nonumber\\
   \frac{M(p)}{F(p)}&=&\frac{\alpha(2+\zeta)}{2\pi} \int_{0}^{\Lambda}dk\ \frac{k^2F(k)M(k)}{k^2+M^2(k)}  \times \nn\\
   &\times&\left[\frac{1}{F(k)}+\frac{1}{F(p)}\right]\left[\frac{\theta(k-p)}{k} +\frac{\theta(p-k)}{p} \right],
\eea

\end{itemize}

\noindent
In Fig.~\ref{fig:critcoupl} we compare the findings of the critical coupling in different gauges for the variants of the truncation described above.



\begin{figure}[h!]
	\centering
		\includegraphics[width=0.9\columnwidth]{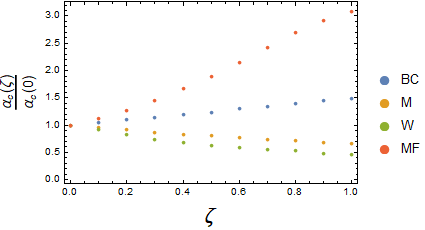}
		\caption{Critical coupling as a function of the covariant gauge parameter for different models.}
		\label{fig:critcoupl}
\end{figure}
At first glance, it might seem surprising that the simplest scenario of neglecting wavefunction renormalization effects and solving the mass function alone yields the least gauge parameter dependent results.
This is better understood from a renormalization group analysis, that exploits the similarities of the $1/N$ approximation in QED$_3$ and RQED itself, as discussed below.

\subsection{Renormalization Group Analysis}

It is remarkable that RQED becomes an infra-red Lorentz invariant fixed-point for a general theory of mixed dimensions for fermions and gauge bosons, RQED$_{d_\gamma,d_e}$~\cite{Gonzalez:1993uz}, because it allows to analyze the critical behavior of the general theory, regarding chiral symmetry breaking,  around such point. The important observation is that as we approach this fixed point, the Fermi velocity of fermion fields  tends to the speed of light in vacuum and at the same time, the coupling constant approaches the fine-structure constant of QED. Therefore, one can exploit these facts to compare the critical structure of RQED and QED$_3$~\cite{Kotikov:2016yrn,critQED3T1,critQED3T2,critQED3T3,critQED3T4,revT}. This is particularly relevant to understand the gauge dependence of the critical numbers associated to the chiral transition. 

In order to make explicit the mapping between QED$_3$ and RQED, we start by expressing the gauge boson propagator in RQED as~\cite{Kotikov:2016yrn}
\be
\Delta_{\mu\nu}^{({\rm RQED})}(q)=\frac{1}{2q}\left(g_{\mu\nu}+(\zeta-1)\frac{q_\mu q_\nu}{q^2} \right),
\ee
where we recall that the gauge fixing parameter $\zeta$ of RQED is half of the corresponding parameter of QED. On the other hand, the leading behavior of the gauge boson propagator~\eqref{eq:fotprop1n} of QED$_3$ in the large  $N$ approximation has the form
\be
\Delta_{\mu\nu}^{({\rm QED_3})}(q)=\frac{8}{Ne^2q}\left(g_{\mu\nu}+(\xi-1)\frac{q_\mu q_\nu}{q^2} \right).
\ee
Thus, identifying
\be
\frac{1}{\pi^2 N}\to \frac{\alpha}{4\pi}, \qquad \zeta \to \frac{\xi}{2},\label{eq:map}
\ee
we can simply map the solutions of the gap equation in RQED to those of QED$_3$ in the $1/N$ approximation. 

For instance, the gauge parameter dependence of the critical number of fermion families for chiral symmetry restoration in QED$_3$ was explored in Refs.~\cite{Kotikov:2016yrn,revT}. The authors find that the critical condition can be written as
\be
1=\frac{16(2+\xi)}{L_c}. 
\ee
Then, from~\eqref{eq:map} we can straightforwardly find that for RQED, this result is translated to the critical coupling
\be
1=16(2+\xi)\frac{\alpha_c}{4\pi},
\ee
which shows a strong gauge dependence of the critical coupling, although it does not depend on the flavor number $N$. This result comes up as a consequence of a partial re-summation of diagrams in the approximation. It can be improved, for instance, by considering a RPA  calculation of the vacuum polarization tensor. In the RPA, the gauge boson propagator has the leading behavior
\be
\Delta_{\mu\nu}^{({\rm RQED})}(q)=\frac{1}{2q(1+N e^2/16)}\left(g_{\mu\nu}+(\zeta-1)\frac{q_\mu q_\nu}{q^2} \right).
\ee
Thus, by redefining the coupling as
\be
\alpha \to \alpha'=\frac{\alpha}{1+e^2N/16},
\ee
the gauge dependence of the critical coupling is now
\be
\alpha_c=\frac{\pi}{2(5+\bar{\xi}) -N^2\pi/4},
\ee
with $\xi=(1+\bar{\xi})/2$ and $N=2$ for graphene. The gauge dependence of $\alpha_c$ is milder in this case.

At next-to-leading order, one observes that the gauge dependence of the critical coupling has the form
\be
\alpha_c=\frac{4\pi}{8(2+\xi)+\sqrt{d(x)}},
\ee
where 
\be
d(\xi)= 8\left(S(\xi)-8\left(4-\frac{112}{3}\xi+9\xi^2 \right) -4N\pi^2\right)
\ee
and
\be
S(\xi)=(1-\xi)R_1 -(1-\xi^2)\frac{R_2}{8}-(7+16\xi-3\xi^2)\frac{P_2}{128},
\ee
with the coefficients
\be
R_1=163.7428, \quad R_2=209.175,\quad P_2=1260.720.
\ee
This is a rather intricate form of the coupling which nevertheless improves upon the leading order gauge parameter dependence. The RPA still helps to reduce the resudial gauge dependence of the critical coupling by redefining 
\be
\alpha \to \alpha'=\frac{\alpha}{1-\frac{\alpha N}{4\pi^2}}.
\ee
Thus, we observe that in this framework, the appropriate re-interpretation of the coupling permits a less severe gauge dependence of the critical value of the coupling to trigger chiral symmetry breaking.


\section{Chiral symmetry breaking in a medium}
\label{sec:medium}

Mixed-dimensional theories allow natural extensions of the QFT formalism to incorporate the effects of (classical and quantum) external agents like a heat bath and/or external electromagnetic, strain or gravitational fields, among others.
Generally speaking, these effects have the potential to enhance or inhibit phase transitions. In the case of the chiral transition, a heat bath is known to have the  effect of restoring chiral symmetry. The imaginary-time formalism of thermal field theory (TFT) is a natural extension to consider in this case. On the contrary, electromagnetic fields promote the breaking this symmetry. In particular, in 3D materials, the configuration of parallel electric and magnetic fields gives rise to a special configuration that can be understood as the presence of a Chern-Simons (CS) term in the electromagnetic Lagrangian, with the added possibility of parity and time reversal symmetry breaking. In 2D systems, the presence of such a term for fermion fields is realized as the possibility of mass terms different from the ordinary Dirac mass.

A Haldane term of this type has interesting effects on the chiral phase transition in ordinary QED$_3$. Curvature effects are also important in modifying the electric and optic properties of the materials and are generally expected to occur in materials with defects. In this section, we review the effect of a thermal bath,  the role of a Chern-Simons term, the influence of strain and the formulation of RQED on curved space. We specialize the discussion in how these effects impact the scenario for chiral symmetry breaking.


\subsection{Finite temperature effects}

 The scenario of chiral symmetry breaking in RQED  has been considered at finite temperature. Let us  revisit the  calculations in~\cite{Gorbar:2001qt,Alves:2013bna,Nascimento:2015ola,Baez:2020dbe}.
 We start our discussion from the imaginary time formalism of TFT. We introduce the fermionic Matsubara frequencies $\omega_n=(2n+1)\pi T$ and replace of any integral over the temporal component of an arbitrary  three-vector $p=(p_0,{\bf p})$ by the summation over frequencies, 
\begin{equation}
    \int dp_0f(p_0,{\bf p}) \to T \sum_{n=-\infty}^\infty f(-i\omega_n,{\bf p}).
\end{equation}
Under this prescription, working in the Landau-like gauge $\zeta=0$ and neglecting wavefunction renormalization effects, the gap equation takes the form~\cite{Nascimento:2015ola}
\begin{eqnarray}
    M_m({\bf p})& =&
    \sum_{n=-\infty}^{\infty} \int_\Lambda \frac{d^2k}{(2\pi)^2}
    \frac{(4\pi\alpha T) M_n({\bf k})}{(2n+1)^2\pi^2T^2+{\bf k}^2+M_n({\bf k})^2}\nn\\
    &&\times
\frac{1}{[4 (m-n)^2\pi^2T^2+({\bf p}-{\bf k})^2 ]^{1/2}},
\label{eq:gap1}
\end{eqnarray}
where  $M_n({\bf p})=M(\omega_n,{\bf p})$ is the mass function associated to every Matsubara frequency and the notation $\int_\Lambda$ indicates that divergent integrals are to be regularized with an ultraviolet cut-off $\Lambda$.

Several considerations are at hand. First of all, as customary in TFT, one would be tempted to perform the sum over Matsubara frequencies and introduce the cut-off in momentum integrals. Nevertheless, because these frequencies appear inside the square-root, the standard techniques to perform such sums are hard to implement. First of all, the convergence of the sum
\bea
U&=& \sum_{n=-\infty}^\infty u_n\nn\\
&=& \sum_{n=-\infty}^\infty  \frac{1}{\sqrt{n^2+A^2}}\frac{1}{(2n+1)^2+B^2}\nn\\
&=& 2 \sum_{n=1}^\infty u_n^{(+)}+u_0\label{sumU}
\eea
is ensured (see Ref.~\cite{Nascimento:2015ola}) provided the sum in the last row of the above expression converges. In Eq.~(\ref{sumU}), it is assumed that $A^2, B^2>0$ and  $u_n^{(+)}=(u_n+u_{-n})/2$. The proof of convergence is established via the auxiliary sum
\be
V=\sum_{n=1}^\infty v_n = \sum_{n=1}^\infty\frac{1}{(n+A)(n+B)},
\ee
which is convergent and satisfies that $v_n\ge u_n^{(+)}\ge 0$ and
\be
\lim_{n=\to\infty} \frac{u_n^{(+)}}{v_n}=0.
\ee
Then, because $V$ is convergent, so it is $\sum_{n}u_n^{(+)}$ and thus, $U$ is also convergent.

Alternatively, one can perform the momentum integrals first and then the sum over Matsubara frequencies. In this case, one can assume that the mass function is the same for all the frequencies such that the gap equation can be written as~\cite{Nascimento:2015ola}
\be
M(p)=\frac{4\alpha T}{\pi}\int_0^\infty dk \ k M(k) I(k),
\ee
where
\bea
I(k)&=& \sum_{n=-\infty}^\infty \frac{1}{(2n+1)^2\pi^2 T^2+k^2+M(k)^2}\nn\\
&&\times \frac{K(x_n)}{\sqrt{(p-k)^2+4n^2\pi^2T^2}},
\eea
$K(x)$ denoting an elliptic function and 
\be
x_n =-\frac{4kp}{(k-p)^2+4n^2\pi^2T^2}.
\ee
Considering only the zeroth Matsubara frequency $n~=~0$, an approximation that is commonly used in high-temperature regime studies, the gap equation reduces to~\cite{Nascimento:2015ola}
\be
M(p)=\frac{4\alpha T}{\pi}\int_0^\infty dk \frac{k}{|k-p|}\frac{K(x_0)M(k)}{\pi^2T^2+k^2+M^2(k)}.\label{gapw0}
\ee
Expanding $k K(x_0)/|k-p|$ for $k\gg p$ and $k\ll p$, the gap equation becomes
\bea
M(p)&=&\frac{2\alpha T}{p}\int_0^p dk \frac{k M(k)}{\pi^2 T^2 + k^2 + M^2(k)} \nn\\
&&+ 2 \alpha T \int_p^\Lambda dk \frac{M(k)}{\pi^2 T^2 + k^2 + M^2(k)}.
\eea
Upon linearization, the resulting integral equation  can be cast in the form of the differential equation
\be
\frac{d}{dp}\left(p^2\frac{dM(p)}{dp} \right)+\frac{2\alpha T}{p}M(p)=0,\label{eqdifw0}
\ee
subject to the boundary conditions
\be
\lim_{p\to \pi T_c}p^2\frac{dM(p)}{dp}=0, \qquad
\lim_{p\to\Lambda}M(p)=0,
\ee
where $T_c$ would correspond to the critical temperature for chiral symmetry restoration, which will be defined from the analytical behavior of the mass function.
The general solution to Eq.~(\ref{eqdifw0}) is expressed as
\be
M(p)=C_1 \sqrt{\frac{\alpha T}{p}}J_1 \left(\sqrt{\frac{8\alpha T}{p}} \right)+C_2 \sqrt{\frac{\alpha T}{p}}Y_1 \left(\sqrt{\frac{8\alpha T}{p}} \right),
\ee
where $J_1(x)$ and $Y_1(x)$ are Bessel functions of the first and second kind. UV boundary condition imply $C_2=0$. At the same time, the IR boundary condition imposes the following relation
\be
2J_1(\xi)+\xi J_0(\xi)-\xi J_2(\xi)=0,
\ee
with $\xi=\sqrt{8\alpha T/\pi T_c}$. The above relation has nontrivial solutions for a given $n$ such that $\{\xi_n\}=\{\xi_0,\xi_1,\ldots\}$ with the hierarchy $\xi_n<\xi_{n+1}$ for any given $n$. The lowest solution $\xi_0\simeq 2.4$ fixes the value of the critical temperature $T_c=0.44\alpha T$. This result establishes that even at low temperatures, the value 
\be
\alpha_c^T\simeq\alpha_c \xi_0^2,
\ee
exceeds the critical value of the coupling in vacuum~\cite{Nascimento:2015ola}. It has its origin in the fact that there is no evidence that the mass function $M(p)$ vanishes at the critical temperature. Therefore, a more refined treatment is called for.

One can improve on the above result recalling that the phenomenon of dynamical chiral symmetry breaking is infrared. Therefore, one can simply neglect the external momentum in the gap equation~(\ref{gapw0}), which then becomes~\cite{Nascimento:2015ola}
\be
1=2\alpha T \int_0^\Lambda \frac{dk}{\pi^2T^2+k^2+m^2(T)},
\ee
where we have introduced the shorthand notation $m(T)=M(0,T)$. Retaining $\Lambda$ as the largest scale of the problem, two solutions are found for the above expression, namely
\be
m(T)=\pm \pi \sqrt{T_c^2-T^2}, \qquad T_c=\frac{\alpha^2 T^2 \Lambda^2}{(2\alpha T+\Lambda)^2}.
\ee
Keeping the positive solution alone, one can derive the critical coupling for chiral symmetry breaking of the form
\be
\alpha_c(T)=\frac{\Lambda}{\Lambda-2T}.
\ee
Nevertheless, this expression is oblivious to the critical value of $\alpha$ at $T=0$.

A different view of the gap equation can be achieved as follows~\cite{Baez:2020dbe}. Let us introduce
 dimensionless quantities,
\begin{equation}
\mathbf{p}=\Lambda\boldsymbol{\sigma},\quad \mathbf{k}=\Lambda\boldsymbol{\rho},\quad
T=\Lambda \Tilde{T},\quad
M_m(\mathbf{k})=\Lambda \tilde{M}_m(\boldsymbol{\rho}),
\end{equation}
such that the gap equation~(\ref{eq:gap1}) can be written as
\bea
\tilde{M}_m (\mathbf{\sigma}) & =&
\alpha \Tilde{T}
\sum_{n=-N_f-1}^{N_f} \int_0^{1}
\frac{d\rho}{\pi}  d\theta\ \rho\nn\\
&&\times
 \frac{ \tilde{M}_n(\mathbf{\rho})}{(2n+1)^2\pi^2 \tilde{T}^2+ \rho^2+\tilde{M}_n^2(\mathbf{\rho})}\nn\\
 &&\times
\frac{1}{\left|4 (m-n)^2\pi^2\tilde{T}^2+ ({\boldsymbol\sigma}-{\boldsymbol\rho})^2  \right|^{1/2}},\label{gapdimless}
\eea
where $\theta$ is the angle between  $\boldsymbol{\sigma}$ and $\boldsymbol{\rho}$~\footnote{In our notation, spatial vectors are written in boldface characters, but their magnitud is detoted by the character alone.}. Notice that in this case, the cut-off is not introduced in momentum integrals, but in the number of frequencies $N_f$ that are summed up. 
%
%
%
Whithin the so-called constant mass approximation~\cite{Ahmad:2015cgh}, all mass functions involved in the gap equation can be replaced  with their values at zero momentum. Denoting $\tilde{M}_m({\bf{\sigma}})\rightarrow S_m$, the gap equation under this approximation simplifies to
\bea
S_m& =&
2\alpha \Tilde{T}
\sum_{n=-N_f-1}^{N_f} \int_0^{1}
d\rho \rho
    \frac{ S_n}{(2n+1)^2\pi^2 \tilde{T}^2+ \rho^2+S_n^2}\nn\\
    &&\times
\frac{1}{\left|4 (m-n)^2\pi^2\tilde{T}^2+ \rho^2  \right|^{1/2}}.
\eea
Fixing the  the cut-off $N_f$ such that 
\bea
(2N_f+1)\pi T_0 =\Lambda& \Rightarrow& (2N_f+1)\pi \tilde{T}_0 =1\nn\\
&
\Rightarrow& \qquad \tilde{T}_0=\frac{1}{(2N_f+1)\pi},
\eea
we are allowed to write all temperature scales as
$T = k \tilde {T} _0,\;k\in \mathbb{N} $. Thus,
performing momentum integration analytically 
 and assuming a linear scaling of $ S_m \approx \gamma_m (T-T_c) $ near criticality, 
\begin{eqnarray}
S_m  &\approx&
2\alpha \Tilde{T} \sum_{n=-N_f-1}^{N_f}
\frac{S_n}{\sqrt{\pi ^2
   \Tilde{T}^2 \left|-4 m^2+8 m n+4 n+1\right|}} 
\nonumber
   \\
&&\hspace{-8mm}\times
\left[\tan ^{-1}
\sqrt{
\frac{4 \pi ^2 \Tilde{T}^2 (m-n)^2+1}
{\pi^2 \Tilde{T}^2 \left|-4 m^2+8 m n+4 n+1\right|}
   } \right.
\nonumber 
   \\
&-&
\left. 
 -  \tan ^{-1} \frac{2  \left| m-n\right| }{\sqrt{ \left|-4 m^2+8
   m n+4 n+1\right|}} \right].
\end{eqnarray}
Specializing on the zeroth Matsubara frequency $m=0$, 
observing that $S_n/S_0\leq 1$, 
and recalling our assumption that near the critical point the temperature-dependent $S_m(\tilde{T})\to 0 $ we have that~\cite{Baez:2020dbe}
\begin{eqnarray}
\frac{1}{\alpha_c } 
&=&
 \sum_{n=-N_f-1}^{N_f}
\frac{2}{ \pi  \sqrt{ \left|4 n+1\right|}} 
\left[\tan ^{-1}
\sqrt{
\frac{4 \pi^2 \Tilde{T}^2_c n^2+1}
{\pi^2 \Tilde{T}^2_c \left|4 n+1\right|}
   } 
   \right.
\nonumber 
   \\
&&
\left. 
   -\tan ^{-1} \frac{2 \left| n \right| }{\sqrt{\left|4 n+1\right|}} \right].\label{ac}
\end{eqnarray}
The sum over $n$ is finite for every value of temperature. This allows to obtain the behavior of the critical coupling, $\alpha_c$, as shown in Fig.~\ref{fig:alphac}


\begin{figure}[h!]
	\centering
		\includegraphics[width=0.9\columnwidth]{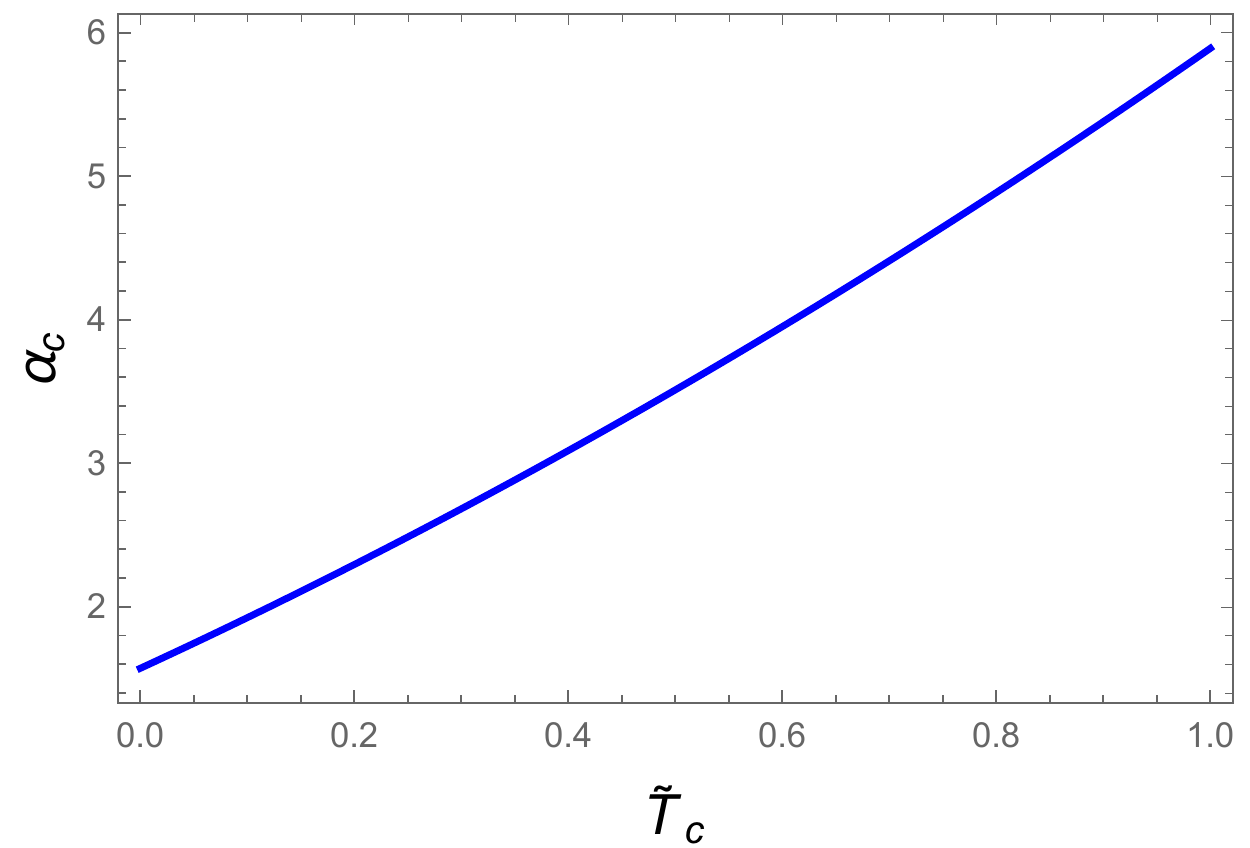}
		\caption{Critical coupling as a function of the critical temperature.}
		\label{fig:alphac}
\end{figure}

%
%
%
%
%
The behavior of the critical coupling as a function of temperature is consistent with 
\be
\alpha_c(\tilde{T})=\frac{\pi}{8} + a \tilde{T} + b \tilde{T}^2,
\ee
with $a, b$ real numbers. This behavior exhibits the correct behavior of $\alpha_c({\tilde{T}}=0)=\pi/8$~\cite{Baez:2020dbe}. 

It is remarkable that within the finite temperature framework, the gauge dependence of the critical temperature and coupling is a pending matter up to date.

\subsection{Chiral symmetry breaking with a Chern-Simons term}

Physics on two spatial dimensions is very interesting in its own right. For electromagnetic interactions, there exists the possibility to include a Chern-Simons (CS) term in the gauge sector through the Lagrangian
\begin{equation}
\mathcal{L}=-i\frac{\vartheta}{4}\epsilon^{\mu\nu\rho}A_\mu\partial_\nu A_\rho-A_\mu J^\mu,\label{eq:CSlag}
\end{equation}
where $\vartheta$ is the CS coefficient~(see, for instance, Ref.~\cite{CSterm}). Such a term induces a topological mass to the gauge bosons. It also induces fractional (anyon) statistics for matter particles as well as parity and time-reversal symmetry breaking of the theory, among other interesting effects. It is through this term that, for instance, the fractional quantum Hall effect can be explained and the emergence of the transverse conductivity without Landau levels can be explained.

The scenario of chiral symmetry breaking in QED$_3$ with a CS has been explored by solving the gap equation in different truncation schemes~\cite{Kondo:1994bt,Kondo:1994cz,christoph1,christoph2}.  The CS coefficient acts as an effective vacuum polarization such that when a large number of fermion families are considered,  besides the existence of a critical number of families $N_c$ that restores chiral symmetry, there exists a critical value for the CS coefficient $\vartheta_c$ that also induces chiral symmetry restoration. The order of the phase transition in this case changes to order one.
	
%
%

In RQED, 
%
the impact of the dimensionless parameter $\vartheta$ has been explored by our group (for a study exploring the effects of a modified Chern-Simons term containing a dimensionful parameter see \cite{Ozela:2021pse} and for other approaches involving time reversal symmetry breaking see \cite{marino5}). The starting point is the addition of the CS term~\eqref{eq:CSlag} in the  RQED Lagrangian. The effective theory under consideration has the structure~\cite{Olivares:2020eko}
\begin{eqnarray} \nonumber
\mathcal{L}_{RQED}^ {CS}&=&-\frac{1}{4}F^{\mu\nu}\frac{2}{(-\Box)^{1/2}}F_{\mu\nu}+\bar{\psi}(i\gamma^\mu\partial_\mu+e\gamma^\mu A_\mu)\psi\\
&+&\frac{1}{2\zeta}(\partial_\alpha A^\alpha)^2 + \frac{-i\vartheta}{4}\varepsilon^{\mu\nu\rho}A_\mu \partial_\nu A_\rho,
\end{eqnarray}
where all the quantities carry their usual meaning. 	
The tree-level gauge boson propagator derived from the above Lagrangian is
\begin{eqnarray}\label{photon_prop_tree}\nonumber
\hat \Delta_{\mu\nu}(q)&=&\frac{1}{2q}\frac{1}{(1+\vartheta^2)}\left(\delta_{\mu\nu}
-\frac{q_\mu q_\nu}{q^2}\right)\\&&+\frac{\zeta}{q^2}\frac{q_\mu q_\nu}{q^2}-\frac{1}{2q^2}\frac{\vartheta}{(1+\vartheta^2)}\epsilon_{\mu\nu\rho}q^{\rho}
\label{photonprop}\;.
\end{eqnarray}
%
From here it is immediate to notice the effect of the CS coefficient as an effective dielectric constant~\cite{Magalhaes:2020nlc} that has the potential to modify the conditions for chiral symmetry breaking. Moreover, 
for fermions, it is convenient to introduce their  {\em right-} and {\em left-}handed projections, $\psi_\pm=\chi_\pm \psi$, defined from the chiral matrix $\tau=[\gamma_3,\gamma_5]/2$. In this form, the chiral projectors are defined such that 
\be
\chi_\pm=\frac12 \left(1\pm\tau\right),
\ee
which fulfill  the following properties 
\be
\chi_\pm^2=\chi_\pm,\quad \chi_+ \chi_-=0, \quad \chi_+ + \chi_-=1.
\ee
The advantage of working with these projectors is the following: In order to explore the general scenario of chiral symmetry breaking and its consequence, the dynamical generation of fermion masses, we consider the possibility of emergence of an ordinary Dirac mass term $m_e\bar{\psi}\psi$, which breaks explicitly chiral symmetry, but is even under partity and time-reversal transformations. On top of that, there is another mass term that potentially can be generated, the Haldane mass term $m_o\bar{\psi}\tau\psi$. Although this term does not break chiral symmetry, it violates parity (and time-reversal) and because the CS is parity violating we expect a relation between them. With these considerations,  the most general Lagrangian we consider for our analysis  is \cite{Dudal:2018pta,Olivares:2020eko}
\begin{eqnarray} \nonumber
\mathcal{L}_{RQED}^ {CS}&=&-\frac{1}{4}F^{\mu\nu}\frac{2}{(-\Box)^{1/2}}F_{\mu\nu} +\frac{1}{2\zeta}(\partial_\alpha A^\alpha)^2\nn\\
&&
+\bar{\psi}(i\gamma^\mu\partial_\mu+e\gamma^\mu A_\mu
+m_e + \tau m_o)\psi\nn\\
&&
 + \frac{-i\vartheta}{4}\varepsilon^{\mu\nu\rho}A_\mu \partial_\nu A_\rho,
\label{massive_Lagrangian}
\end{eqnarray}
These masses however do not correspond to poles in the fermion propagator. The operators $\chi_\pm$ project the upper and lower two-component spinors, making explicit that the would-be poles of each fermion species correspond to combination of Dirac and Haldane mass terms. Explicitly, upon acting with the chiral projectors, the matter Lagrangian can be written as
\begin{equation}
\mathcal{L}_F = \bar\psi_+ (i\slashed\partial + M_+)\psi_+ + \bar\psi_- (i\slashed\partial + M_-)\psi_-,
\end{equation}
where $M_+=m_e+m_o$ and $M_-=m_e-m_o$.

As mentioned earlier, the presence of a CS term generates a mass for the photon in QED$_3$, where $\vartheta$ has mass dimension one. Such a mass is known as topological mass (in this case no real topology is involved and the terminology comes from historical reasons). In QED$_3$, radiative corrections may give a contribution to this mass. For the case of QED$_3$ incremented with a CS term, the Coleman-Hill theorem states that only one-loop radiative corrections may be non-vanishing, with the other corrections being identically zero to all orders. We note that the argument to demonstrate this theorem is valid for a theory with massive fermions. Remarkably, it was shown that the Coleman and Hill theorem is also valid for RQED \cite{Dudal:2018mms} in spite of the fact that the CS coefficient is dimensionless. Therefore, the only possible correction to the photon propagator comes from the one-loop diagram Fig.\ref{fig:1loop_diagrams}(a).

In order to address the issue of chiral symmetry and parity breaking, in Ref.~\cite{CarringtonRQED} a robust truncation of the SDE was incorporated. To this end, the possibility of Fermi velocity renormalization by splitting the fermion momentum $P=(p_0,{\bf p})$ into its temporal $p_0$ and spatial ${\bf p}$ components such that the propagator reads
\bea
S(p_0,{\bf p})&=& -\frac{F^0_+(p)\gamma^0 p_0+F_+(p) \boldsymbol{\gamma}\cdot {\bf p}+M_+(p)}{{(F^0_+)}^2(p)p_0^2+F_+^2(p)v_F{{\bf p}}^2v_F^2+M^2_+(p)}\chi_+\nn\\
&&\hspace{-8mm}
-\ \frac{F^0_-(p)\gamma^0 p_0+F_-(p) v_F{\boldsymbol{\gamma}}\cdot {\bf p}+M_-(p)}{{(F^0_-)}^2(p)p_0^2+F_-^2(p){{\bf p}}^2v_F^2+M^2_-(p)}\chi_-.
\eea
Tha gap equation is truncated including a gauge-boson--fermion vertex dressing model and vacuum polarization effects. The former is based in the central Ball-Chiu vertex, which in this formalism takes the explicit form
\bea
\Gamma_\nu(P,K)&=& \frac{1}{4}\left[H_{\nu\sigma}^+(P) +H_{\nu\sigma}^+(K)\right]\gamma_\sigma (1+\gamma_5)\nn\\
&&\hspace{-5mm}+
\frac{1}{4}\left[H_{\nu\sigma}^-(P) +H_{\nu\sigma}^-(K)\right]\gamma_\sigma (1-\gamma_5),
\eea
where
\be
H^\pm(P)=\left( \begin{array}{ccc}
\frac{1}{F_\pm^{0}(P)} & 0 & 0 \\ 0 & \frac{1}{v_F F_\pm(P)} & 0 \\ 0 & 0 & \frac{M_\pm(P)}{F_\pm(P)}
 \end{array} \right),
\ee
whereas the latter are included modelling a Coulomb-like approximation. Noticing that the most general form of the gauge boson propagator can be written as
\bea
\Delta_{\mu\nu}(Q)&=&G_L(Q) P_{\mu\nu}^3+G_T (P_{\mu\nu}^1-P_{\mu\nu}^3) + G_D P_{\mu\nu}^6\nn\\
&&+ G_E (P_{\mu\nu}^{10}-P_{\mu\nu}^{11}),
\eea
where the involved tensors are
\bea
P_{\mu\nu}^1 &=& g_{\mu\nu}-\frac{Q_\mu Q_\nu}{Q^2}\,,\nn\\
P_{\mu\nu}^3&=& \frac{n_\mu n_\nu}{n^2}\;,\nn\\
P_{\mu\nu}^6&=&\epsilon_{\mu\nu\alpha}Q^\alpha\;,\nn\\
P_{\mu\nu}^{10}&=& -\epsilon_{\mu\alpha\beta}Q^\alpha n^\beta n_\nu \frac{Q^2}{{\bf q}^2}\;,\nn\\
P_{\mu\nu}^{11}&=& -\epsilon_{\nu\alpha\beta}Q^\alpha n^\beta n_\mu \frac{Q^2}{{\bf q}^2},
\eea
with
\be
n_\mu = g_{\mu 0} -\frac{q_0 Q_\mu}{Q^2}\;.
\ee
It turns out that these vacuum polarization effects enter into the fermion gap equation only through the combinations $G_L$ and $G_{DE}=G_D+G_E$. Then, the gap equation becomes a system of equation for the six unknown functions $F^0_\pm(P)$, $F_\pm(P)$ and $M_\pm(P)$, which depend on the value of the coupling constant $\alpha$ and the CS coefficient $\vartheta$. Different models of $G_L$ and $G_{DE}$ are proposed that are consistent with a Coulomb-like interaction.
The main conclusions that are drawn from this work are
\begin{itemize}
    \item The gap equation supports solutions which break parity, associated to the generation of a Haldane mass.
    \item Non-trivial parity-preserving solutions are found that, nevertheless break chiral symmetry.
    \item The Dirac dynamical mass  reduces as $\theta$ increases. This confirms that the CS coefficient behaves as an effective dielectric constant, as was proposed in~\cite{RQEDcurved}.
\end{itemize}

A point-wise evolution of the dynamical mass in terms of the coupling and the CS term was considered by some of us in \cite{Olivares:2020eko}. Assuming that Fermi velocity renormalization has already been carried out and thus no need to split the spatial and temporal components of the momentum, a rainbow-ladder truncation of the gap equation was adopted  with the additional simplification of neglecting the wavefunction renormalization effects setting $F_\pm(p)=1$ in Landau gauge. The gap equation is found equivalent to the following uncoupled system of non-linear integral equations for the left and right handed mass functions
%
%
\begin{eqnarray} \nonumber
M_\pm(p)&=&2\pi\alpha\int\frac{d^3k}{(2\pi)^3}\Big[\frac{2M_{\pm}(k)}{k^2+M^2_{\pm}(k)}\frac{1}{q(1+\vartheta^2)}\\
&\mp&\frac{1}{q^2}\frac{\vartheta}{1+\vartheta^2}\frac{k\cdot q}{k^2+M^2_{\pm}(k)}\Big].
\label{mpm}
\end{eqnarray}
Inspired by the reasoning of the previous section, the gap equation can be linearized and reduced to the following differential equation
\begin{eqnarray} \nonumber
p^2M_\pm''(p)+2pM_\pm'(p)+\frac{2\alpha}{\pi(1+\vartheta^2)}M_\pm(p)\\
=\mp\frac{2\alpha\vartheta}{\pi(1+\vartheta^2)}\Big[\frac{5p}{3}-\frac{6p^2}{\kappa}+\frac{\Lambda^3}{9p^2}-\frac{p}{9}\Big],
\label{diff_eq}
\end{eqnarray}
with the boundary conditions
\begin{equation}
p^2{M'}_\pm(p)\Bigg|_{p\to \kappa} =0, \qquad (p {M'}_\pm(p)+M_\pm(p))\Bigg|_{p=\Lambda}  =0,
\end{equation}
where $\kappa$ is the IR regulator introduced as a scale that quantifies the amount of mass being dynamically generated.
The general solution to eq.\eqref{diff_eq} is
\begin{eqnarray}
M_\pm(p)&=&p^{-\frac{1}{2}\sqrt{1-\frac{\alpha}{\alpha_c}}-\frac{1}{2}} \left(c_\pm^{(2)} p^{\sqrt{1-\frac{\alpha}{\alpha_c}}}+c_\pm^{(1)}\right)\nonumber\\
&+&f(\vartheta, p)
\label{SD_solution}
\end{eqnarray}
where 
\begin{equation}
\alpha_c=\frac{\pi}{8}(1+\vartheta^2)\label{crit}
\end{equation}
and
\begin{eqnarray} \nonumber
f(\vartheta, p)&=&\Bigg[\pi  \left(\vartheta ^2+1\right) \left(\mp \frac{2\alpha  \vartheta }{\vartheta ^2+1}\right) \bigg(\kappa \left(\alpha +3 \pi  \left(\vartheta ^2+1\right)\right)\\ \nonumber
&&\left(\Lambda ^3+14 p^3\right)-54 p^4 \left(\alpha + \pi  \left(\vartheta ^2+1\right)\right)\bigg)\Bigg]\\
&\times&\frac{1}{18 \kappa p^2 \left(\alpha + \pi  \left(\vartheta ^2+1\right)\right) \left(\alpha +3 \pi  \left(\vartheta ^2+1\right)\right)}.
\label{f}
\end{eqnarray}
We notice that when $\vartheta=0$, $f(\vartheta,p)=0$ and we obtain precisely the solution discussed in~\eqref{eq:MirRQED} which  demands  $\alpha>\alpha_c=\pi/8$.  For $\vartheta\ne 0$, we observe the role of the CS coefficient $\vartheta$  as an effective dielectric constant. The function $f(\vartheta,p)$ describes deviations from the Miransky scaling by virtue of the CS coefficient. 

The numerical solution to the non-linear equations~\eqref{mpm} are shown in Figs.~\ref{fig:Mminus_p} and~\ref{fig:Mplus_p}
%
for $\vartheta=\eta \vartheta_c$ with varying  $0\leq\eta\leq1$ but keeping $\alpha=1.07 \alpha_c$ fixed. These numbers are consistent with a critical value of $\vartheta_c=6.6\times 10^{-14}$. 
For small values of $p$, we observe that both masses are finite and approximately flat as $p\to 0$. For larger momentum, the height of the mass functions decreases. This happens for different values of $\vartheta$ provided $\alpha>\alpha_c$. Furthermore,  Figs.~\ref{fig:m_plus} and~\ref{fig:m_minus} show that the dynamical masses depend on the magnitude of the CS coefficient in opposite forms for right- and left-handed fermions. While $M_-(0)$ increases  $M_+(0)$ decreases as $\vartheta$ grows bigger. At the critical value $\vartheta_c$, $M_+(0)$ flips sign abruptly, signaling a first order phase transition. 



\begin{figure}
\includegraphics[width=8cm]{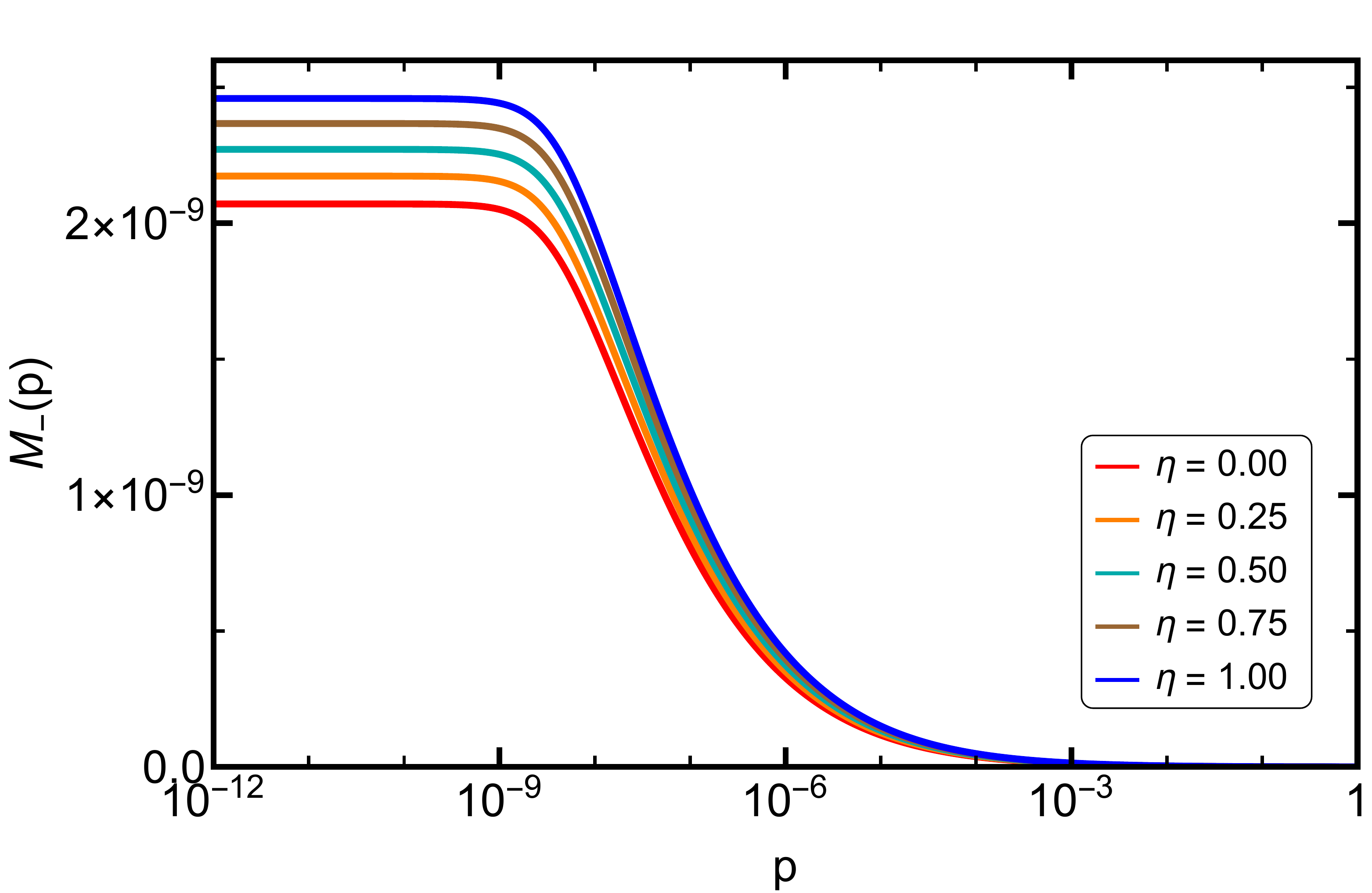}
\caption{Mass function $M_-(p)$ plotted as a function of momentum. The different curves correspond to different values of the CS parameter $\vartheta=\eta \vartheta_c$ with $0\leq\eta\leq1$ and fixed value of the coupling $\alpha=1.07 \alpha_c$.}
\label{fig:Mminus_p}
\end{figure}



\begin{figure}
\includegraphics[width=8cm]{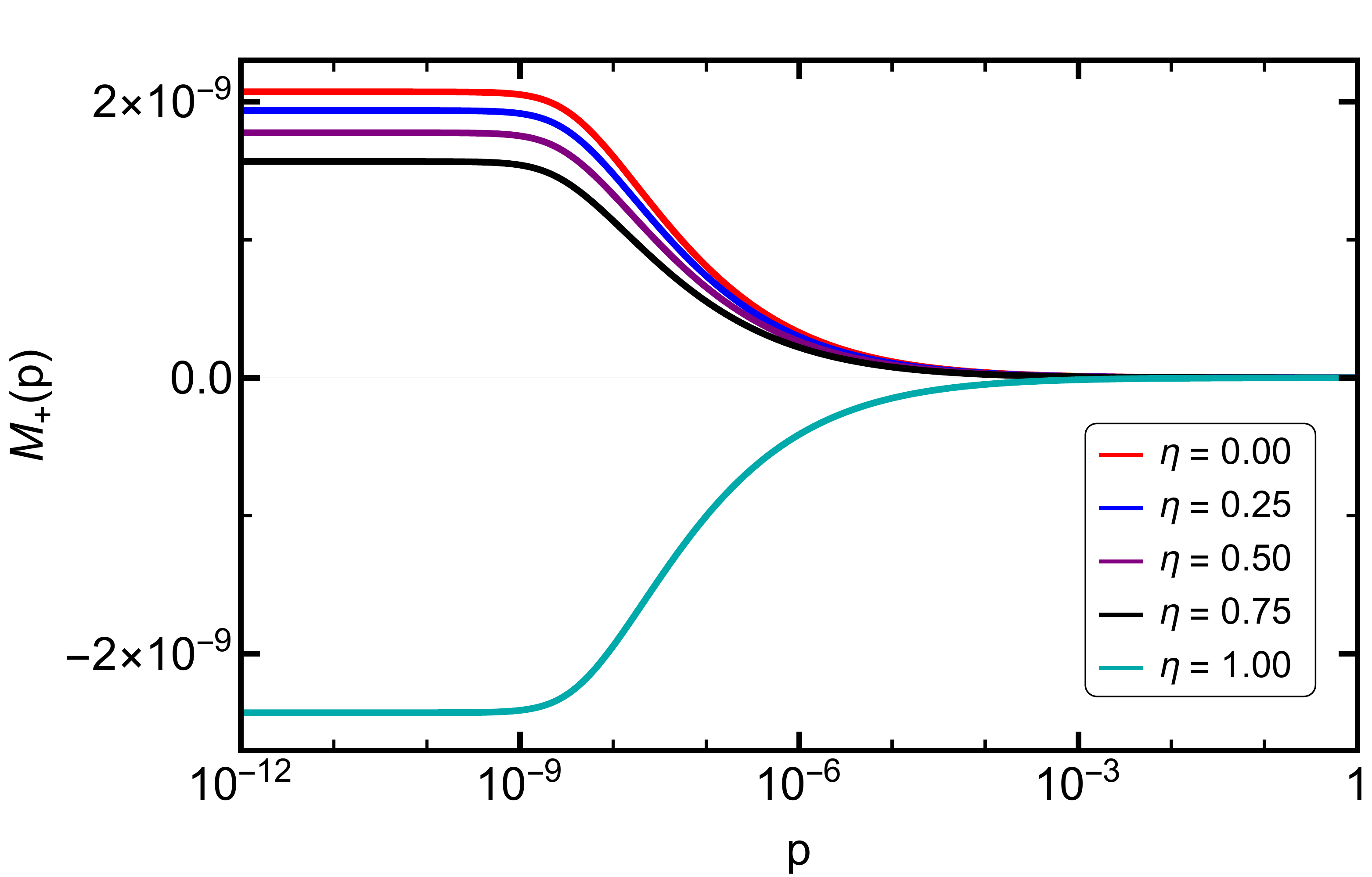}
\caption{Mass function $M_+(p)$ plotted as a function of momentum. The different curves correspond to different values of the CS parameter $\vartheta=\eta \vartheta_c$ with $0\leq\eta\leq1$ and fixed value of the coupling $\alpha=1.07 \alpha_c$.}
\label{fig:Mplus_p}
\end{figure}



\begin{figure}
\includegraphics[width=8cm]{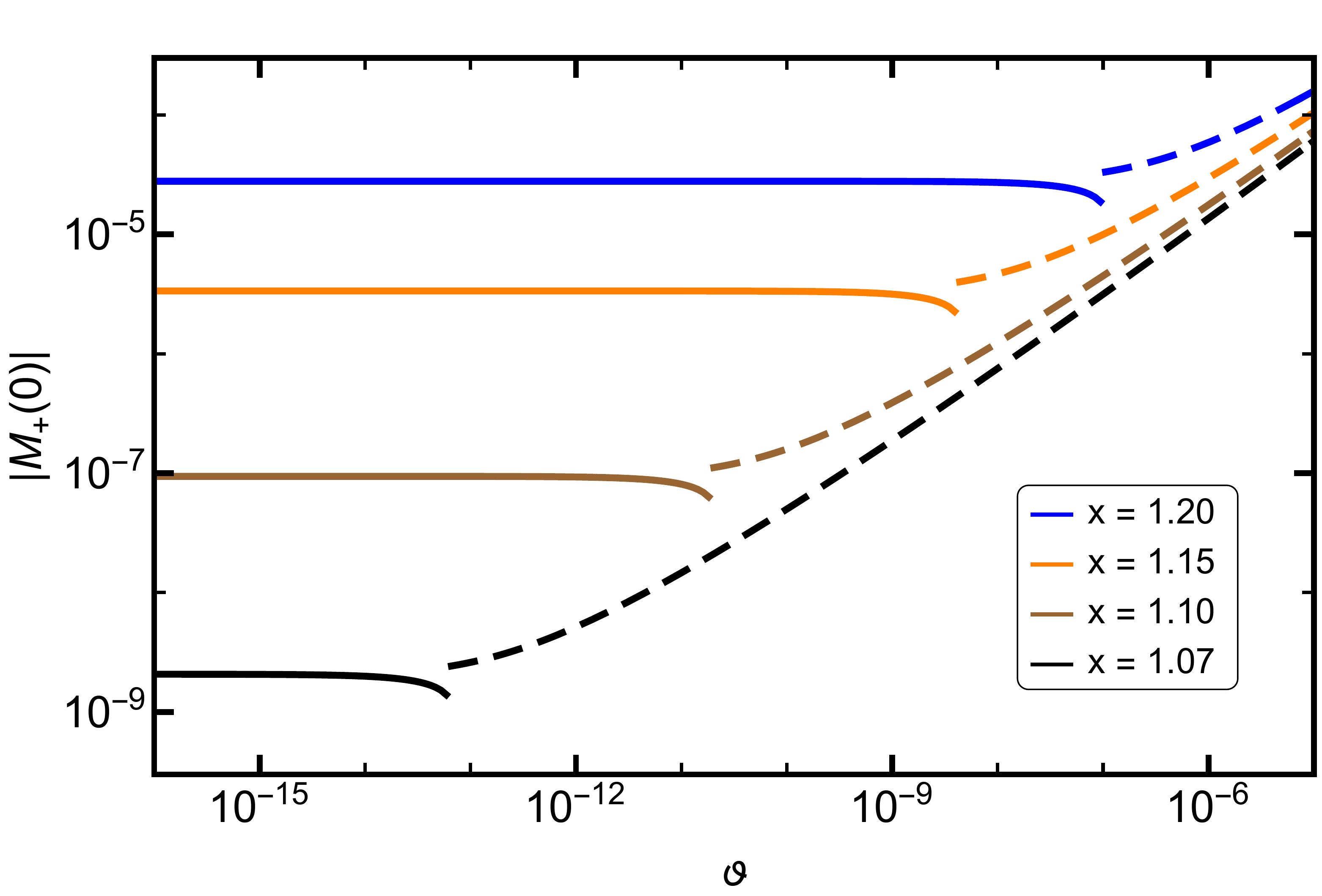}
\caption{The absolute value of the height of the right-handed fermion mass function $M_+(0)$ plotted as a function of the CS parameter $\vartheta$. Above $\vartheta_c$ the mass $M_+(0)$ becomes negative and here we flipped its sign for a better visualization. The different curves correspond to different values of the coupling $\alpha=x \alpha_c$.}
\label{fig:m_plus}
\end{figure}



\begin{figure}
\includegraphics[width=8cm]{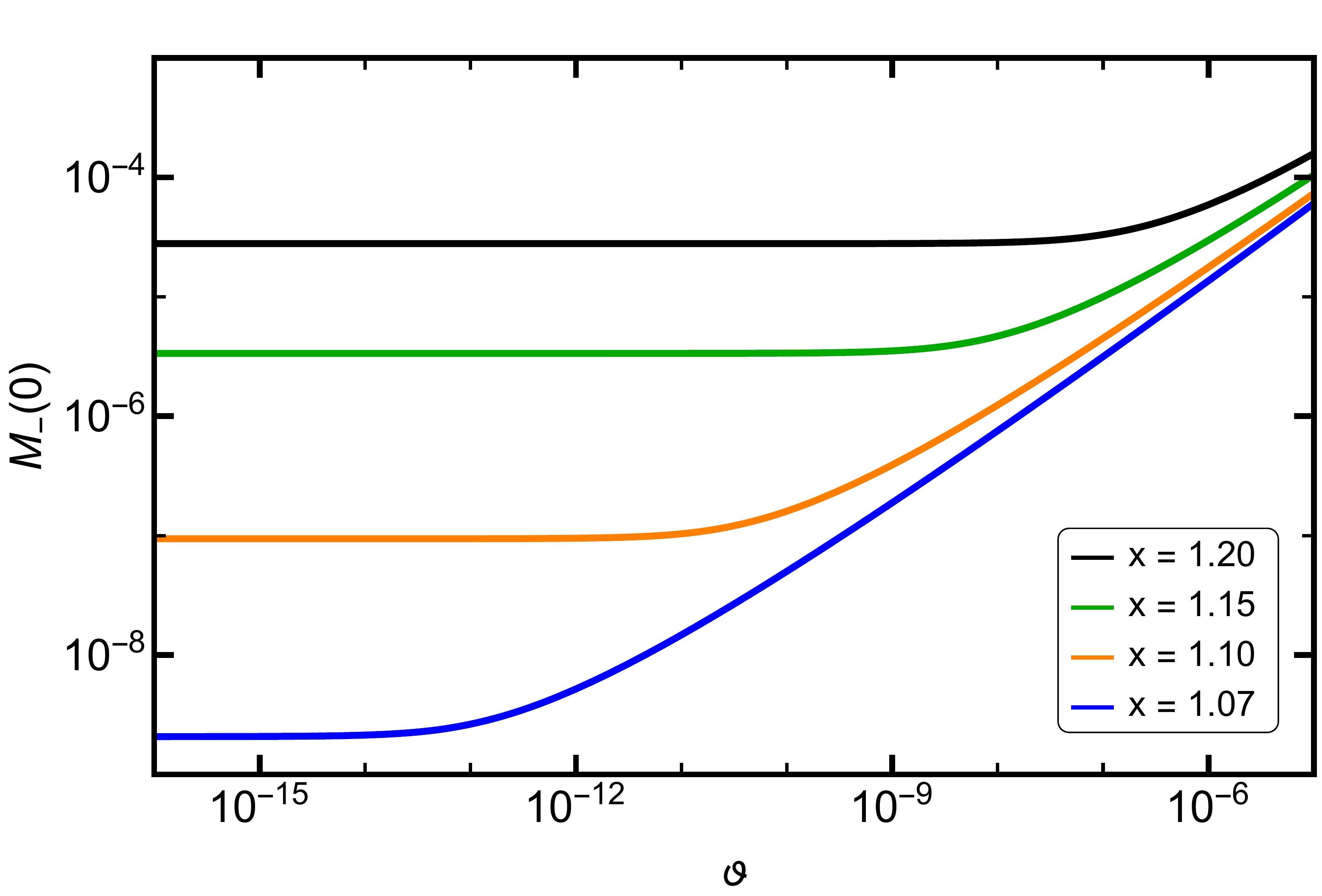}
\caption{Height of the left-handed fermion mass function $M_-(0)$ plotted as a function of the CS parameter $\vartheta$. The different curves correspond to different values of the coupling $\alpha=x \alpha_c$.}
\label{fig:m_minus}
\end{figure}



\begin{figure}
\includegraphics[width=8cm]{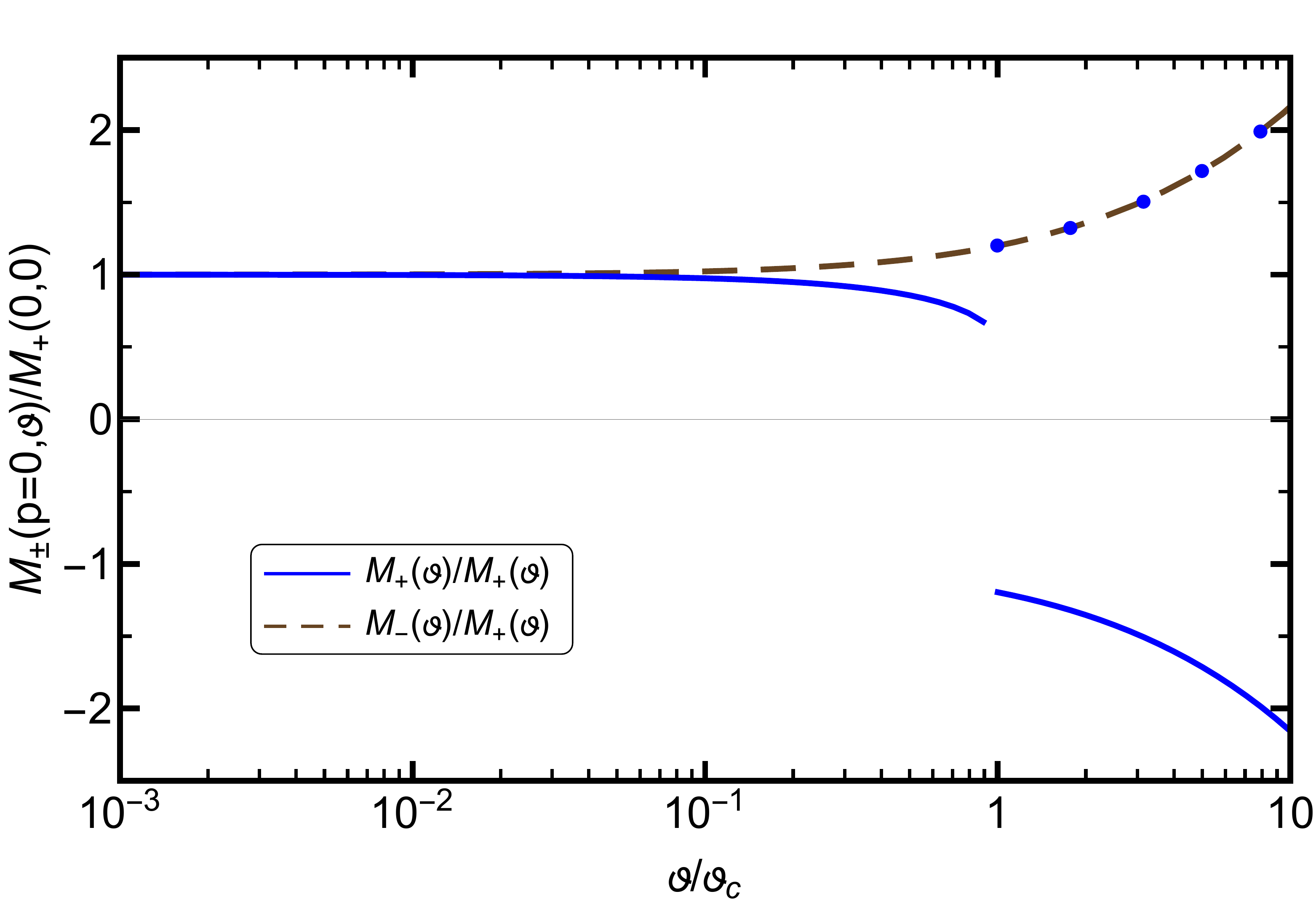}
\caption{$M_\pm(0)$ re-scaled by their value at $\vartheta=0$. The solid line corresponds to $M_+(0)$ and dashed line $M_-(0)$. The red dots are the absolute value $|M_+(0)|$ above the critical $\vartheta$.}
\label{fig:scaled_mass}
\end{figure}

Notice that the heights $M_\pm(0)$ of the mass functions parametrize how much mass has been generated. These hights are depicted in Figs. \ref{fig:m_plus} and \ref{fig:m_minus} for a constant value of $\alpha=x\alpha_c$. Above $\vartheta_c$, $M_+(0)$ undergoes a discontinuity and jumps assuming negative values, as shown in Fig.\ref{fig:scaled_mass}, where we have flipped the sign after the discontinuity to lead the eye.  The blue dots stand for the absolute value of $M_+$ above $\vartheta_c$, where we observe that the two masses become the mirror image of one another. We notice that since the curves are re-scaled, they coincide for any value of $x>1$.


The Dirac and Haldane masses that appear in the Lagrangian (\ref{massive_Lagrangian}) can be obtained summing up or subtracting $M_\pm(0)$, respectively. From  Figs.~\ref{fig:m_even} and~\ref{fig:m_odd}, we notice that both these masses are non-vanishing for a large range of values of $\vartheta$, clearly showing that CSB can occur within this model. For small $\vartheta$, the Dirac mass $m_e$, associated to chiral symmetry breaking, presents a plateau, leaping at a critical $\vartheta$ to a value several orders of magnitude smaller. After that,  $m_e$ increases again and finally drops to zero. We interpret the discontinuity as an attempt to restore chiral symmetry, associated to the first piece in Eq.(\ref{mpm}). That would be the critical point if $f(\vartheta,0)$ was absent. However, as one moves towards the right of the discontinuity, the CS contribution $f(\vartheta,0)$ becomes dominant and $m_e$ starts to be significant again. For even larger $\vartheta$ the Dirac mass drops and the real critical $\vartheta$, for which chiral symmetry is restored, is around $\vartheta=1$. In turn, parity breaking is encoded in $m_o$. As the CS parameter becomes dominant, the Haldane mass increases, never restoring the symmetry. 

As in the finite temperature case, the gauge dependence of these findings represent  an open avenue to explore~\cite{christoph1,christoph2}. Furthermore, feedback effects of the dynamical mass and the CS coefficient $\vartheta$ need to be incorporated by solving the SDE for the fermion and photon propagators simultaneously. 



\begin{figure}
\includegraphics[width=8cm]{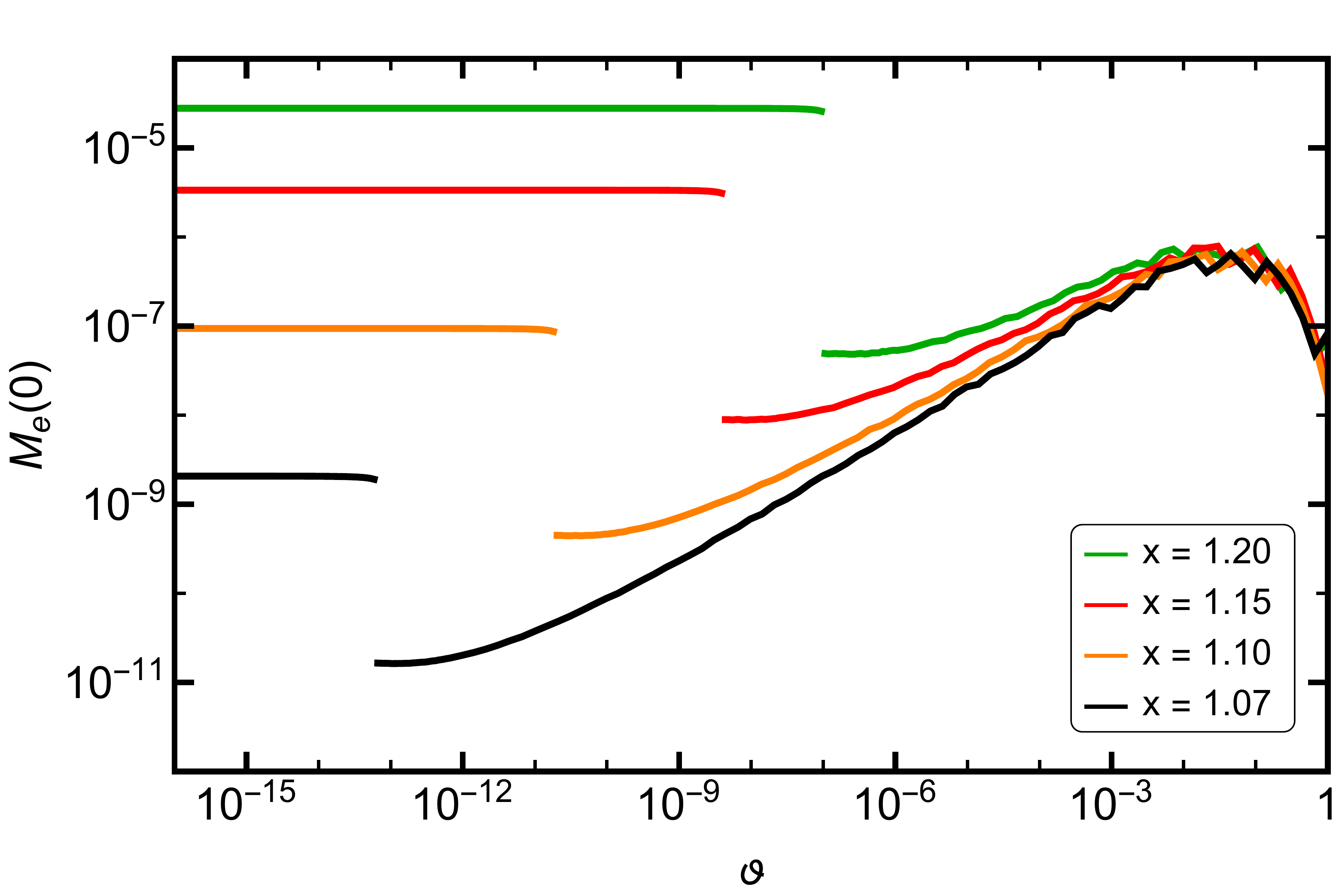}
\caption{Dynamical mass term, even under parity transformation, as a function of the CS parameter $\vartheta$.}
\label{fig:m_even}
\end{figure}



\begin{figure}
\includegraphics[width=8cm]{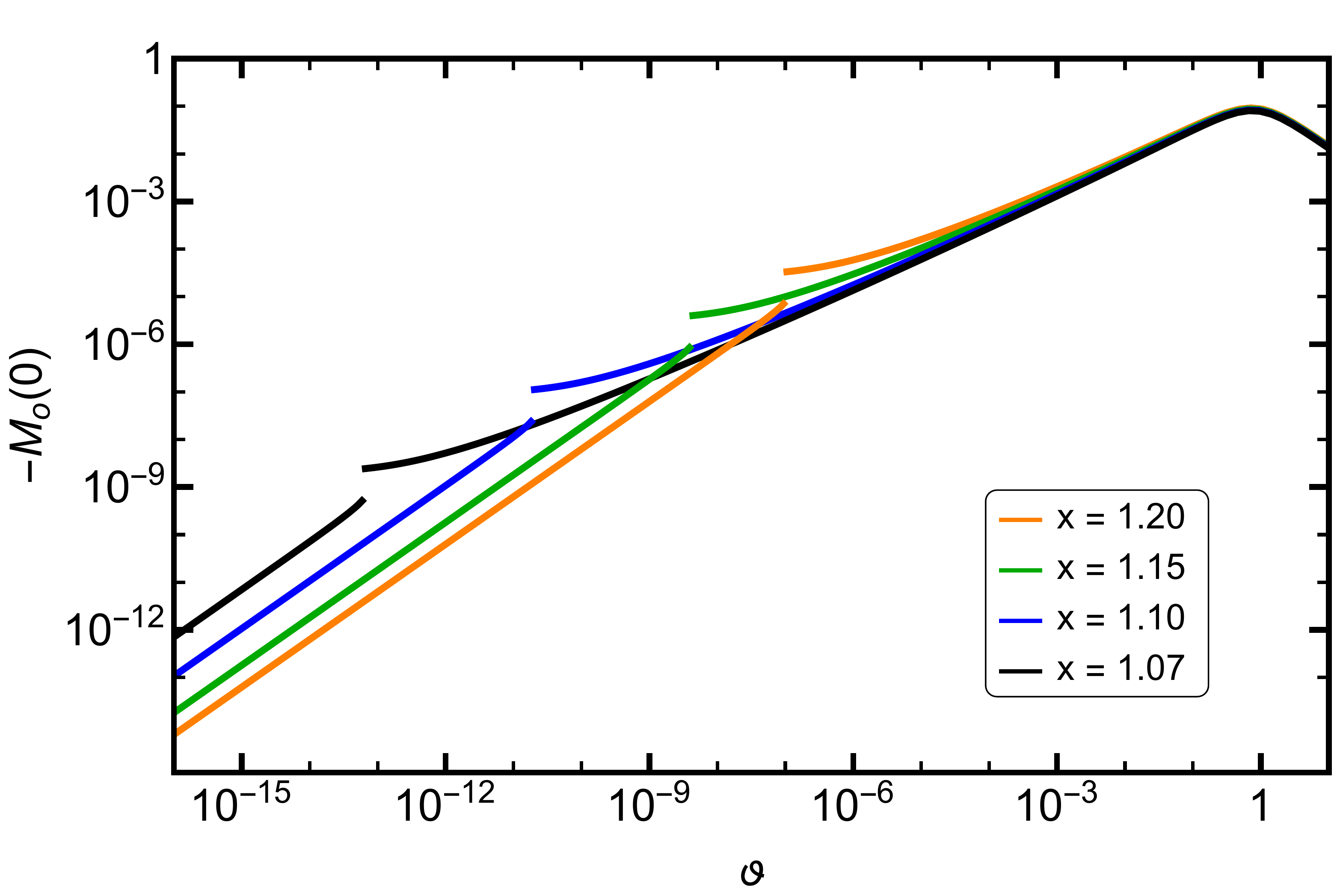}
\caption{Dynamical mass term, odd under parity transformation, as a function of the CS parameter $\vartheta$}
\label{fig:m_odd}
\end{figure}

\subsection{Anisotropy effects on chiral symmetry breaking}

Straintronics has emerged as the field in which the electrical properties of graphene and other materials are manipulated through mechanical deformations (see, for instance, Ref.~\cite{Naumis}). Effects of anisotropy (strain) in the gap equation have been recently considered in Ref.~\cite{Carrington:2020qfz}. By writing the fermion propagator of RQED with strain as
\be
S(p)=-\left[\gamma_\mu M_{\mu \nu}p_\nu \right]^{-1},
\ee
with
\be
M_{\mu \nu}=\left( \begin{array}{ccc}
1 & 0 & 0 \\ 0 & v_1 & 0 \\ 0 & 0 & v_2
\end{array}\right),
\ee
the gap equation is considered as a function of the anisotropy parameter $\eta=v_1/v_2$. Including one-loop vacuum polarization effects and a suitable generalization of the Ball-Chiu vertex, the authors of Ref.~\cite{Carrington:2020qfz} solved the gap equation and searched for the effect of $\eta$ on the critical coupling. They observe that the net effect is to increase the value of $\alpha_c$ for chiral symmetry breaking.

\subsection{Chiral symmetry breaking in curved space}

It is well known that for 2D materials, electric and optical properties are modified by impurities and deformation. The latter can be introduced 
in a  framework in which these mechanical deformations can be incorporated in terms of curved geometry of the underlying space-time in the equations of motion. In this regard, the dynamics of gauge bosons and electrons in mixed space-time dimensions has been considered in a curved-space background for the latter (and flat extra dimensions for the former)~\cite{RQEDcurved}. The starting point is the action
\bea
S&=& \int d^{d_\gamma}x \ \sqrt{-q}\left[-\frac{1}{4}F_{\mu\nu} F^{\mu\nu}-\frac{1}{2\xi}\left( \nabla_\mu A^\mu\right)^2 \right]\nn\\
&&\hspace{-5mm}
+ \int d^{d_e}x \sqrt{-H}\bar\psi i\gamma^\mu(x)\left( \partial_\mu+\Omega_\mu+ieA_\mu\right) \psi,
\eea
where $F_{\mu\nu}=\nabla_\mu A_\nu-\nabla_\nu A_\mu=\partial_\mu A_\nu-\partial_\nu A_\mu$, due to the cancellation of connection terms in the derivatives $\nabla_\mu$, $\gamma^\mu(x)=e^\mu_\nu \gamma^\nu$, $(\Omega_\mu)^\alpha_\beta=1/2 \omega_\mu^{ab} (J_{ab})^\alpha_\beta$, where  $(J_{ab})^\alpha_\beta$ represent the Lorentz generators in spinor space and $\omega_{\mu \phantom{m}b}^{a}=e_b^\nu (-\delta_\lambda^\nu \partial_\mu + \Gamma^\nu_{\lambda \mu}e^a_\lambda)$ stands for the spin connection, related to the Christoffel connection through the metricity condition namely $\nabla_\mu e^a_\nu = \partial_\mu e^a_\nu-\Gamma^\lambda_{\mu\nu}e^a\lambda+{(\omega_\mu)^a}_b e^n_\nu=0$. In all the above, the {\em vielbein} $e^a_\lambda$ verifies $\eta_{ab}e^a_\mu e^b_\nu=g_{\mu\nu}$. The induced metric on the brane (boundary space) is denoted by $H_{\alpha \beta}$ and $g_{\mu\nu}$ is the bulk space metric. For applications in graphene, it is considered that
\be
g_{\mu\nu}dx^\mu dx^\nu=dt^2-dz^2-h_{ij}dx^i dx^j,
\ee
with $i,j=1,2$ and thus
\be
\int
d^{d_\gamma}x \sqrt{-g}=\int dt\ dz \  dx^1 \ dx^2\ \sqrt{h}.
\ee
Considering the representation
\be
\tilde{e}^a_\mu(x)=\left\{ 
\begin{array}{cc}
e^a_\mu(x)\delta(x^{d_\gamma-d_e}), &  a, \mu = \mu_e \\
0 & a, \mu = d_e,\ldots d_\gamma -1 \end{array}
\right.
\ee
where in the case of graphene $x^{d_\gamma-d_e}$ corresponds to the third spatial dimension. In fact, assuming that all extra dimensions are flat, the action of this theory is very similar to the corresponding action of QED is curved space~\cite{RQEDcurved}, namely,
\bea
S&=&\int d^{d_\gamma}x\sqrt{-g}\Bigg[-\frac{1}{4}F_{\mu\nu}F^{\mu\nu}-\frac{1}{2\xi}\left(\nabla_\mu A^\mu \right)^2 \nn\\
&&\bar\psi {\bar{\gamma}}^\mu(x)\left(\partial_\mu +\Omega_\mu +ieA_\mu \right)\psi
\Bigg],
\eea
where ${\bar{\gamma}}^\mu(x)=e^a_\mu(x) \gamma^a$. In the local momentum representation, the propagators of the theory have the form
\bea
S_0(p)&=& \frac{{\not \! p}}{p^2-M_e^2},\\
\Delta_{\mu\nu}^{(0)}(q)&=&\frac{1}{q^2-M_\gamma^2}\left( g_{\mu\nu}+(\xi-1)\frac{q_\mu q_\nu}{q^2-M_\gamma^2}\right),\nn
\eea
where
\be
M_e^2=\frac{R(x')}{12},\qquad M_\gamma^2=-\frac{R(x')}{6},
\ee
where $R(x')$ is the Ricci scalar. For graphene, $M_\gamma^2=0$ and $M_e$ is a constant.

The one-loop renormalization of the fermion propagator has been discused in~\cite{RQEDcurved}. All divergent terms are the same as in curved-space ordinary QED. Curvature effects are only seen in the finite parts of the self-energies. As for the polarization tensor, such a term is found finite, hence making the beta function null at the one-loop level. Furthermore, the curvature effects in the case of graphene show up as an effective chemical potential of the form $\mu=\sqrt{M_e^2v_F^4}$, where $v_F$ is the effective Fermi velocity.

\section{Landau-Khalatnokov-Fradkin transformations}
\label{sec:LKF}

Gauge symmetry lays at the very foundation of the description of fundamental interactions. It can manifest in different ways at many levels. For the Green functions of QED, different sets of relations among them can be derived from the fundamental symmetry of the theory, as for example Ward-\cite{ward},  Ward-Green-Takahashi- \cite{ward,green,takahashi} and Transverse
Ward-identities~\cite{twi1,twi2,twi3,twi4,twi5} which relate $(n+1)$-point to $n$-point functions in constructions resembling divergence and curl of currents. In a different setting, the so-called Landau-Khalatnikov-Fradkin transformations (LKFT)~\cite{Landau:1955zz,Fradkin:1955jr} which have been analyzed in different versions of QED and extended to non-Abelian gauge theories like QCD, have been derived from different arguments in the past decades~\cite{LKF1,LKF2,LKF3,LKF4,LKF5}. These transformations are non-perturbative in nature and hence have the nature to address the issue of gauge invariance in perturbative and non-perturbative studies of field theories. 

A widely used example of LKFT has been in the case of the fermion propagator. It has been used to establish the multiplicative renormalizability of spinor and scalar QED in different space-time dimensions~\cite{Fernandez-Rangel:2016zac,Ahmadiniaz:2015kfq,Kizilersu:2009kg,Bashir:2001vi}. The main lesson is that this transformation  fixes some of the coefficients of the perturbative expansion of the fermion propagator at all orders. In the case of RQED, given that WTI are satisfied, it turns interesting that the LKFT serves as a tool to establish general features of multi-loop calculations as well as non-perturbative calculations in connection with chiral symmetry breaking.
In what follows of this section, we review the LKFT in RQED.

\subsection{LKFT for the fermion propagator in equal dimensional QED}~\label{QED}

We start from the most general form of the Dirac fermion propagator in arbitrary space-time dimensions and in momentum space and relate it to its coordinate space representation
%
%
\begin{equation}
\label{eqn:SxSpace}
S(x;\xi)={\not \! x}X(x;\xi) + Y(x;\xi).
\end{equation}
%
through a Fourier
transformation. We have additionally added the gauge parameter label $\xi$ as we are interested in the form of these propagators for different gauges. 
%

For theories of fermions and bosons in the same space-time dimensions, the momentum space free gauge boson propagator $\Delta_{\mu\nu}^{(0)}(q)$ has the general form (in analogy with~\eqref{eq:photon_split})
\begin{equation}
\label{eqn:photond}
\Delta_{\mu\nu}^{(0)}(q)=\Delta_{\mu\nu}^{T}(q)+\xi\frac{q_\mu q_\nu}{(q^2)^2}\,.
\end{equation}
such that the longitudinal portion of the propagator, inversely proportional to $q^4$,  describes how this Green function changes in different gauges, being proportional to the gauge fixing parameter.  The LKFT emerges precisely from this part of the propagator~\cite{Landau:1955zz,Fradkin:1955jr,LKF1,LKF2,LKF3,LKF4,LKF5}. These transformations are more clearly written in coordinate space. For the fermion propagator, the LKFT states that the
fermion propagator in an arbitrary covariant gauge $S(x;\xi)$ is related to its form in Landau gauge $S(x;0)$ through the transformation
\begin{equation}
\label{eqn:LKFTQEDd}
S(x;\xi)=S(x;0)\me^{-\mi[\Delta_{d}(0)-\Delta_{d}(x)]}\;,
\end{equation}
where
%
\begin{equation}
\label{eqn:DeltadQED}
\Delta_{d}(x)=-\mi\xi e^{2}\mu^{4-d}\int\dfd{q}{d}\frac{\me^{-\mi q\cdot x}}{q^{4}},
\end{equation}
 $e$ denoting  the electric charge, and $\mu$ is an energy scale that renders $e$ dimensionless in $d=4$, but yields a dimensionful coupling $e^2$ in QED$_3$. 
Performing the momentum integration, $\Delta_{d}(x)$ is explicitly given by~\cite{Bashir:2002sp}

\begin{equation}
\label{eqn:DeltadQEDExpl}
\Delta_{d}(x)=-\frac{\mi\xi\alpha}{4\pi^{\frac{d-2}{2}}}\Gamma\left(\frac{d-4}{2}\right)
(\mu x)^{4-d},
\end{equation}

\noindent where $\alpha=e^2/(4\pi)$ is the coupling constant, and $\Gamma(z)$ is the Euler Gamma function.

To taste a feeling of the structure of LKFT, let us take $F(p;0)=1$ and $M(p;0)=0$, namely, the massless fermion propagator in Landau gauge as a seed of LKFT. 
%
%
For $d=3$, the transformation reveals that the wavefunction renormalization in an arbitrary covariant gauge is (see, for instance, Ref.~\cite{Bashir:2000,Bashir:2002sp})
\begin{eqnarray}
\label{eqn:SLKF3d}
F(p;\xi)&=&1-\frac{\bar{\alpha}\xi}{2p}\arctan{\left(\frac{2p}{\bar{\alpha}\xi} \right)},
\end{eqnarray}
\noindent with $\bar{\alpha}=e^2/(4\pi)$ which has mass dimension one. We see directly that a weak coupling expansion fixes all terms of the form $(\bar{\alpha}\xi)^j$ at any given order in perturbation theory. This is a major asset of  the LKFT. Below we discuss in detail the progress achieved so far regarding the  LKFT in RQED.

\subsection{The LKF transformation for RQED}~\label{RQED}

Let us consider the 
%
%
free gauge boson propagator over bulk dimensions in RQED. It is of the same form as in~\eqn{eqn:photond},  but when considered reduced to the $d_e$-dimensional brane, it becomes~\cite{Teber:2012de,Teber:2014hna}
\begin{equation}
\label{eqn:photondebrane}
\Delta_{{\mu_e}{\nu_e}}(q)= D(q^2)\left(g_{{\mu_e}{\nu_e}}-\frac{q_{\mu_e} q_{\nu_e}}{q^2}\right)+\tilde{\xi} D(q^2)\frac{q_{\mu_e} q_{\nu_e}}{q^2},
\end{equation}
\noindent  with,
\begin{eqnarray}
D(q^2)& =& \frac{i}{(4\pi)^{\varepsilon_e}}\frac{\Gamma(1-\varepsilon_e)}{(-q^2)^{1-\varepsilon_e}},
\end{eqnarray}
\noindent where $\varepsilon_{e}=(d_{\gamma}-d_{e})/2$ and $\tilde{\xi}=(1-\varepsilon_e)\xi$. From here we see that
the longitudinal part of the propagator changes the form of the LKFT for the fermion propagator from the equal dimensional theory.
Considering that the propagator changes from gauge to gauge according to
\begin{equation}
\label{eqn:LKFTGraphene}
S_{d_e}(x;\xi)=S_{d_e}(x;0)\me^{-\mi[\tilde{\Delta}_{d_e}(0;\varepsilon_{e})-\tilde{\Delta}_{d_e}(x;\varepsilon_{e})]},
\end{equation}
\noindent we define 
\begin{eqnarray}
\label{eqn:DeltaRQED}
\tilde{\Delta}_{d_{e}}(x;\varepsilon_{e})&=&-\mi f(\varepsilon_{e})\xi e^{2}\mu^{4-d_{\gamma}}\int\dfd{q}{d_{e}} 
\frac{\me^{-\mi q\cdot x}}{q^{4-2\varepsilon_{e}}} \\
&=&
\label{eqn:DeltaRQEDl2}
-\mi f(\varepsilon_{e})\xi e^{2}\frac{\Gamma(\frac{d_{e}-a}{2})}{2^{a}\pi^{d_{e}/2}\Gamma(\frac{a}{2})}(\mu x)^{a-d_{e}},\nonumber\\
\end{eqnarray}
with 
\be
f(\varepsilon_{e})=\frac{\Gamma(1-\varepsilon_{e})(1-\varepsilon_{e})}{(4\pi)^{\varepsilon_{e}}}, \qquad  a=4-2\varepsilon_{e}.
\ee
This is the 
general form of LKFT for the fermion propagator in RQED$_{d_\gamma,d_e}$ first presented in~\cite{AftabLKFT}. Note that when $d_{\gamma}=d_{e}=d$, ~\eqn{eqn:DeltaRQED} matches the usual form of the LKFT~\eqn{eqn:LKFTQEDd} 
in any dimension $d$. 
For  graphene and other 2D materials,
%
\begin{equation}
\label{eqn:DeltaGrapheneExpl}
\tilde{\Delta}_{3}\left(x,\frac{1}{2}\right)=\frac{-\mi\xi e^{2}}{16\pi^{2}}
\Gamma\left(\frac{1-2\epsilon}{2}\right)(\mu x)^{2\epsilon-1},
\end{equation}
where the limit $\epsilon\to 1/2$ is understood. Expanding~\eqn{eqn:DeltaGrapheneExpl} around $\epsilon=1/2$, defining $\delta=\epsilon-1/2$,  we get
\begin{equation}
\label{eqn:DeltaGrapheneExplExp}
\tilde{\Delta}_{3}\left(x,\frac{1}{2}\right)=\frac{\mi\xi e^{2}}{16\pi^{2}}\left[
\frac{1}{\delta}+ \gamma_{E}+2\ln(\mu x)
+ \mathcal{O}(\delta)\right].
\end{equation}
\noindent  Since the transformation function, \eqn{eqn:DeltaGrapheneExplExp}, diverges at $x=0$, we introduce a cutoff $x_{min}$ and consider
\begin{equation}
\label{eqn:DeltaGrapheneFinal}
-\mi\left[\tilde{\Delta}_{3}\left(x_{min},\frac{1}{2}\right)-\tilde{\Delta}_{3}\left(x,\frac{1}{2}\right)\right]
= \ln\left(\frac{x}{x_{min}}\right)^{-2\nu},
\end{equation}
\noindent where we have defined $\nu=\xi\alpha/(4\pi)$, and the dimensionless coupling constant $\alpha= e^{2}/(4\pi)$.

With \eqn{eqn:DeltaGrapheneFinal}, we can obtain the non-perturbative structure of the fermion propagator in RQED starting with the tree-level massless case in Landau gauge. In coordinate space, we have

\begin{eqnarray}
\label{eqn:XAnyGaugeGraphene}
X(x;\xi)&=&X(x;0)\me^{-\mi\left[\tilde{\Delta}_{3}\left(x_{min},\frac{1}{2}\right)-\tilde{\Delta}_{3}\left(x,\frac{1}{2}\right)\right]} \nn \\
&=&-\frac{x_{min}^{2\nu}}{4\pi}x^{-2\nu-3}.
\end{eqnarray}

\noindent whereas in momentum space, 

\begin{equation}
\label{eqn:FAnyGaugeGrapheneFinal}
F(p,\xi)=
\frac{\sqrt{\pi}}{2}\frac{\Gamma(1-\nu)}{\Gamma(\frac{3}{2}+\nu)}
\left(\frac{p^{2}}{\Lambda^{2}}\right)^{\nu}.
\end{equation}

\noindent This is the non-perturbative form of the fermion propagator in RQED in any covariant gauge $\xi$. The power-law behavior of this two-point function is consistent with the
multiplicative-renormalizable character of the theory. We notice that a similar behavior is found for the propagator in ordinary QED.

Next, expanding this result in powers of $\alpha$, we get

\begin{equation}
\label{eqn:FAnyGaugeGrapheneWC}
F(p,\xi)=1+\frac{\xi\alpha}{4\pi}F_{1} 
+ \left(\frac{\xi\alpha}{4\pi}\right)^{2}F_{2}+\mathcal{O}(\alpha^{3}),
\end{equation}

\noindent with 

\begin{eqnarray}
\label{eqn:FAnyGaugeGrapheneWCF1Redef}
F_{1}&=& \ln\left(\frac{p^{2}}{\Lambda^{2}}\right) - \gamma_{E}-\psi\left(\frac{3}{2}\right) \nn \\
&=&\ln\left(\frac{p^{2}}{\Lambda^{2}}\right)+2\gamma_{E}+\ln(4)-2, \\
\label{eqn:FAnyGaugeGrapheneWCF2Redef}
F_{2}&=&\frac{1}{2}\left[\left(\ln\left(\frac{p^{2}}{\Lambda^{2}}\right) 
-\gamma_{E}-\psi(3/2)\right)^{2}-2\zeta(2)+4 \right] \nn \\
&=&\frac{1}{2}\left[\left(\ln\left(\frac{p^{2}}{\Lambda^{2}}\right)+2\gamma_{E}+\ln(4)
-2\right)^{2}-2\zeta(2)+4\right]\nonumber\\
\end{eqnarray}

\noindent where $\psi(z)$ is the digamma function, $\zeta(s)$ is the Riemann Zeta function, and we have made use of the identity 
$\psi(3/2)=-\gamma_{E}-\ln(4) +2$.

We make a comparison against the exact one-loop calculation of the wavefunction renormalization\cite{Kotikov:2013eha,Teber:2014hna},
\begin{multline}
\label{eqn:SPertRQED43}
F(p;\xi)= 1 + 
\frac{\alpha}{4\pi}\left[\frac{4}{9}-\frac{1-3\xi}{3}\overline{L}\right] \\
+ \left(\frac{\alpha}{4\pi}\right)^{2}
\left[\frac{(1-3\xi)^{2}}{18}(\overline{L}^{2}-2\zeta(2)+4)\right. \\
+ \left.4\frac{(3\xi+7)\overline{L}+48\zeta(2)}{27}
-8\zeta(2)(\overline{L}+2-\ln(4))-\frac{280}{27}\right],
\end{multline}

\noindent where

\begin{equation}
\label{eqn:Lbar}
\overline{L}=\ln\left(-\frac{p^{2}}{\mu^{2}}\right)+\ln(4)-2,
\end{equation}

\noindent and $\mu$ is the renormalization mass scale.
This correction has a contribution 
that is proportional to the gauge parameter and one that remains finite in Landau gauge. Thus, we can only compare against the former from the LKFT result.  $F_1$ defined in  \eqn{eqn:FAnyGaugeGrapheneWCF1Redef} is equivalent to
$\overline{L}$ in \eqn{eqn:Lbar} provided we identify

\begin{eqnarray}
\label{eqn:iden1}
\ln\left(\frac{p^{2}}{\Lambda ^{2}}\right)+2\gamma_{E}&\to& \ln\left(-\frac{p^{2}}{\mu^{2}}\right).
\end{eqnarray}
 
At ${\cal O}(\alpha^{2})$, the propagator has terms independent,  linear and quadratic $\xi$. From Eqs.~(\ref{eqn:FAnyGaugeGrapheneWC}) and (\ref{eqn:FAnyGaugeGrapheneWCF2Redef}), we observe
the LKFT only gives terms proportional $\xi^{2}$ at order $\alpha^{2}$.  This means that we can only compare our $\alpha^{2}$ result, $F_{2}$, defined by \eqn{eqn:FAnyGaugeGrapheneWCF2Redef} with the coefficient of $(\alpha\xi/(4\pi))^{2}$ in the perturbative result, \eqn{eqn:SPertRQED43}.  We observe that
\begin{equation}
F_{2}\to \frac{9}{18}\left(\overline{L}^{2}-2\zeta(2)+4\right), 
\end{equation}

\noindent with the identification \eqn{eqn:iden1}, as expected.
Thus, we have shown that  there is full consistency between our LKFT result, \eqn{eqn:FAnyGaugeGrapheneWC}, and the perturbative
result, \eqn{eqn:SPertRQED43}, up to order $\alpha^{2}$. We hence predict the form of all the coefficients of the form $(\alpha\xi)^j$ in the all order
perturbative expansion from our LKFT result~\eqn{eqn:FAnyGaugeGrapheneFinal}.

\subsection{Extensions of the LKF transformation}

It is interesting to notice that the LKF transformation can be represented directly in momentum space. In Refs.~\cite{TeberLK1,TeberLK2,TeberLK3,TeberLK4} it is discussed the form in which the wavefunction renormalization is related in two gauges labeled by the parameters $\xi$ and $\eta$, respectively. Assuming that in gauge $\eta$ the multiloop expansion of the wavefunction renormalization is
\be
F(p,\eta)=\sum_{m=0}^\infty a_m(\eta)\alpha^m \left( \frac{\tilde{\mu}^2}{p^2}\right)^{m\epsilon},
\ee
with $\epsilon=2-d/2$ and $\tilde{\mu}$ is a renormalization scale, the LKFT implies that
\be
F(p,\xi)=\sum_{m=0}^\infty a_m(\xi) \alpha^m \left( \frac{\tilde{\mu}^2}{p^2}\right)^{m \epsilon},
\ee
 where the new coefficients
\bea
a_m(\xi)&=&a_m(\eta)\frac{\Gamma(2-(m+1)\epsilon)}{\Gamma(1+m\epsilon)}\nn\\
&& \times\sum_{l=0}^\infty \Bigg[\frac{\Gamma(1+(m+l)\epsilon)\Gamma^l(1-\epsilon)}{l! \Gamma(2-(m-l+1)\epsilon)}\nn\\
&& \times \left(\frac{(\eta-\xi)\alpha}{-\epsilon}\right)^l \left( \frac{\tilde{\mu}^2}{p^2}\right)^{l\epsilon}\Bigg].
\eea
With the appropriate choice of the renormalization scale $\tilde{\mu}$, the multiloop expansion of the fermion propagator can be obtained in an easier form.

Moreover, taking advantage of the above mentioned multiplicative renormalizability of the theory, by factorizing
\be
F(p;\xi)=Z_\psi(\alpha,\xi)F_r(p;\xi),
\ee
one can see from LKFT that
\be
\log Z_\psi(\alpha,\xi)=\log Z_\psi (\alpha, \eta) -\frac{(\xi-\eta)\alpha}{\epsilon}.
\ee
Thus, for the fermion anomalous dimension
\be
\gamma_\psi(\alpha,\xi)=-\beta(\alpha) \frac{\partial \log Z_\psi(\alpha,\xi)}{\partial \alpha}-\xi \alpha \frac{\partial \log Z_\psi(\alpha,\xi)}{\partial \xi},
\ee
where $\beta(\alpha)$ is the beta function and $\gamma(\alpha)$ the gauge boson anomalous dimension, given explicitly as
\be
\beta(\alpha)=-2\epsilon \alpha+\sum_{l=0}^\infty \beta_l \alpha^{l+1}, \qquad \gamma(\alpha)=-\sum_{l=0}^\infty \beta_l \alpha^l,
\ee
where $\beta_l=0$ for graphene. From here its is evident that the fermion anomalous dimension in two gauges, namely, 
\be
\gamma_\psi(\alpha,\xi)=\gamma_\psi(\alpha,\eta)-2\alpha(\xi-\eta),
\ee
i.e., the gauge dependence is encoded in the one-loop correction alone, whereas higher loops are gauge invariant.

The two-loop correction to the mass anomalous dimension $\gamma_m$ has been explored very recently in~\cite{revT}. In this work, the authors consider a theory in which the gauge boson lives in 4 dimensions, but the fermion fields are allowed to live in a general dimension $d_e$. From the multiplicative renormalizability of the fermion mass,
\be
m=Z_m m_r,
\ee
where the renormalization constant $Z_m$ is obtained from the relation 
\be
\frac{1+\Sigma_S}{1-\Sigma_V} Z_m = const.
\ee
where $\Sigma_S$ and $\Sigma_V$ are the scalar and vector projections of the fermion self-energy. Furthermore, the mass anomalous dimension runs with the energy scale $\mu$ as
\be
\gamma_m=\frac{d \log Z_m}{d\log \mu}.
\ee
Using the $\bar{{\rm MS}}$-regularization scheme, it is observed that in the particular case of RQED,
\bea
\gamma_m &=& \frac{32}{3}  \left(\frac{\alpha_r}{4\pi}\right)- 64\left(\frac{\alpha_r}{4\pi}\right)^2\left(N \zeta_2 -\frac{8}{27}\right)\nn\\
&&+ {\cal O}\left(\left(\frac{\alpha_r}{4\pi}\right)^3 \right) ,
\eea
 where $N$ is the number of fermion families and $\zeta_2$ is the Euler function. This mass anomalous function is completely gauge invariant, and thus the critical coupling and/or critical number of fermion flavor derived from $\gamma_m$ are also gauge invariant.

\section{Outlook}
\label{sec:outlook}

In this review article we have revisited several aspects of a mixed-dimensional theory of electron and gauge boson interactions aiming to describe the new family of 2D Dirac material like graphene and its cousins. Quantum field theories in mixed dimensions have been discussed theoretically for over 4 decades in literature, but the isolation of graphene has indeed boosted the interest of these highly non-local yet well behaved theories. The fact that RQED, the theory we have specialized in, is a fixed point in the renormalization group allows to extend general conclusions from the behavior of the Green's functions in RQED. From the historical point of view, we have revisited the formulation of the theory and the most fundamental aspects that render the theory suitable for phenomenological description of the properties of a variety of Dirac materials. The fact that the theory allows a standard $1/r$ fall-off  for static Coulomb interactions to the observation of the scale invariance of the theory makes it already interesting to explore. Fundamental aspects of causality and unitarity have been revisited along with perturbation theory studies of the propagators and vertices.

An interesting aspect of this formulation is the possibility to carry out non-perturbative calculations in particular those connected with the chiral symmetry of the theory which might be broken when the coupling is strong enough. Schwinger-Dyson equations and renormaization group analysis exploit the structure of the gap equation as compared to other formulations of QED$_3$ including vacuum polarization effects allow to extract critical properties of the chiral transition in RQED. Still a number of improvements and refining of the truncations to the infinite tower of SDE in particular in connection with the gauge invariance of the predictions still awaits for a formal answer. 

The influence of external agents has also been considered in this theory. Finite temperature effects indicate a relation of the transition temperature competing against the strong coupling which by virtue of the heat bath diminishes, hence restoring chiral symmetry. Findings of different groups are still controversial, hence pointing toward the need of a careful revision of the working hypothesis. Moreover, the influence of a chemical potential is still required to have a more realistic description of the physical system and eventual test in the laboratory. In this sense, an effort was made in \cite{Dudal:2021ret} to address the role of chemical potential in a configuration with spin-orbit coupling. The phase diagram in the plane of temperature and chemical potential of RQED, similar to that of QCD, is a pending assignment for the interested community.

The possibility of parity and time reversal symmetry breaking has been addressed by different groups by including the CS term in the fundamental Lagrangian. The existence of parity and time reversal breaking solution is remarkable in this theory. The effect of the CS coefficient theta as an effective dielectric constant is seen to impact the chiral symmetry breaking pattern inasmuch as the effective screening might restore chiral symmetry but still break the discrete symmetries in the theory.

An interesting feature is the modeling of elastic deformation and anisotropy in relation to the dynamics of fermions in curved space. The effects of curvature are seen as density effects through a definition of a position dependent chemical potential.

LKF tranformations are a key ingredient from gauge invariance that have a fundamental role in perturbative and non-perturbative studies in RQED. In the former case, the closed connection between RQED and QED$_3$ in the large $N$ limit serves as a benchmark for perturbative calculations in improving the calculation of vacuum polarization effects.

Many features are still pending to address in this theory. Some of which include the influence of magnetic and electric field in different configurations and potential connections with other phenomenology are important features that perhaps the community will address in the future.

\section*{Acknowledgements}
We benefited from discussions with L. A. Rangel, S. Henrn\'andez-Ortiz, K. Raya, J.~C. Rojas, J. B\'aez-Cuevas. JACO and AR acknowledge financial support from Consejo Nacional de Ciencia y Tecnolog\'{\i}a (M\'exico) under grant FORDECyT-PRONACES/61533/2020. AJM acknowledge financial support from FAPESP under grant 2016/12705-7. 




\begin{thebibliography}{99}
\bibitem{Miransky:2015ava} 
V.~A.~Miransky and I.~A.~Shovkovy,
{\em Quantum field theory in a magnetic field: From quantum chromodynamics to graphene and Dirac semimetals,}
Phys.\ Rept.\  {\bf 576}, 1 (2015)
\url{https://doi.org/10.1016/j.physrep.2015.02.003}



\bibitem{Wallace} P. R. Wallace,
 {\em The Band Theory of Graphite} 
 Phys. Rev. {\bf 71} (9), (1947), 622-634. \url{https://doi.org/10.1103/PhysRev.71.622}


 \bibitem{Semenoff}
G. W. Semenoff, {\em Condensed-Matter Simulation of a Three-Dimensional Anomaly} Phys. Rev. Lett. {\bf 53}, (1984) 2449.
\url{https://doi.org/10.1103/PhysRevLett.53.2449}

\bibitem{Haldane:1988zza}
F. D. M. Haldane,
{\em Model for a Quantum Hall Effect without Landau Levels: Condensed-Matter Realization of the Parity Anomaly}.
Phys. Rev. Lett. {\bf 61}, (1988) 2015-2018.
\url{https://doi.org/10.1103/PhysRevLett.61.2015}


\bibitem{supercond1} N Dorey, and N. E. Mavromatos {\em QED3 and two-dimensional superconductivity without parity violation}  Nucl Phys B {\bf 386}  (1992) 614. \url{https://doi.org/10.1016/0550-3213(92)90632-L}

\bibitem{supercond2} K. Farakos and N. E.  Mavromatos {\em Dynamical gauge symmetry breaking and superconductivity in three-dimensional systems} 
Mod Phys Lett A {\bf 13} (1998) 1019. \url{https://doi.org/10.1142/S0217732398001091} 

\bibitem{supercond3} M. Franz and Z. Tesanovic {\em Algebraic Fermi Liquid from Phase Fluctuations: ``Topological'' Fermions, Vortex ``Berryons,'' and 
QED$_3$
 Theory of Cuprate Superconductors}  Phys Rev Lett {\bf 87} (2001) 257003. \url{https://doi.org/10.1103/PhysRevLett.87.257003}.
 
\bibitem{supercond4} I. F. Herbut, {\em QED$_3$
 theory of underdoped high-temperature superconductors}
 Phys Rev B {\bf 66} (2002) 094504. \url{https://doi.org/10.1103/PhysRevB.66.094504}

\bibitem{supercond5} M. Franz, Z. Tesanovic, O Vafek  {\em QED$_3$ theory of pairing pseudogap in cuprates: From $d$-wave superconductor to antiferromagnet via an algebraic Fermi liquid} Phys Rev B {\bf 66} (2002) 054535. \url{https://doi.org/10.1103/PhysRevB.66.054535}

\bibitem{graphene1} 
K. S. Novoselov, A. K. Geim, S. V. Morozov, D. Jiang, Y. Zhang, S. V. Dubonos, I. V Grigorieva and A. A. Firsov,
{\em Electric field effect in atomically thin carbon films}, Science {\bf 306}, (2004) 666. \url{https://doi.org/10.1126/science.1102896
}

\bibitem{graphene3} 
Zhang Y, Tan Y-W, Horst L S and Kim P,
{\em Experimental observation of the quantum Hall effect and Berry's phase in graphene}
Nature {\bf 438} (2005), 201.
\url{https://doi.org/10.1038/nature04235}

\bibitem{relativistic}
K. S. Novoselov, et al.,
{\em Two-dimensional gas of massless Dirac fermions in graphene} Nature {\bf 438}, (2005) 197-200. \url{https://doi.org/10.1038/nature04233}

\bibitem{rise} A. Geim, and K. Novoselov, {\em The rise of graphene}, Nature Mater {\bf 6}, (2007) 183. \url{https://doi.org/10.1038/nmat1849}

\bibitem{craig} C. J. Burden, J. Praschifka, C. D. Roberts, {\em Photon polarization tensor and gauge dependence in three-dimensional quantum electrodynamics}, Phys. Rev. D {\bf 46}(6), (1992) 2695-2702. \url{https://doi.org/10.1103/physrevd.46.2695}

\bibitem{Marino}
E.~Marino, {\it Quantum Field Theory Approach to Condensed Matter Physics}, (Cambridge University Press, Cambridge, UK, 2017).

\bibitem{Marino:1992xi}
E.~Marino,
{\em Quantum electrodynamics of particles on a plane and the Chern-Simons theory,}
Nucl. Phys. B {\bf408}, (1993) 551-564.
\url{https://doi.org/10.1016/0550-3213(93)90379-4}

\bibitem{Gonzalez:1993uz}
J. Gonzalez, F. Guinea and M. A. H. Vozmediano,
{\em Non-Fermi liquid behavior of electrons in the half filled honeycomb lattice (A Renormalization group approach)},
Nucl. Phys. B {\bf 424} (1994), 595-618.
\url{https://doi.org/10.1016/0550-3213(94)90410-3}

\bibitem{QHE_tong}
D.~Tong, {\it The Quantum Hall Effect}, Lecture Notes, https://www.damtp.cam.ac.uk/user/tong/qhe/qhe.pdf

\bibitem{doAmaral:1992td}
R. L. P. G. do Amaral and E. C.~Marino,
{\em Canonical quantization of theories containing fractional powers of the d'Alembertian operator},
J. Phys. A {\bf 25} (1992), 5183-5200
\url{https://doi.org/10.1088/0305-4470/25/19/026}

\bibitem{Marino:2014oba}
E. C. Marino, L. O. Nascimento, V. Alves and C. M. Smith,
{\em Unitarity of theories containing fractional powers of the d\textquoteright{}Alembertian operator,''}
Phys. Rev. D {\bf90} (2014) no.10, 105003
\url{https://doi.org/10.1103/PhysRevD.90.105003}

\bibitem{RG2}
J. Gonzalez, F. Guinea and M. A. H. Vozmediano,
{\em Unconventional Quasiparticle Lifetime in Graphite},
Phys. Rev. Lett. {\bf 77} (1996), 3589-3592
\url{https://doi.org/10.1103/PhysRevLett.77.3589}

\bibitem{RG3}
J. Gonzalez, F. Guinea and M. A. H. Vozmediano,
{\em Marginal-Fermi-liquid behavior from two-dimensional Coulomb interaction},
Phys. Rev. B {\bf 59} (1999), 2474-2477
\url{http://doi.org/10.1103/PhysRevB.59.R2474}

\bibitem{RG4}
J. Gonzalez, F. Guinea and M. A. H. Vozmediano,
{\em Electron-electron interactions in graphene sheets}
Phys. Rev. B {\bf 63} (2001) 134421
\url{http://doi.org/10.1103/PhysRevB.63.134421}


\bibitem{Vozmediano:2010fz}
M. A. H. Vozmediano,
{\em Renormalization group aspects of graphene,}
Phil. Trans. Roy. Soc. Lond. A {\bf 369} (2011), 2625-2642
\url{http://doi.org/10.1098/rsta.2010.0383}


\bibitem{Fernandez:2020wne}
L.~Fern\'andez, V.~Alves, L.~O.~Nascimento, F.~Pe\~na, M.~Gomes and E.~C.~Marino,
Phys. Rev. D \textbf{102} (2020) no.1, 016020
doi:10.1103/PhysRevD.102.016020
[arXiv:2002.10027 [hep-th]].

\bibitem{Gorbar:2001qt} 
E.~V.~Gorbar, V.~P.~Gusynin and V.~A.~Miransky,
{\em Dynamical chiral symmetry breaking on a brane in reduced QED,}
Phys.\ Rev.\ D {\bf 64}, (2001) 105028 
\url{https://doi.org/10.1103/PhysRevD.64.105028}
  
  
\bibitem{Teber:2012de} 
  S.~Teber,
 {\em Electromagnetic current correlations in reduced quantum electrodynamics,}
  Phys.\ Rev.\ D {\bf 86} (2012) 025005,
\url{https://doi.org/10.1103/PhysRevD.86.025005}
  
  \bibitem{Kotikov:2013eha} 
A.~V.~Kotikov and S.~Teber,
{\em Two-loop fermion self-energy in reduced quantum electrodynamics and application to the ultrarelativistic limit of graphene,}
Phys.\ Rev.\ D {\bf 89}, no. 6, (2014) 065038 
\url{https://doi.org/10.1103/PhysRevD.89.065038}


\bibitem{Teber:2014hna} 
S.~Teber,
{\em Two-loop fermion self-energy and propagator in reduced QED$_{3,2}$,}
Phys.\ Rev.\ D {\bf 89}, no. 6, (2014) 067702, 
\url{https://doi.org/10.1103/PhysRevD.89.067702}
 
\bibitem{Teber:2018goo}
S.~Teber and A.~Kotikov,
{\em Field theoretic renormalization study of reduced quantum electrodynamics and applications to the ultrarelativistic limit of Dirac liquids,}
Phys. Rev. D \textbf{97}, no.7, (2018) 074004.
\url{https://doi.org/10.1103/PhysRevD.97.074004}
  
\bibitem{Alves:2013bna}
V. Alves, W. S. Elias, L. O. Nascimento, V. Juri\u{c}i\'{c} and F. Pe\~na,
{\em ``Chiral symmetry breaking in the pseudo-quantum electrodynamics,''}
Phys. Rev. D \textbf{87}, no.12, (2013) 125002.
\url{https://doi.org/10.1103/PhysRevD.87.125002}

\bibitem{Kotikov:2016yrn}
A.~Kotikov and S.~Teber,
{\em Critical behaviour of reduced QED$_{4,3}$ and dynamical fermion gap generation in graphene,}
Phys. Rev. D \textbf{94}, no.11, (2016) 114010.
\url{https://doi.org/10.1103/PhysRevD.94.114010}

\bibitem{Nascimento:2015ola}
L. O. Nascimento, V. Alves, F. Pe\~na, C. M. Smith and E. Marino,
{\em Chiral-symmetry breaking in pseudoquantum electrodynamics at finite temperature,}
Phys. Rev. D \textbf{92}, (2015) 025018.
\url{https://doi.org/10.1103/PhysRevD.92.025018}

\bibitem{Baez:2020dbe} 
J. B\'aez-Cuevas, A. Raya and J. C. Rojas, {\em Chiral symmetry restoration in reduced QED at finite temperature in the supercritical coupling regime},
Phys. Rev. D {\bf 102} (2020), 056020.
\url{https://doi.org/10.1103/PhysRevD.102.056020}

\bibitem{CarringtonRQED} M. E. Carrington, {\em Effect of a Chern-Simons term on dynamical gap generation in graphene}, Phys. Rev. B {\bf 99} (2019) 115432. \url{https://doi.org/10.1103/PhysRevB.99.115432}

\bibitem{Olivares:2020eko}
J.~A.~C.~Olivares, L.~Albino, A.~J.~Mizher and A.~Raya,
{\em Influence of a Chern-Simons term in the dynamical fermion masses in reduced or pseudo QED,}
Phys. Rev. D \textbf{102} (2020) no.9, 096023
\url{https://doi.org/10.1103/PhysRevD.102.096023}
[arXiv:2009.08484 [hep-ph]].

\bibitem{Magalhaes:2020nlc}
G.~C.~Magalh\~aes, V.~Alves, E.~C.~Marino and L.~O.~Nascimento,
{\em Pseudo Quantum Electrodynamics and Chern-Simons theory Coupled to Two-dimensional Electrons,}
Phys. Rev. D \textbf{101} (2020) no.11, 116005
\url{https://doi.org/10.1103/PhysRevD.101.116005}.

\bibitem{Carrington:2020qfz}
M.~E.~Carrington, A.~R.~Frey and B.~A.~Meggison,
{\em Effect of anisotropy on phase transitions in graphene},
Phys. Rev. B \textbf{102} (2020) no.12, 125427.
\url{https://doi.org/10.1103/PhysRevB.102.125427}

\bibitem{RQEDcurved} P. I. C. Caneda and G. Menezes, {\em Reduced quantum electrodynamics in curved space }, Phys. Rev. D{\bf 103}, (2021) 065010. \url{https://doi.org/10.1103/PhysRevD.103.065010}

\bibitem{Landau:1955zz}
  L. D. Landau and I. M. Khalatnikov,
  {\em The gauge transformation of the Green function for charged particles,}
  Sov.\ Phys.\ JETP {\bf 2} (1956) 69
   [Zh.\ Eksp.\ Teor.\ Fiz.\  {\bf 29} (1955) 89].
 
\bibitem{Fradkin:1955jr} 
 E. S. Fradkin,
{\em Concerning some general relations of quantum electrodynamics},
  Zh.\ Eksp.\ Teor.\ Fiz.\  {\bf 29}, 258 (1955)
  [Sov.\ Phys.\ JETP {\bf 2}, 361 (1956)].


\bibitem{Bashir:2000} A. Bashir, {\em Non-perturbative Fermion Propagator for the Massless Quenched QED3} Phys. Lett. B{\bf 491}, (2000) 280.
\url{https://doi.org/10.1016/S0370-2693(00)01043-1}

\bibitem{Bashir:2002sp} 
A.~Bashir and A.~Raya,
{\em Landau-Khalatnikov-Fradkin transformations and the fermion propagator in quantum electrodynamics,}
Phys.\ Rev.\ D {\bf 66}, (2002) 105005.
\url{https://doi.org/10.1103/PhysRevD.66.105005}
  
    
\bibitem{Bashir:2004rg} 
A.~Bashir and A.~Raya,
{\em Fermion propagator in quenched QED3 in the light of the Landau-Khalatnikov-Fradkin transformation,}
Nucl.\ Phys.\ Proc.\ Suppl.\  {\bf 141}, (2005) 259 
\url{https://doi.org/10.1016/j.nuclphysbps.2004.12.039}

\bibitem{AftabLKFT} A. Ahmad, J. J. Cobos-Mart\'{\i}nez, Y. Concha-S\'anchez, and A. Raya, {\em Landau-Khalatnikov-Fradkin transformations in reduced quantum electrodynamics}, Phys. Rev. D {\bf 93}, (2016) 094035. 
\url{https://doi.org/10.1103/PhysRevD.93.094035}

\bibitem{Dudal:2018pta}
D. Dudal, A. J. Mizher and P. Pais,
{\em Exact quantum scale invariance of three-dimensional reduced QED theories},
Phys. Rev. D {\bf 99} (2019) no.4, 045017-
\url{https://doi.org/PhysRevD.99.045017}

\bibitem{Bollini:1991fp}
C. G. Bollini and J. J. Giambiagi,
{\em Arbitrary powers of D\textquoteright{}Alembertians and the Huygens' principle,}
J. Math. Phys. {\bf 34} (1993), 610-621
\url{https://doi.org/10.1063/1.530263}

\bibitem{Coste:1989wf}
A.~Coste and M.~Luscher,
{\em Parity Anomaly and Fermion Boson Transmutation in Three-dimensional Lattice QED},
Nucl. Phys. B {\bf 323} (1989), 631-659
\url{https://doi.org/10.1016/0550-3213(89)90127-2}

\bibitem{CraigReviewSDE}
C. D. Roberts and A. G. Williams
{\em Dyson-Schwinger Equations and the Application to Hadronic Physics} Prog. Part. Nucl. Phys. {\bf 33}  (1994) 477. \url{https://doi.org/10.1016/0146-6410(94)90049-3}

\bibitem{Mike} M. R. Pennington {\em Swimming with Quarks}, J. Phys. Conf. Ser. {\bf 18} (2005) 1. 
\url{https://doi.org/10.1088/1742-6596/18/1/001}

\bibitem{ward}  J.~C. Ward, {\em An Identity in Quantum Electrodynamics}, Phys. Rev.  {\bf 78} (1950) 182  \url{https://doi.org/10.1103/PhysRev.78.182}

\bibitem{green} H.~S. Green, {\em A Pre-renormalized quantum electrodynamics}, Proc. Phys. Soc. (London) A{\bf 66}, (1953) 873. \url{https://doi.org/10.1088/0370-1298/66/10/303}

\bibitem{takahashi}  Y. Takahashi, {\em On the generalized ward identity}, Nuovo Cimento {\bf 6}, (1957) 371. \url{https://doi.org/10.1007/BF02832514}

\bibitem{twi1}  Y. Takahashi, {\em Canonical quantization and generalized Ward relations: Foundation of nonperturbative
approach}, Print-85-0421 (Alberta), (1985).

\bibitem{MiranskyCPT} V. Miransky and K. Yamawaki, {\em Conformal phase transition in gauge theories} Phys. Rev. D {\bf 55}, (1997) 5051. \url{https://doi.org/10.1103/PhysRevD.55.5051}.

\bibitem{BRH} A. Bashir, A. Huet, and A. Raya {\em Gauge dependence of mass and condensate in chirally asymmetric phase of quenched three-dimensional QED }
Phys. Rev. D {\bf 66}, (2002) 025029. \url{https://doi.org/10.1103/PhysRevD.66.025029}.

\bibitem{multiple} K. Raya, A. Bashir, S. Hern\'andez-Ortiz, A. Raya, and C. D. Roberts, 
{\em Multiple solutions for the fermion mass function in QED3}, Phys. Rev. D{\bf 88}, (2013) 096003. \url{https://doi.org/10.1103/PhysRevD.88.096003}

\bibitem{appel1} T. Appelquist, M. J. Bowick, E. Cohler, and L. C. R. Wijewardhana {\em Chiral-symmetry breaking in 2+1 dimensions}
Phys. Rev. Lett. {\bf 55} (1985) 1715. \url{https://doi.org/10.1103/PhysRevLett.55.1715}.

\bibitem{appel2} T. W. Appelquist, M. Bowick, D. Karabali, and L. C. R. Wijewardhana {\em Spontaneous chiral-symmetry breaking in three-dimensional QED}
Phys. Rev. D {\bf 33} (1986) 3704. \url{https://doi.org/10.1103/PhysRevD.33.3704}.

\bibitem{appel3} T. Appelquist, D. Nash, and L. C. R. Wijewardhana, {\em Critical Behavior in (2+1)-Dimensional QED } Phys. Rev. Lett. {\bf 60} (1988) 2575. \url{https://doi.org/10.1103/PhysRevLett.60.2575}

\bibitem{oldsaul} A. Bashir, A. Raya, S. S\'anchez-Madrigal and C.D. Roberts,  {\em Gauge Invariance of a Critical Number of Flavours in QED3.} Few-Body Syst {\bf 46}  (2009) 229. \url{https://doi.org/10.1007/s00601-009-0069-9}.

\bibitem{critQED3T1} 
A. V. Kotikov and S. Teber, {\em Critical Behavior of (2+1)-Dimensional QED: $1/N$ Expansion}, Particles 2020, 3, 345-354. 
\url{https://doi.org/10.3390/particles3020026}

\bibitem{critQED3T2} A. V. Kotikov and Sofian Teber, {\em Critical behaviour of (2+1)-dimensional QED: 1/N-corrections}, EPJ Web of Conferences {\bf 138}  (2017), 06005. 
\url{https://doi.org/10.1051/epjconf/201713806005}

\bibitem{critQED3T3} A. V. Kotikov and S. Teber, {\em Critical behavior of (2+1)-dimensional QED: $1/N_f$ corrections in an arbitrary nonlocal gauge}, Phys. Rev. D{\bf 94}, (2016) 114011. \url{https://doi.org/10.1103/PhysRevD.94.114011}

\bibitem{critQED3T4} A. V. Kotikov, V.I. Shilin and S. Teber, {\em Critical behavior of (2+1)-dimensional QED: $1/N_f$ corrections in the Landau gauge},  Phys.Rev.D 94 (2016) 5, 056009. \url{https://doi.org/10.1103/PhysRevD.94.056009}

\bibitem{Juan} J. A. Casimiro-Olivares {Generaci\'on din\'amica de Masas en R(P)QED} M. Sc. Thesis (In Spanish) UMSNH (2019).

\bibitem{BallChiu} J. S. Ball and T.-W. Chiu {\em Analytic properties of the vertex function in gauge theories. II}
Phys. Rev. D {\bf 22} (1981) 2550. \url{https://doi.org/10.1103/PhysRevD.22.2550}


\bibitem{revT} S. Metayer and S. Teber {\em Two-loop mass anomalous dimension in reduced quantum electrodynamics and application to dynamical fermion mass generation} e-Print: 2107.07807 [hep-th] \url{https://arxiv.org/abs/2107.07807}

\bibitem{Ahmad:2015cgh}
A.~Ahmad, A.~Ayala, A.~Bashir, E.~Guti\'errez and A.~Raya,
{\em QCD Phase Diagram and the Constant Mass Approximation,}
J. Phys. Conf. Ser. \textbf{651}, no.1, (2015) 012018 
\url{https://doi.org/10.1088/1742-6596/651/1/012018}

\bibitem{CSterm}  A. Khare, {\em Fractional Statistics and Quantum Theory}
(World Scientific, Singapore, 2005), 2nd ed.

\bibitem{Kondo:1994bt}
K. I. Kondo and P. Maris,
{\em First-order phase transition in three-dimensional QED with Chern-Simons term}
Phys. Rev. Lett. {\bf 74}(1995) 18.
\url{https://doi.org/10.1103/PhysRevLett.74.18}

\bibitem{Kondo:1994cz}
K. I. Kondo and P. Maris,
{\em Spontaneous chiral symmetry breaking in three-dimensional QED with a Chern-Simons term,}
Phys. Rev. D \textbf{52}, (1995) 1212.
\url{https://doi.org/10.1103/PhysRevD.52.1212}

\bibitem{christoph1}
C. P. Hofmann, A. Raya and Sa\'ul S\'anchez-Madrigal, {\em Confinement in Maxwell-Chern-Simons planar quantum
electrodynamics and the $1/N$ approximation}
Phys. Rev. D {\bf 82},  (2010) 096011. \url{https://doi.org/10.1103/PhysRevD.82.096011}

\bibitem{christoph2} S. S\'anchez Madrigal, C. P. Hofmann and A Raya, {\em Dynamical Mass Generation and Confinement in Maxwell-Chern-Simons Planar Quantum
Electrodynamics} J. Phys.Conf. Ser. {\bf 287} (2011) 012028. 
\url{https://doi.org/10.1088/1742-6596/287/1/012028}

\bibitem{Ozela:2021pse}
R.~F.~Ozela, V.~Alves, G.~C.~Magalh\~aes and L.~O.~Nascimento, {\em Effects of the pseudo-Chern-Simons action for strongly correlated electrons in a plane}
Phys. Rev. D \textbf{105} (2022) no.5, 056004
\url{https://doi.org/10.1103/PhysRevD.105.056004}

\bibitem{marino5}
E. C. Marino and Leandro O. Nascimento and Van Sérgio Alves and N. Menezes and C. Morais Smith, 
{\em Quantum-electrodynamical approach to the exciton spectrum in transition-metal dichalcogenides}, 
2D Materials, {\bf 5}, 4 (2018).
\url{https://doi.org/10.1088%2F2053-1583%2Faacc3f}



\bibitem{Dudal:2018mms}
D.~Dudal, A.~J.~Mizher and P.~Pais,
{\em Remarks on the Chern-Simons photon term in the QED description of graphene} 
Phys. Rev. D {\bf 98} (2018) 065008.
\url{https://doi.org/10.1103/PhysRevD.98.065008}

\bibitem{Naumis} G. Garc\'{\i}a Naumis, {\em Electronic properties of 2D materials and its heterostructures: a minimal review }, Rev. Mex. Fis. {\bf 67} (5), (2021) 1. 
\url{https://doi.org/10.31349/RevMexFis.67.050102}

\bibitem{twi2} K.-I. Kondo, {\em Transverse Ward-Takahashi Identity, Anomaly and Schwinger-Dyson Equation}, Int. J. Mod. Phys. A{\bf 12}, (1997) 5651. \url{https://doi.org/10.1142/S0217751X97002978}

\bibitem{twi3} H.-X. He, F. Khanna and Y. Takahashi, {\em Transverse Ward-Takahashi identity for the fermion-boson vertex in gauge theories}, Phys. Lett. B{\bf 480}, (2000) 222. \url{https://doi.org/10.1016/S0370-2693(00)00353-1}

\bibitem{twi4}  H.-X. He, {\em Transverse vector vertex function and transverse Ward-Takahashi relations in QED}, Commun. Theor. Phys. {\bf 46}, (2006) 109. \url{https://doi.org/10.1088/0253-6102/46/1/025}

\bibitem{twi5} H.-X. He, {\em Transverse ward-takahashi relation for the fermion-boson vertex function in four-dimensional abelian gauge theory}, Int. J. Mod. Phys. A{\bf 22}, (2007) 2119. 
\url{https://doi.org/10.1142/S0217751X07036257}

\bibitem{LKF1} K. Johnson and B. Zumino, {\em Gauge Dependence of the Wave-Function Renormalization Constant in Quantum Electrodynamics}, Phys. Rev. Lett.{\bf 3}, (1959) 351. \url{https://doi.org/10.1103/PhysRevLett.3.351}

\bibitem{LKF2} B. Zumino, {\em Gauge properties of propagators in quantum electrodynamics}, J. Math. Phys. {\bf 1},  (1960) 1. \url{https://doi.org/10.1063/1.1703632}

\bibitem{LKF3} S. Okubo, {\em The gauge properties of Green's functions in quantum electrodynamics}, Nuovo Cim. {\bf 15}, (1960) 949. \url{https://doi.org/10.1007/BF02860201}


\bibitem{LKF4} I. Bialynicki-Birula, {\em On the Gauge Properties of Green's Functions}, Nuovo Cim. {\bf 17}, (1960) 951. \url{https://doi.org/10.1007/BF02732140}

\bibitem{LKF5} H. Sonoda, {\em On the gauge parameter dependence of QED}, Phys. Lett. B {\bf 499} (2001) 253.
\url{https://doi.org/10.1016/S0370-2693(01)00030-2}

\bibitem{Fernandez-Rangel:2016zac} 
 L.~A.~Fernandez-Rangel, A.~Bashir, L.~X.~Gutierrez-Guerrero and Y.~Concha-Sanchez,
 {\em Constructing Scalar-Photon Three Point Vertex in Massless Quenched Scalar QED,}
 Phys.\ Rev.\ D {\bf 93}, no. 6, (2016) 065022.
 \url{https://doi.org/10.1103/PhysRevD.93.065022}
 
\bibitem{Ahmadiniaz:2015kfq}
N.~Ahmadiniaz, A.~Bashir and C.~Schubert,
 {\em Multiphoton amplitudes and generalized LKF transformation in Scalar QED,}
 Phys.\ Rev.\ D {\bf 93} (2016) 4,  045023
 \url{https://doi.org/10.1103/PhysRevD.93.045023}
 
\bibitem{Kizilersu:2009kg} 
 A.~Kizilersu and M.~R.~Pennington,
{\em Building the Full Fermion-Photon Vertex of QED by Imposing Multiplicative Renormalizability of the Schwinger-Dyson Equations for the Fermion and Photon Propagators,}
Phys.\ Rev.\ D {\bf 79}, (2009) 125020.
\url{https://doi.org/10.1103/PhysRevD.79.125020}

\bibitem{Bashir:2001vi} 
 A.~Bashir and A.~Raya,
{\em Constructing the fermion boson vertex in QED(3),}
Phys.\ Rev.\ D {\bf 64}, (2001) 105001.
\url{https://doi.org/10.1103/PhysRevD.64.105001}



\bibitem{TeberLK1} A.V. Kotikov and S. Teber {\em Landau-Khalatnikov-Fradkin transformation and the mystery of even 
$\zeta$-values in Euclidean massless correlators}
Phys. Rev. D {\bf 100}, (2019) 105017. \url{https://doi.org/10.1103/PhysRevD.100.105017}

\bibitem{TeberLK2} A. James, A. V. Kotikov, and S. Teber {\em Landau-Khalatnikov-Fradkin transformation of the fermion propagator in massless reduced QED }
Phys. Rev. D {\bf 101} (2020) 045011. \url{https://doi.org/10.1103/PhysRevD.101.045011}

\bibitem{TeberLK3} A. V. Kotikov and S. Teber, {\em Landau-Khalatnikov-Fradkin Transformation and Hatted $\zeta$-Values}
Phys. Part.Nucl. {\bf 51},  (2020) 562. \url{https://doi.org/10.1134/S1063779620040425}

\bibitem{TeberLK4} V. P. Gusynin, A. V. Kotikov, and S. Teber, {\em Landau-Khalatnikov-Fradkin transformation in three-dimensional quenched QED}
Phys. Rev. D {\bf 102}, (2020) 025013 \url{https://doi.org/10.1103/PhysRevD.102.025013}


\bibitem{Dudal:2021ret}
D.~Dudal, F.~Matusalem, A.~J.~Mizher, A.~R.~Rocha and C.~Villavicencio, {\em Half-integer anomalous currents in 2D materials from a QFT viewpoint}
Sci. Rep. \textbf{12} (2022) no.1, 5439
\url{https://doi.org/10.1038/s41598-022-09483-4}






  



 
\end{thebibliography}
\end{document}